\documentclass[preprint,authoryear,12pt]{elsarticle}

\usepackage{amsmath}

\usepackage{bm}

\newcommand{\ds}{\displaystyle}	

\usepackage{graphicx}
\usepackage[font=footnotesize]{caption}
\usepackage[font=footnotesize]{subcaption}
\usepackage{placeins}

\usepackage[table]{xcolor}
\usepackage{booktabs}
\usepackage{tabu}
\usepackage{threeparttable}
\usepackage{multirow}

\usepackage{xpatch}
\makeatletter
\chardef\TPT@@@asteriskcatcode=\catcode`*
\catcode`*=11
\xpatchcmd{\threeparttable}
  {\TPT@hookin{tabular}}
  {\TPT@hookin{tabular}\TPT@hookin{tabu}}
  {}{}
\catcode`*=\TPT@@@asteriskcatcode
\makeatother

\usepackage{array}
\newcolumntype{P}[1]{>{\centering\arraybackslash}p{#1}}

\usepackage{algorithm}
\usepackage[noend]{algpseudocode}
\usepackage{setspace}

\usepackage{fullpage}
\usepackage{ulem}
\usepackage{outlines}
\usepackage{enumitem}
\setenumerate[1]{label=\Roman*.}
\setenumerate[2]{label=\Alph*.}
\setenumerate[3]{label=\roman*.}
\setenumerate[4]{label=\alph*.}

\usepackage[%
breaklinks=true,%
colorlinks=true,%
pdfauthor={First Author et al.},%
pdftitle={Template for manuscripts in Advances in Space Research}%
]{hyperref}

\journal{Advances in Space Research}

\begin{document}

\begin{frontmatter}

\title{Assessing and Minimizing Collisions in \\ Satellite Mega-Constellations}

\author{Nathan Reiland\corref{cor}}
\cortext[cor]{Corresponding author}
\ead{nreiland@email.arizona.edu}
\author{Aaron J. Rosengren}
\ead{ajrosengren@email.arizona.edu}
\address{Aerospace and Mechanical Engineering, The University of Arizona, Tucson, AZ 85721, USA}
\author{Renu Malhotra}
\ead{renu@lpl.arizona.edu}
\address{Lunar and Planetary Laboratory, The University of Arizona, Tucson, AZ 85721, USA}
\author{Claudio Bombardelli}
\ead{claudio.bombardelli@upm.es}
\address{Space Dynamics Group, Technical University of Madrid, 2040 Madrid, Spain}

\begin{abstract}

We aim to provide satellite operators and researchers with an efficient means for evaluating and mitigating collision risk during the design process of mega-constellations. We first establish a baseline for evaluating various techniques for close-encounter prediction and collision-probability calculation (Hoots et al. 1984, Gronchi 2005, JeongAhn and Malhotra 2015) by carrying out brute-force numerical simulations and using a sequence of filters to greatly reduce the computational expense of the algorithm. Next, we estimate conjunction events in the orbital environment following the anticipated deployments of the OneWeb LEO and SpaceX Starlink mega-constellations. As a final step, we investigate Minimum Space Occupancy (MiSO) orbits (Bombardelli et al. 2018), a generalization of the well-known frozen orbits that account for the perturbed-Keplerian dynamics of the Earth- Moon-Sun-satellite system. We evaluate the ability of MiSO configurations of the proposed mega-constellations, as suggested by Bombardelli et al. 2018, to reduce the risk of endogenous (intra-constellation) collisions. The results indicate that the adoption of the MiSO orbital configuration can significantly reduce risk with nearly indistinguishable adjustments to the nominal orbital elements of the constellation satellites.									

\end{abstract}

\begin{keyword}
Mega-Constellations;
Satellite conjunction; 
Space debris; 
Frozen orbits; 
Dynamical evolution and stability
\end{keyword}

\end{frontmatter}

\parindent=0.5 cm

\section{Introduction}

One of the foremost space science and engineering issues facing society today is conquering Earth's space junk problem, being paramount to managing the increasing orbital traffic in near-Earth space and safeguarding satellite operations \citep{aW18}. This pressing problem is fundamentally connected with the modern fields of space situational awareness (SSA) and space traffic management (STM), which integrate many traditional areas of space research into a single focused topic \citep{NRC, AFSPC}. A major challenge is predicting with sufficient accuracy the location and collision risks of all significant resident space objects (RSOs), a problem that has been compounded in recent years with the launch of numerous small satellites by many nations and the proliferation of orbital debris. The ``Kessler syndrome'' of collisional cascading, whereby random collisions are predicted to produce new debris at a rate that is greater than the removal rate due to orbital decay, is a more realistic scenario now than when it was first proposed in the late 1970s \citep{dK78}. Another emerging concern is the robustness of the current debris mitigation guidelines developed by the Inter-Agency Space Debris Coordination Committee \citep{bB16}, which were based on the continuation of space traffic at the rates observed in the 1990s. Space traffic, mostly driven by geopolitical and economic factors, has always been subject to considerable fluctuations, but all indications point to a significant increase of traffic in low-Earth orbit (LEO), the most densely populated orbital lanes. OneWeb, SpaceX, and Amazon, in particular, have each submitted ambitious plans to place thousands of satellites in LEO to provide low-latency broadband internet to the world. The full deployment of these ``mega-constellations'' represents hitherto unknown challenges to the Earth's most congested and contested orbital environment (see, e.g., Fig.~\ref{fig:leodistrib} in Section~\ref{sec:setup}).

While previous techniques were sufficient to handle past SSA needs, future demands will require new algorithms to build and maintain an expanded space object catalog (SOC), both in quiescent operations and in the presence of a debris-generating event. The only instance of a satellite-satellite collision has been the famed Iridium-Cosmos; given accurate states and detailed maneuver histories of all LEO satellites pre collision, however, the natural question is whether the existing techniques could have actually pinpointed the doom of Iridium. Quite likely not, since previous methods have largely resorted to handwavy probability calculations and simplified dynamical models, being developed at a time when high-fidelity orbit propagations were too computationally expensive. For example, the satellites of proposed mega-constellations, such as OneWeb, are placed in nearly intersecting orbits by design; yet, classic analytical, collision-probability techniques, like those of \citet{eO51} and \citet{gW67}, will indicate that these closely-spaced objects are at high risk \citep{cR17, sL18, cP20}. Whether there is a true conjunction or collision risk requires the use of accurate orbit and state uncertainty propagation of these satellites; an otherwise brute-force approach.  

The space object catalog currently contains around 20,000 objects, and when the planned ``space fence'' radar network becomes operational this number is expected to exceed one hundred thousand \citep{NRC, AFSPC}. It has often been assumed that even the current SOC is too large to permit the application of a highly accurate, brute-force approach to close-approach prediction and collision-probability estimation. Previous similar algorithms, although extremely efficient, such as T.S. Kelso's \texttt{SOCRATES}\footnote{\url{https://www.celestrak.com/SOCRATES/}}, have instead relied on antiquated simplified analytical orbit propagators that were developed at the time of punch cards and limited computing power \citep{fH04, tsK09}. A more recent tool, the ``Conjunction Streaming Service Demo'' contained in Moriba Jah's \texttt{ASTRIAGraph}\footnote{\url{http://astria.tacc.utexas.edu/AstriaGraph/}} also currently relies on these simplified propagators, which, together with the two-line elements themselves, cannot yield the required accuracy needed. While a systematic and more rigorous study also accounting for the uncertainty in the state-estimation of each object would represent a formidable task with significant computational requirements,\footnote{A task that is likely being carried out by LeoLabs, although the inner workings of their conjunction assessment service is not entirely known \citep{mN17}.} the problem becomes tractable if modern developments in astrodynamics (i.e., regularization, nonsingular orbital element formulations, perturbed collision-probability algorithms, etc.) are properly leveraged with sophisticated computing resources. Such an approach, moreover, can serve as a truth model for assessing close encounters and collision probabilities.

Here, we demonstrate how several new astrodynamical tools could be used by satellite operators and other researchers to investigate the ``safety'' (i.e., collision risk) of mega-constellations or any arbitrary set of artificial satellites and debris. First, we develop a new brute-force, close-approach prediction algorithm utilizing regularized equations of motion to establish ``truth'' for the number and severity of endogenous and exogenous satellite-satellite and satellite-debris close approaches experienced by the nominal orbital planes within the OneWeb LEO and SpaceX Starlink constellations.\footnote{It is important to note that the ``real'' initial placement of these mega-constellation satellites will likely be quite different than those listed in the FCC reports and consequently the goal of this work is not to criticize a ``dummy'' constellation or even the actual (ever changing and often unknown) orbital designs of the real constellation.} The performance of the \citet{fH84} and \citet{rM15, rM17} close-approach-probability algorithms are then evaluated against this baseline. As a final step, and as suggested by \citet{cB18}, we consider Minimum Space Occupancy (MiSO) variants of the nominal\footnote{The nominal configurations are those detailed in the most recent FCC filings (no. SAT-LOI-20160428-00041 and SAT-MOD-20181108-00083, respectively).} mega-constellations, which consist of redistributing satellites of different orbital planes in non-overlapping MiSO shells with an altitude separation of 600 meters. The effectiveness of these new design solutions in reducing the number of critical conjunctions is evaluated using the aforementioned algorithms.

\section{Problem formulation}

\subsection{THALASSA, a numerical tool for accurate and rapid orbit propagation}
\label{sec:thalassa}

Earth-satellite dynamics is best described by the perturbed two-body problem, which, in Cartesian coordinates, can be stated as $\ddot{\bm{r}} = -(\mu/r^3) \bm{r} + \bm{F}$, where $-(\mu/r^3) \bm{r}$ is the primary (Keplerian) acceleration and $\bm{F}$ is the vector sum of perturbing accelerations due to the non-sphericity of the Earth's gravitational field, the gravity of external ``third'' bodies (i.e., lunisolar perturbations), atmospheric drag, solar radiation pressure, etc.

Accurate numerical solutions to the perturbed-Kepler problem are often generated through Cowell's method; that is, the integration of the equations of motion in Cartesian coordinates with a numerical solver (the most basic formulation of special perturbation theory). This approach, while simple and robust, is computationally inefficient. In particular, the presence of a singularity causes large oscillations in the magnitude of the right-hand side, which are aggravated with increasing eccentricity and unstable error propagation characteristics \citep{vrB96}. These disadvantages can be mitigated or eliminated altogether by employing equations of motion (or formulations of the perturbed two-body problem) that have been regularized.

\begin{figure}[ht!]
	\centering
	\includegraphics[width=1.75in]{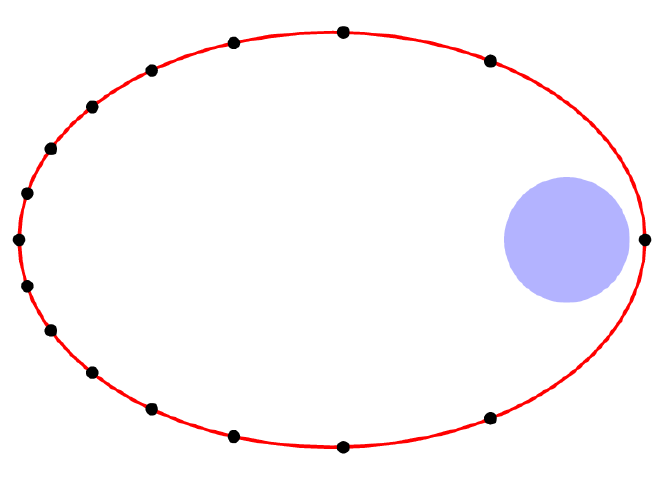}
	\hspace{3pt}
	\includegraphics[width=1.75in]{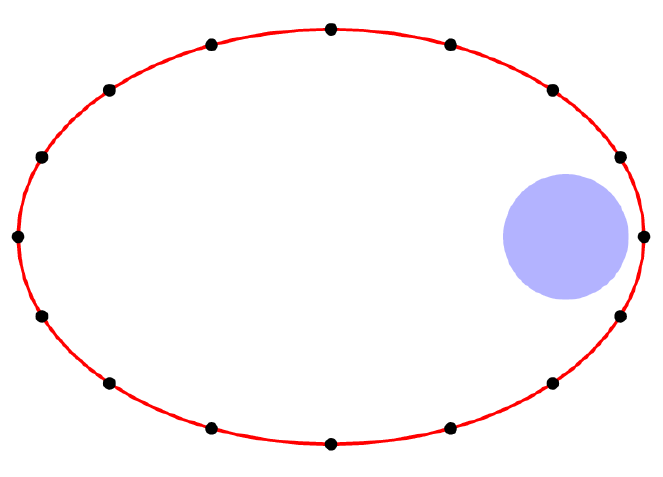}
	\caption{Uniform spacing of points along an orbit in physical ({\it left}) and fictitious ({\it right}) time coinciding with the eccentric anomaly. 
	Adapted from \citet{mB02}.}
	\label{fig:ellipses}
\end{figure}

In regularized formulations, the independent variable is transformed from the physical time to a fictitious time through the generalized Sundman transformation \citep{mB02}. Using fictitious time as the independent variable gives an immediate advantage: since the fictitious time is an angle-like quantity, meshing the orbit uniformly results in a distribution of points whose density can be adjusted by appropriately choosing numerical parameters. One can select, in particular, a uniform distribution rather than one that is densest at apoapsis, as in Fig.~\ref{fig:ellipses}. Regularized equations are also stable with respect to the propagation of numerical error, unlike the Cowell method \citep{jR17}, and can be linearized without expressing the perturbations explicitly, thus with no need to truncate expansions in perturbation parameters \citep{eSgS71}. Variation of parameters or projective decomposition can also be employed to obtain regularized, nonsingular sets of orbital elements, which are particularly advantageous for weak perturbations \citep{jP07, gB15}.

A collection of regularized formulations is contained in the \texttt{THALASSA} Earth-satellite orbit propagation tool \citep{dA18_luna3, dA18, dA19}, which is freely available through a GitLab repository.\footnote{\textsc{URL:} \url{https://gitlab.com/souvlaki/thalassa}.} \texttt{THALASSA} uses the variable step-size and order (up to 12\textsuperscript{th}) \texttt{LSODAR} solver to numerically integrate the equations of motion, which automatically selects the solution algorithm between the implicit Adams-Bashforth-Moulton and backwards differentiation formulas. Even when using such a sophisticated, adaptive solver, regularized formulations have been shown by \citet{dA19} to radically improve computational efficiency for long-term propagations. 

\begin{figure}[ht!]
	\centering    
	\includegraphics[width=4.75in]{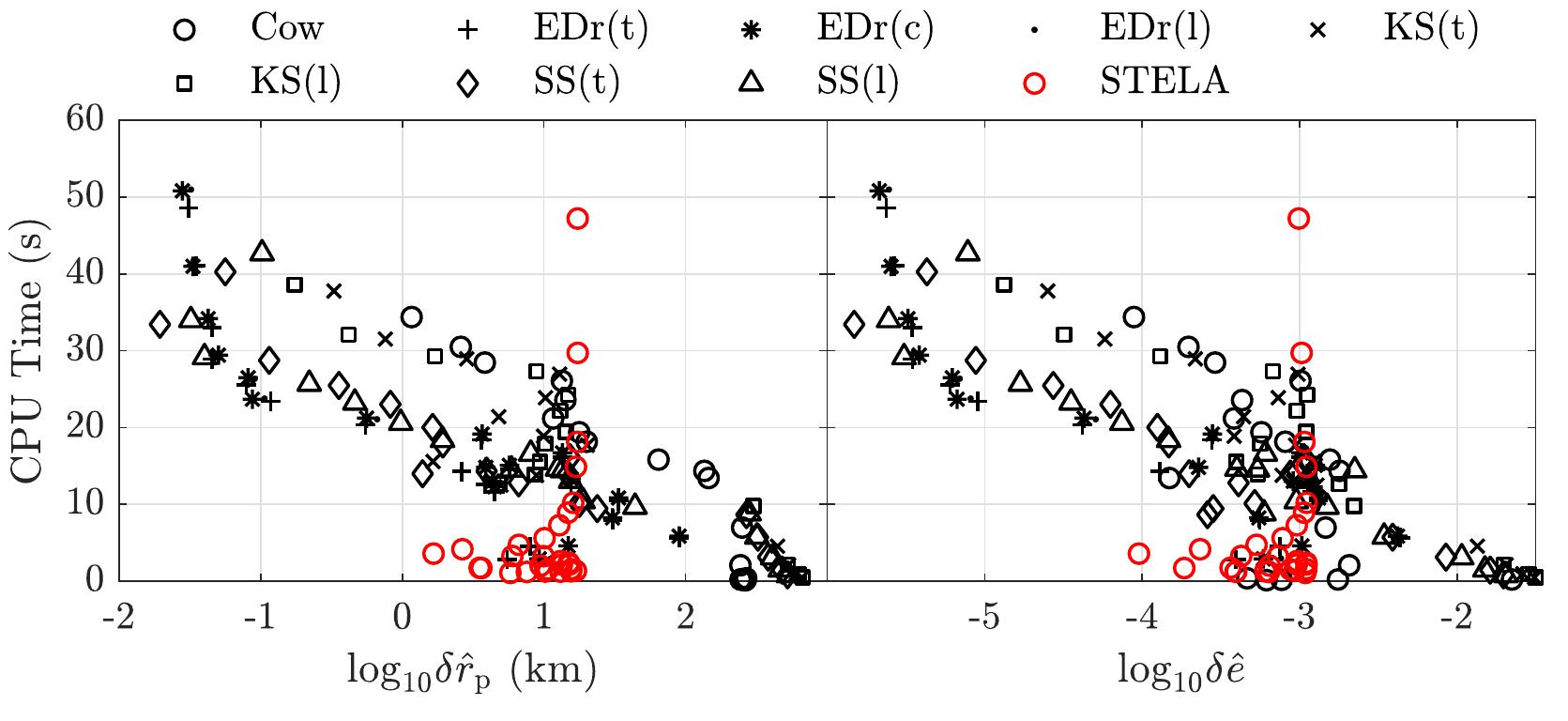}
	\caption{CPU time for 18-year integrations of a LEO against error in orbital elements. Black and red circles denote Cowell's method and \texttt{STELA} propagations, respectively, and other symbols represent various regularized formulations. Adapted from~\cite{dA19}, 
	to which we refer for further details.}
	\label{fig:efficiency}
\end{figure}

Figure~\ref{fig:efficiency} shows that regularized formulations significantly improve performance with respect to Cowell's in the integration of LEOs, while nearly matching the speed of semi-analytical methods. Moreover, semi-analytical methods, such as the widely-used orbit integration software package, \texttt{STELA}, are intrinsically limited in their accuracy due to the approximations introduced in the averaging process \citep{cL14, dA19}. Accordingly, brute-force approaches based on semi-analytical or even less accurate analytical propagators, like \texttt{SGP4} and \texttt{SDP4} for use with the two-line element sets \citep{fH04, cLwM11}, cannot achieve the requisite precision needed to accurately and reliably predict conjunctions.  

The numerical orbit propagation engine in \texttt{THALASSA} implements the Kustaanheimo-Stiefel~\citep[ch. 2]{eSgS71}, Stiefel-Scheifele~\citep[ch. 5]{eSgS71}, Dromo~\citep{gB13,jP07}, and EDromo~\citep{gB15} regularized formulations. These formulations are chosen due to their optimal performance in a wide range of dynamical configurations~\citep{dA19,jR17}. The equations of motion are integrated with adaptive numerical solvers in modern Fortran. 

\subsection{The Rapid-Integrations for Conjunction Assessment (RICA) Algorithm}

\label{sec:bruteForce}
The developed brute-force algorithm leverages the \texttt{THALASSA} orbit propagator~\citep{dA19}, with the EDromo(l) formulation of the equations of motion~\citep{gB15}, shown to be the one of the most efficient and precise for the LEO regime. We adapt this high-fidelity astrodynamics tool to be highly parallelizable, enabling rapid and accurate propagation of thousands of RSOs. 
	
A given set of ``target objects'' and potentially impacting ``field objects'' are passed through three stages of filters that compare the trajectories in order to determine the occurrence of close approaches within some specified distance, $\tau$. First the set $\mathcal{S}$ containing all combinations of potentially colliding objects is defined. Next, the vis-viva energy equation, 
\begin{equation}
    \label{eq:visviva}
    v^2 = \mu\left( \frac{2}{r} - \frac{1}{a} \right),
\end{equation}
is used to calculate the maximum possible relative velocity, $v_\text{max}$, between all objects of the set using the initial osculating orbital elements, where $\mu$ is the gravitational parameter of the primary body (Earth), $r$ and $v$ are the satellite's relative position and velocity, respectively, and $a$ is the semi-major axis (Algorithm~\ref{alg:maxVelocity_pseudo}).

The integration time step is then chosen based on the selected close-approach distance between the target and field objects, $\tau$, according to 
\begin{equation}
    \label{eq:selectTstep}
    t_\text{step} = \frac{1}{f_s}\left( \frac{\tau}{v_\text{max}}\right),
\end{equation}
where $f_s$ is a factor of safety (typically a value of 2 is used). This simple step, whose effect can be seen in Fig.~\ref{fig:exampleOutput}, ensures that no close approaches greater than $\tau$ will be neglected.

The initial states of the target and field objects are then propagated forward over the time span of interest (Algorithm~\ref{alg:propagate_pseudo}) and the Cartesian separation distance (Euclidean norm), $r_\text{sep}$, between each set of objects in $\mathcal{S}$ at each time step is calculated. Objects with trajectories satisfying $r_\text{sep} < \tau$ are passed on to the next stage (Algorithm~\ref{alg:filterStage_pseudo}). Before beginning propagation of the set of objects not eliminated by the previous filter, a new time step of integration is calculated according to Eq.~\ref{eq:visviva} to determine the new $v_{max}$ of the filter stage and a significantly smaller value of $\tau$ is chosen. Performing the computations with this structure greatly increases the efficiency of the algorithm without sacrificing accuracy. 

The structure outlined in Algorithm~\ref{alg:fica_pseudo} was implemented using C programming and the MPI message passing interface for parallelization. The resulting program, \texttt{RICA} (``Rapid Integrations for Collision Assessment''), was run on the University of Arizona's High Performance Computing (HPC) cluster on over 200 CPUs. This structure allows for the efficient parallelization of the propagation and trajectory-comparison portions of the code.

\subsection{Simple Example}

\begin{figure}[htp!]
	\centering
    \includegraphics[trim = 3.6in 1.7in 5.55in 1.7in, clip,scale=0.5,width=0.49\textwidth]{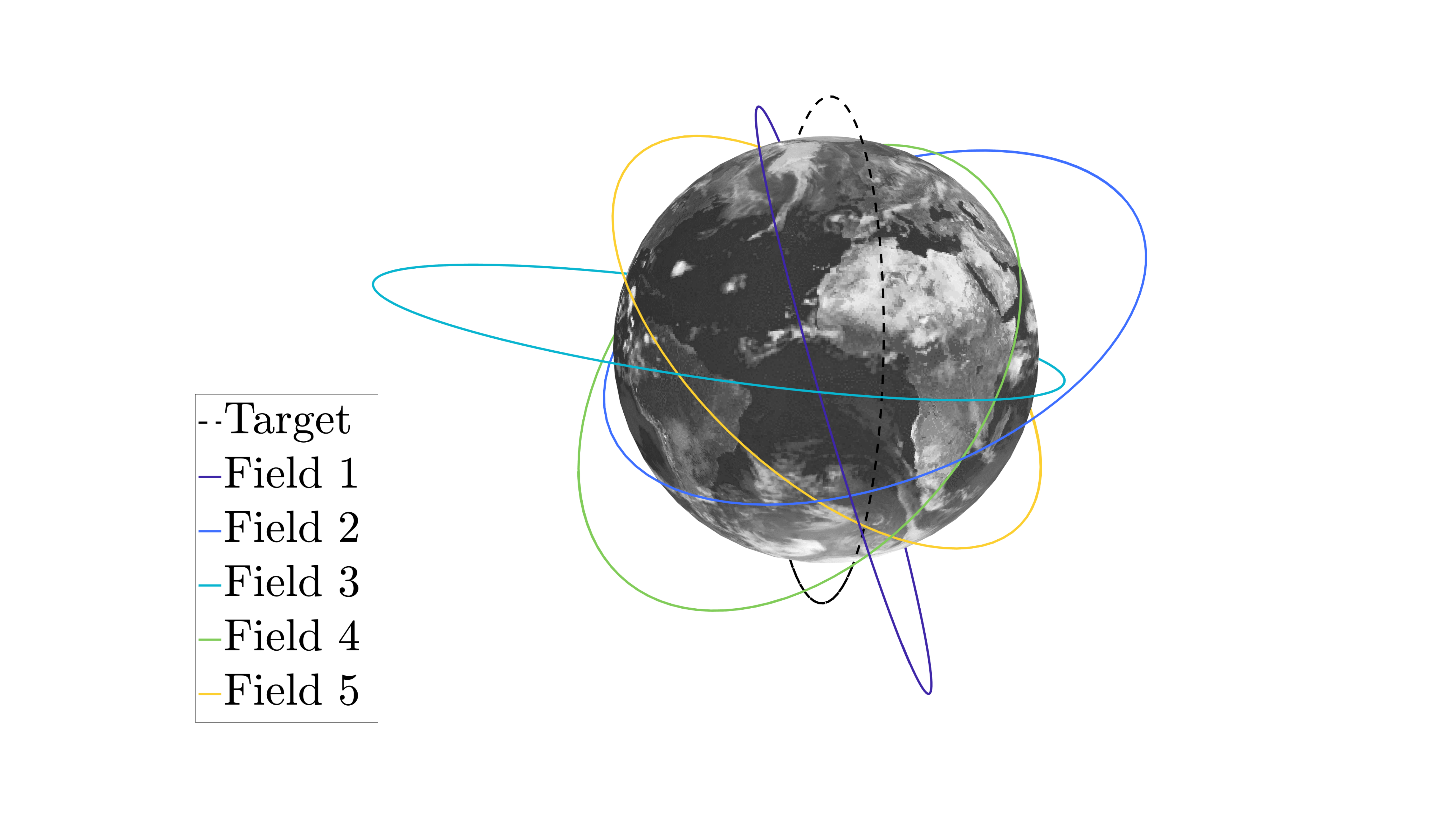}
    \includegraphics[trim = 6.85in 1.7in 2.5in 1.7in, clip,scale=0.5,width=0.49\textwidth]{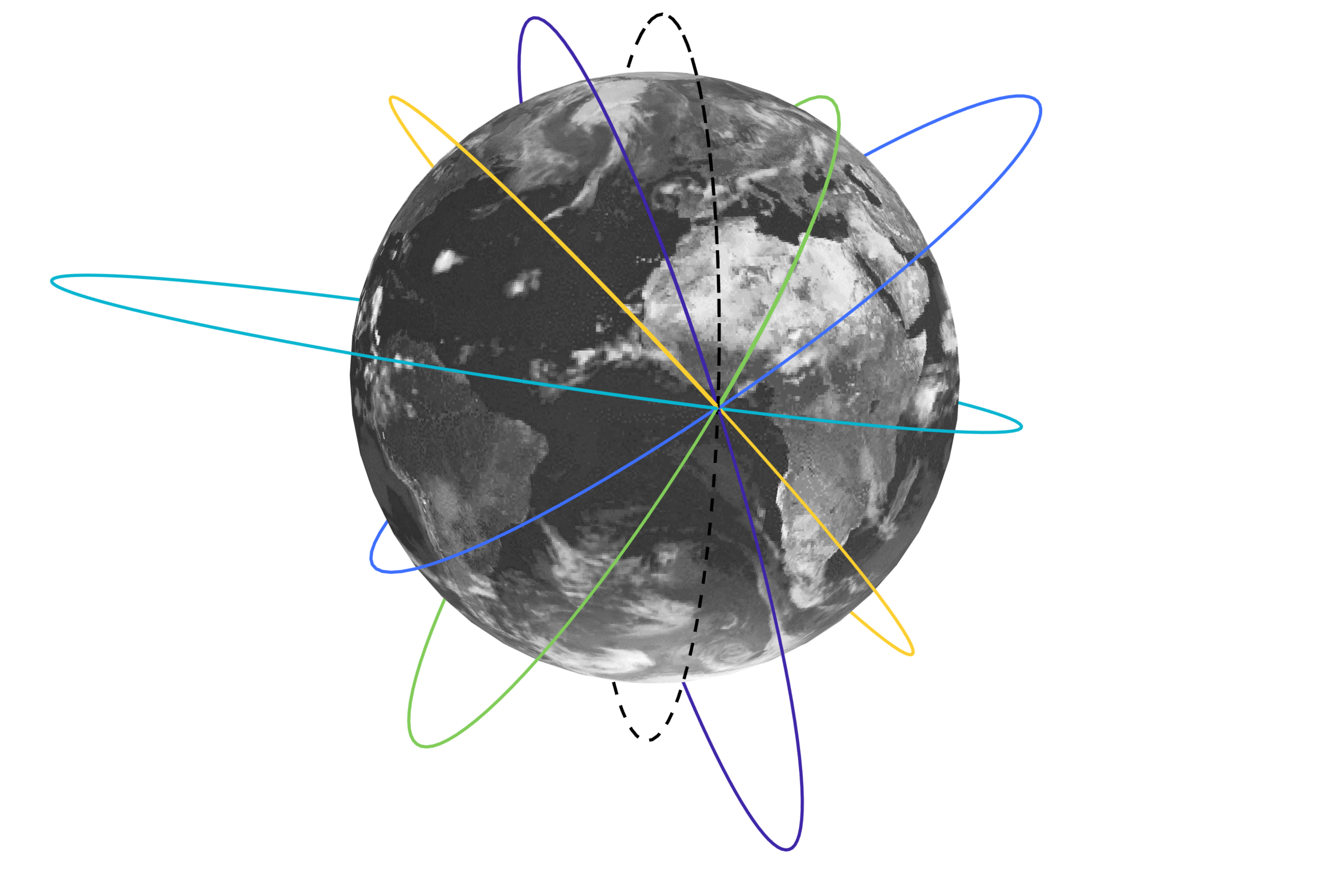}
	\caption{The initial orbits ({\it left}) at the epoch MJD $58834$ and final orbits ({\it right}) at MJD $58849$, the time of manufactured collision, for the \texttt{RICA} example objects of Table~\ref{tab:exampleFica_ICs}.}
	\label{fig:exampleOrbits}
\end{figure}

\begin{figure}[t!]
	\centering
    \includegraphics[trim = 2.0in 0.05in 2.3in 0.9in, clip,scale=0.5,width=0.45\textwidth]{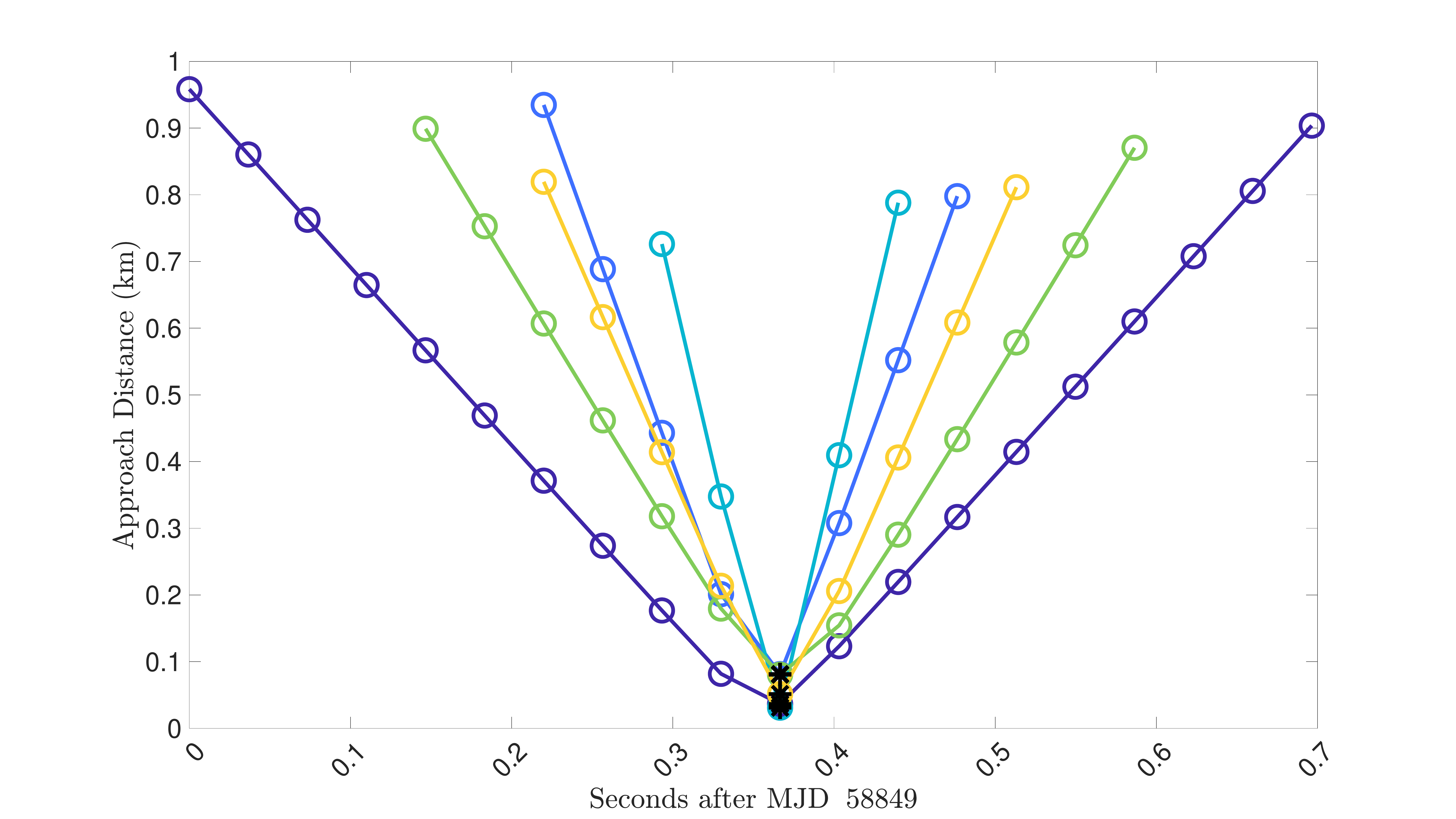}
    \includegraphics[trim = 1.8in 0.05in 2.3in 0.9in, clip,scale=0.5,width=0.45\textwidth]{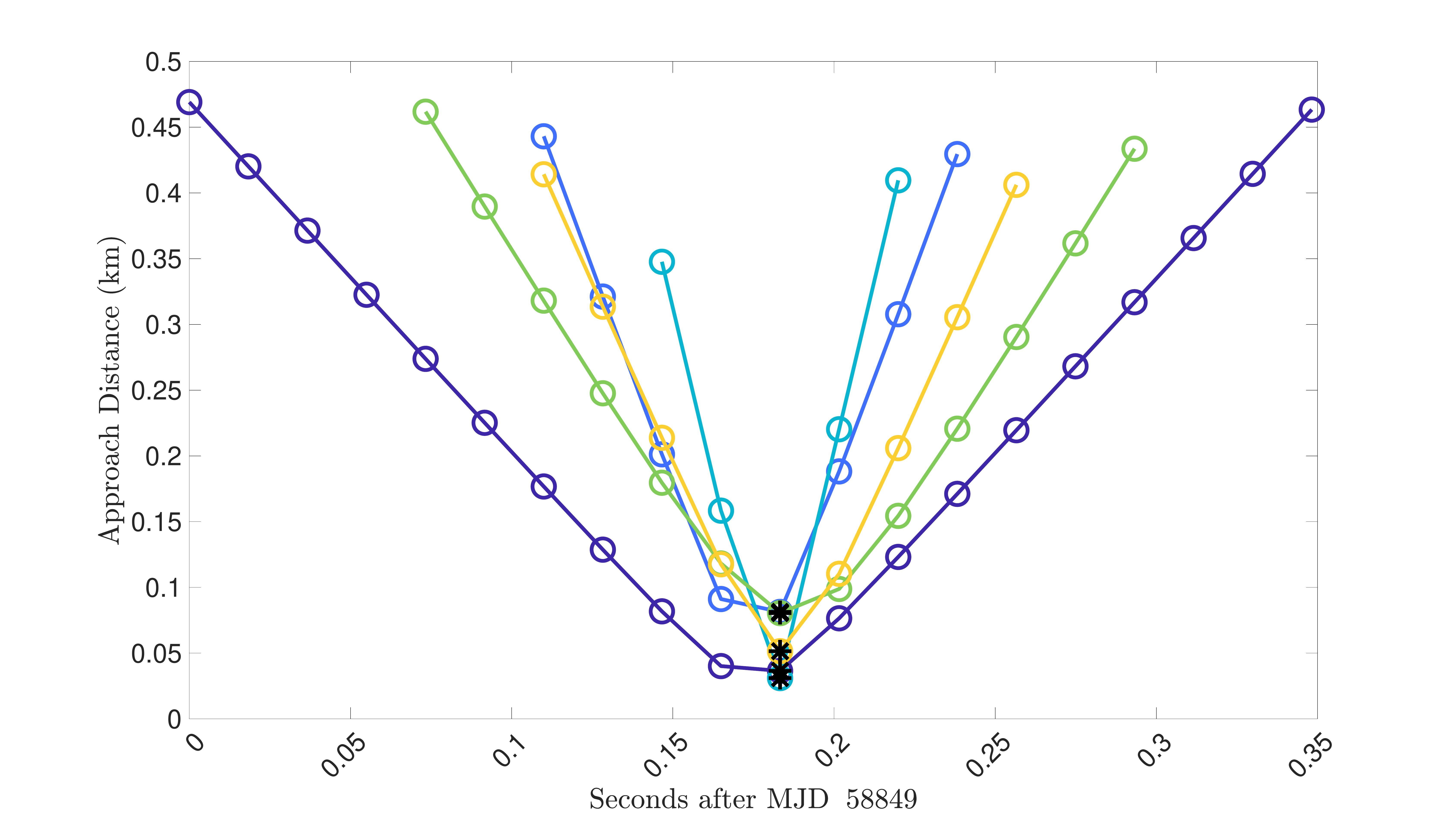}
    \includegraphics[trim = 1.7in 0.05in 2.3in 0.9in, clip,scale=0.5,width=0.45\textwidth]{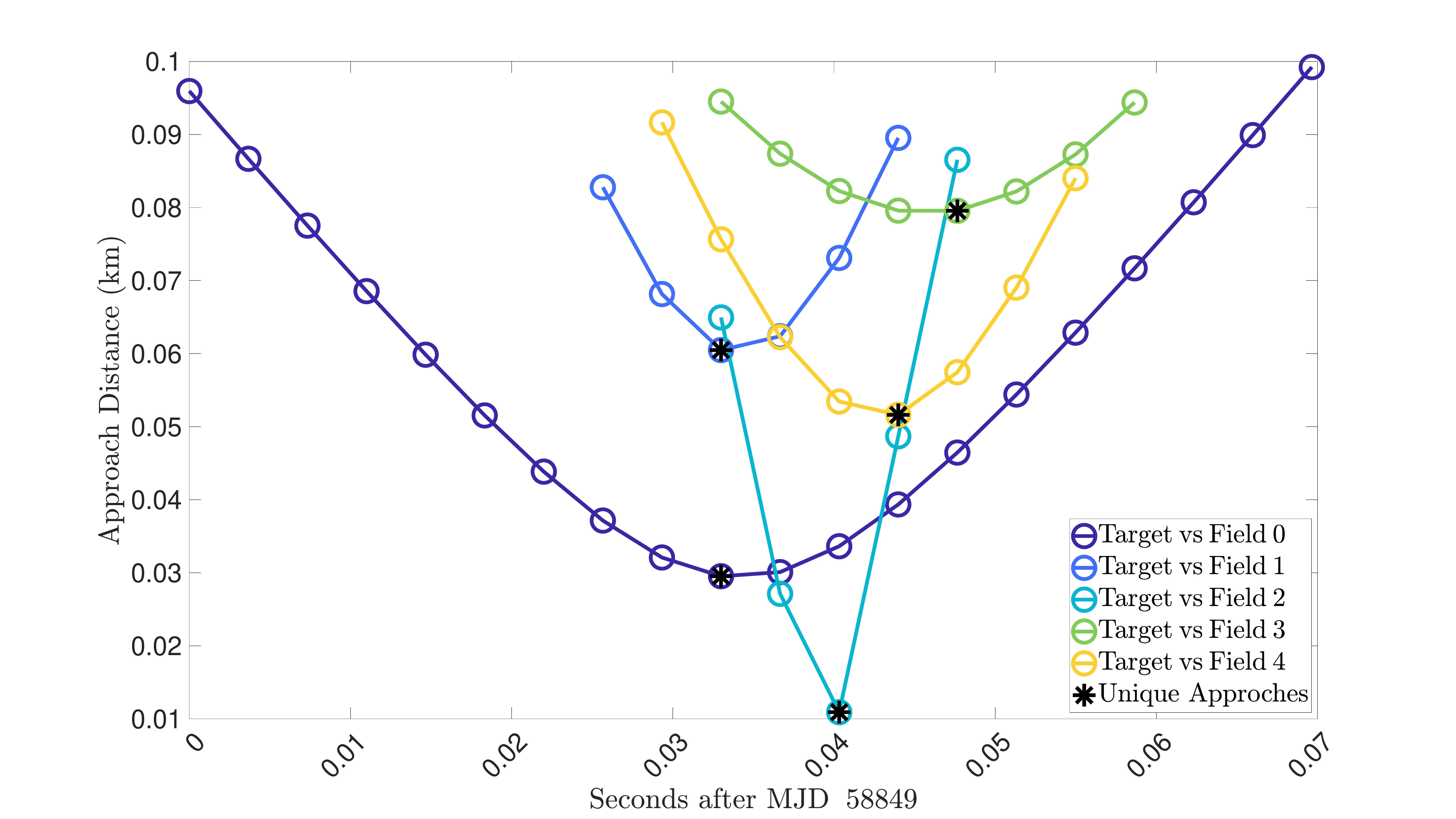}

	\caption{Integration output time steps of $0.0367$ seconds ({\it top left}) and $0.0183$ seconds ({\it top right}) are able to resolve a close approach of $0.0310$ km, while an output time step of $0.0037$ seconds ({\it bottom}) can resolve a close approach of $0.0109$ km.}
	\label{fig:exampleOutput}
\end{figure}

\begin{figure}[h!]
	\centering
    \includegraphics[trim = 1in 0.15in 2in 0.1in, clip,scale=0.5,width=0.45\textwidth]{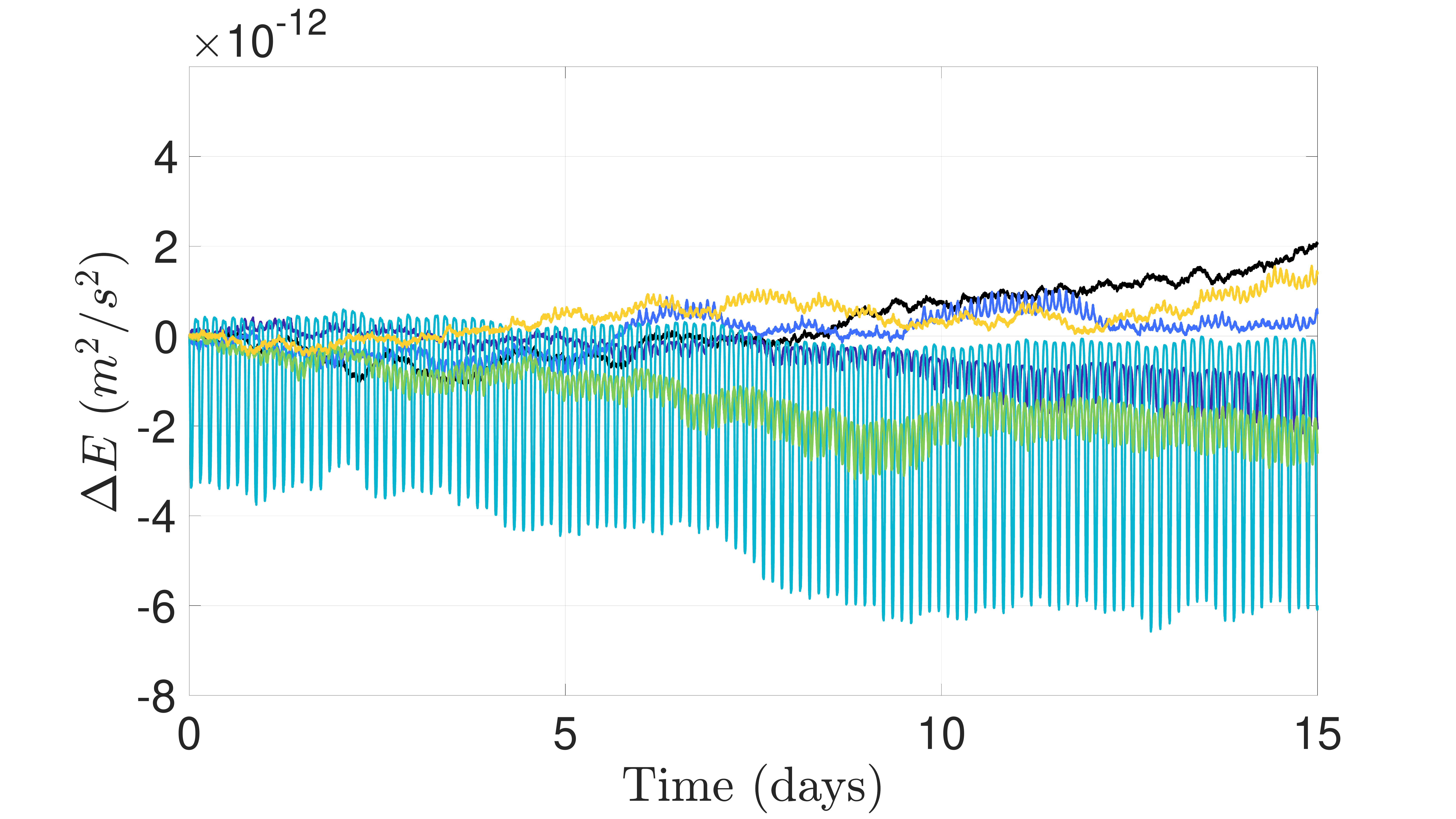}
    \includegraphics[trim = 0.9in 0.15in 2.1in 0.1in, clip,scale=0.5,width=0.45\textwidth]{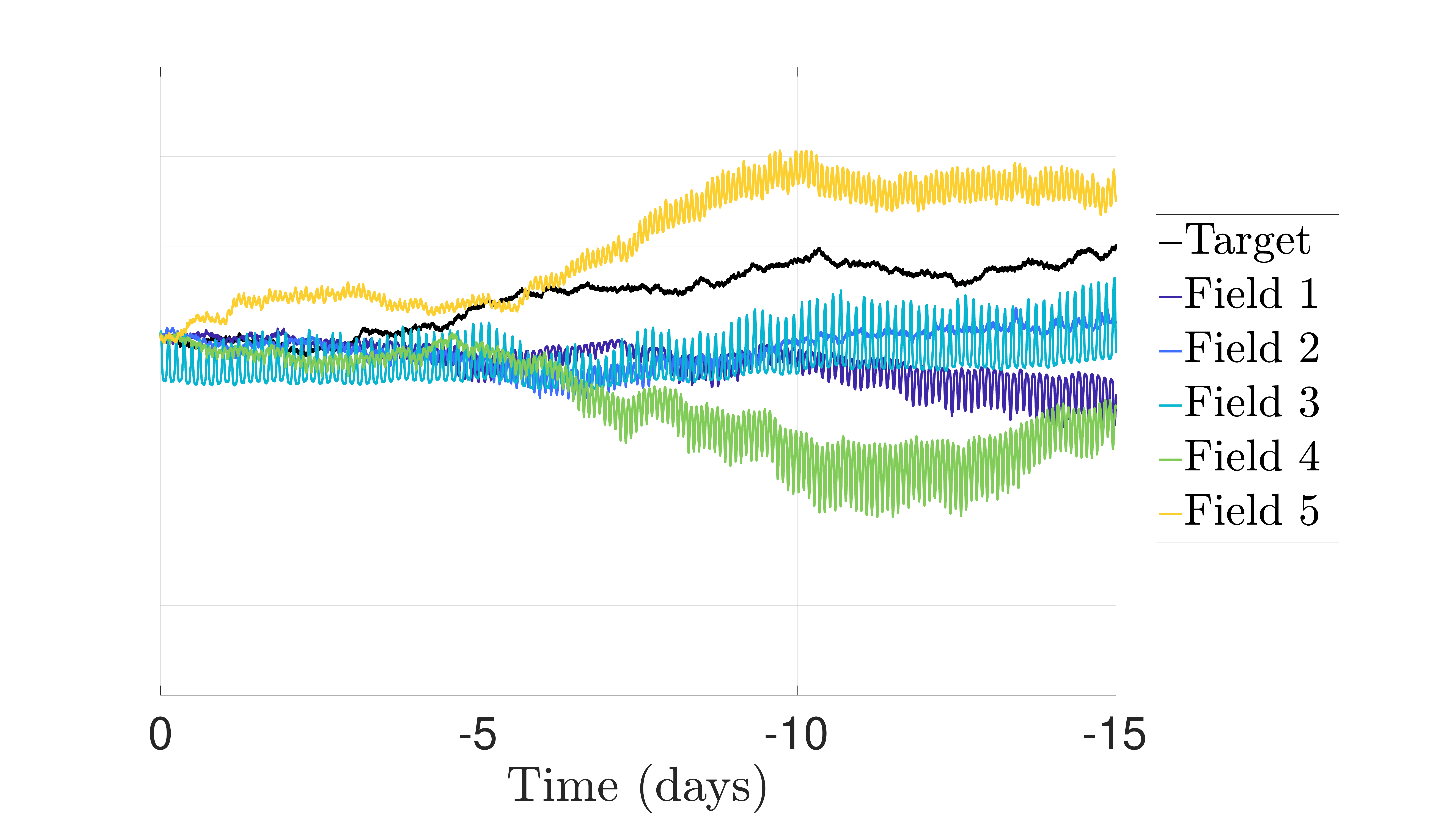}

	\caption{Deviation in the ``Hamiltonian'' (orbital energy) from its initial value for backwards ({\it left}) and forwards ({\it right}) propagation using \texttt{THALASSA} with a force model consisting of only the zonal gravity field terms to degree seven.}
	\label{fig:exampleEnergyError}
\end{figure}

To better illustrate the functionality of \texttt{RICA}, we investigate a simple test case consisting of an induced artificial collision between five different satellites in a variety of different orbital configurations. As can be seen in Table~\ref{tab:exampleFica_ICs}, the initial epoch of the colliding objects is the modified Julian date (MJD) $58834$. The collisions are manufactured by backwards propagating with \texttt{THALASSA} the target and field objects from the same initial position and epoch (MJD $58849$) for a duration of 15 days (Fig.~\ref{fig:exampleOrbits}). These initial conditions (ICs) are then passed to \texttt{RICA} using the same force model and a forward integration time span of 15.1 days (to account for any numerical errors), with close approach values set to $\tau_1 = 1.0$ (km), $\tau_2 = 0.5$ (km), and $\tau_3 = 0.1$ (km), respectively. The effect of selecting a time step according to Equation~\ref{eq:selectTstep} can be seen clearly in Fig.~\ref{fig:exampleOutput}, where relative distances much less than the specified $\tau$ between objects are detected. In theory, the approach distance between the target object and each field object should go to zero at a MJD of exactly 58849; however, the presence of numerical error, particularly with backwards propagation, shifts the time of closest approach as well as the minimum approach distance (Fig.~\ref{fig:exampleEnergyError}).

\begin{table}[h!]
\centering
\caption{Initial Keplerian orbital elements at MJD $58834$ of a target and set of field objects for use in demonstrating the \texttt{RICA} algorithm. The initial conditions were obtained by backwards propagating these objects from the same position at MJD $58849$.}
\label{tab:exampleFica_ICs}
\begin{tabular}{@{}lllllll@{}}
\toprule
ID                  & $a$ (km)  & $e$       & $i$ (${}^\circ$) 
                    & $\Omega$ (${}^\circ$) 
                                & $\omega$ (${}^\circ$) 
                                            & $M$ (${}^\circ$)
                    \\ \midrule
Target              & 7577.6    & 0.001     & 87.9 
                    & 3.0       & 45.2      & 126.8 \\
$\text{Field}_1$    & 9209.8    & 0.218     & 106.3            
                    & 347.2     & 56.7      & 255.5 \\
$\text{Field}_2$    & 8486.6    & 0.189     & 34.7             
                    & 48.7      & 224.3     & 188.5 \\
$\text{Field}_3$    & 10434.8   & 0.313     & 171.4            
                    & 327.5     & 256.3     & 291.2 \\
$\text{Field}_4$    & 7988.8    & 0.143     & 57.1
                    & 38.4      & 59.9      & 142.7 \\
$\text{Field}_5$    & 7958.9    & 0.047     & 132.2            
                    & 313.4     & 316.1     & 188.3 \\
                    \bottomrule
\end{tabular}
\end{table}

\subsection{The JeongAhn-Malhotra Approach adapted for circumterrestrial space}
\label{sec:JMapproach}

The study of orbital collision probability has its roots within Solar System dynamics and was pioneered by \citet{eO51} and \citet{gW67}, respectively, to study collisions within the asteroid belt. The theory begins with the calculation of the collision probability for two intersecting Keplerian orbits, $P_i (\tau, \textbf{\oe}_1, \textbf{\oe}_2)$, which is a function of the orbital elements of the target and field objects, $\textbf{\oe}_1$ and $\textbf{\oe}_2$, respectively, and the collision distance, $\tau$. Here, being Keplerian, the orbits are fixed in space and the mean anomalies are assumed to be independent. Over a long period of time the objects will have a well-defined collision probability at their intersection. In \citet{rM17}, a simplified but equivalent derivation of the collision probability of two objects in Keplerian orbits is developed: 
\begin{equation}
    \label{eq:renuColProb}
    P_i = \frac{\pi \tau U}{2|{\bm v}_1 \times {\bm v}_2|T_1T_2},
\end{equation}
where $P_i$ is the collision probability, $\tau$ is the collision distance, ${\bm v}$ is the velocity at the point of closest approach, $U$ is the relative velocity of the objects at the point of closest approach, and $T$ is the orbital period. This theory is then further expanded to account for tangential encounters, which generally have a much higher collision probability. This probability can be computed according to:
\begin{equation}
    \label{eq:renuColProb_tangential}
    P_i = 1.7 \cdot T_1 T_2 \frac{\sqrt{(1 - k)\tau}}{(1 + k)g \sin{\alpha}},
\end{equation}
where $k = {|{\bm v}_2|} / |{\bm v}_1|$ if the objects orbit in the same direction, i.e., both objects are either prograde or retrograde, and $k = -|{\bm v}_2| / |{\bm v}_1|$ if they orbit in the opposite sense. Here, $\alpha$ is the angle between the common direction of the two bodies and the positive $x$-axis, where 
\begin{equation}
    \sin{\alpha} = 
    \frac{1 + e \sin{f}}{\sqrt(1 + 2 e \cos{f} + e^2)},
\end{equation}
and $e$ and $f$ are the eccentricity and true anomaly, respectively, of either object at the time of closest approach. The determination of whether or not an encounter is tangential is covered in great detail in \citet{rM17} and can be seen in Algorithm~\ref{alg:probability_pseudo}.

To be applicable to the near-Earth space environment, the probability with respect to an ensemble of fields is required. Past studies using techniques that descended from \citet{eO51} and \citet{gW67} have in large part been limited to the field of Solar System dynamics, where the orbital planes of objects typically change at much slower rates and extremely large sets of objects are considered (the asteroid belt, for example). As such, the semi-major axis ($a$), eccentricity ($e$), and inclination ($i$) of the target and field objects are assumed to be fixed, while the right ascension of the ascending node ($\Omega$), argument of perigee ($\omega$) and $\tau$ are assumed to be random stochastic variables. Of course, when considering artificial, Earth-orbiting satellites, such an approach is not suitable as the orbital parameters can change at extremely fast rates and, accordingly, fixing $a$, $e$, and $i$ fails to capture the dynamics of the circumterrestrial problem. Following \citet{rM15}, our solution is to create a distribution of clones by propagating the states of the field and target objects forwards with the full dynamics and randomly sampling the resultant trajectories. In order to keep the problem computationally manageable, only the field objects are cloned, however, the target objects are forward propagated and their trajectories are randomly sampled (Algorithm~\ref{alg:generateClones_pseudo}). Using the \citet{gG05} algorithm, the MOID for each pair of target and field objects is calculated. The total collision probability, ($P_\text{total}$) of the field and target sets is then computed by summing the individual collision probabilities (\ref{eq:renuColProb}) of all pairs of objects whose MOID is less than or equal to the specified approach distance (Eq.~\ref{eq:totalProb}), 
\begin{equation}
    \label{eq:totalProb}
    P_\text{total} = \frac{1}{N_c}\sum P_i(\textbf{\oe}_1, \textbf{\oe}_2),
\end{equation}
where $N_c$ is the multiplicity of field object clones. The complete structure of this procedure, which has been named, the ``JM Approach'', or simply \texttt{JM}, can be seen in Algorithm~\ref{alg:jmColProb_pseudo}.

\subsection{Frozen Orbits and The Minimum Space Occupancy (MiSO)}

Frozen orbits correspond to equilibria for the averaged equations of motion, or, as \citet{sC94} candidly remarked, for a system fabricated to represent the averaged orbital behavior of the satellite. Such secular equilibria, under various perturbative environments, have attracted a lot of attention in Earth-satellite missions \citep{gC66, sC94, pGmL13} and planetary satellite and small body orbiters \citep{dS12, tNpG18}. In near-Earth space, where the dominant perturbations arise from planetary oblateness, the existence of frozen orbits is attributed to the dynamical balancing of the secular effects of the even zonal harmonics with the long-periodic perturbations of the odd zonal harmonics \citep{gC66}. These types of orbits with stationary perigee and eccentricity, on average, are of special interest because they minimize altitude variations using only the natural dynamics. Accordingly, as they reduce station-keeping requirements and maintain the relative configuration of clusters of satellites \citep{pGmL13}, frozen orbits would be ideal in mega-constellation design. 

Nominal operational orbits are realized by their osculating elements, not the mean elements used in the determination of frozen-orbit conditions, and consequently the short-period effects must be readmitted to obtain the precise initial conditions. This readmission of the small fluctuations of short period that the averaging process had removed is done using the transformation from mean-to-osculating elements \citep[e.g.,][]{dB59}. The question arises as to whether the secular equilibria will ``unfreeze'' when short-term variations are included or when the averaging is pushed to higher order, or when other perturbing effects are taken into account. Although the frozen orbit definition is tied to the averaged equations of motion, these stationary solutions when recast in osculating space can also be identified as periodic orbits in the meridian plane of the satellite, and as quasi-periodic in the three-dimensional space \citep{rB94}. The direct computation of frozen, periodic orbits can thus be performed directly from the non-averaged equations, and, when accounting for other perturbations, must be done using an optimization routine.

Strictly speaking, the tesseral harmonics in the geopotential coupled with non-gravitational perturbations destroy the frozen orbit conditions of the zonal-only problem. This realization has lead to a significant generalization of frozen orbits that we designate the Minimum Space Occupancy (MiSO) orbits \citep{cB18}. In particular, we use the \texttt{THALASSA} orbit propagator with a sufficiently complex force model to search for the perturbed-Keplerian orbits that trace out the least amount of volume in three-dimensional space. We note that if we adopt only the zonal-harmonics model in our algorithm, then the MiSO solutions degenerate to the classical frozen orbits. 

\section{Numerical Experimental Setup} 
\label{sec:setup}

\subsection{Mega Constellations in the Context of the Current Satellite-Debris Field}

\begin{figure}[htp!]
	\centering
	\includegraphics[width=0.49\textwidth]
	{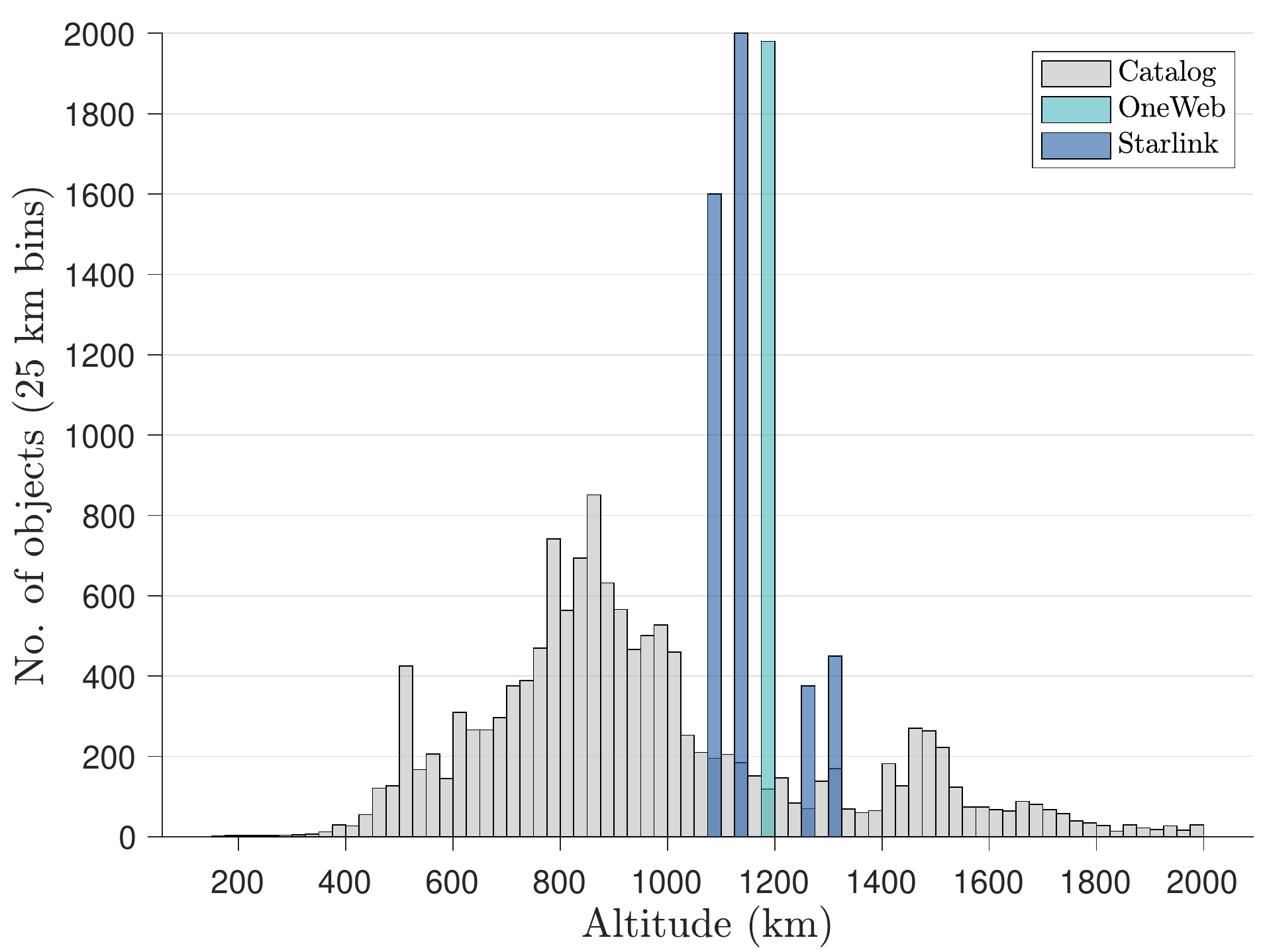}
	\includegraphics[width=0.49\textwidth]
	{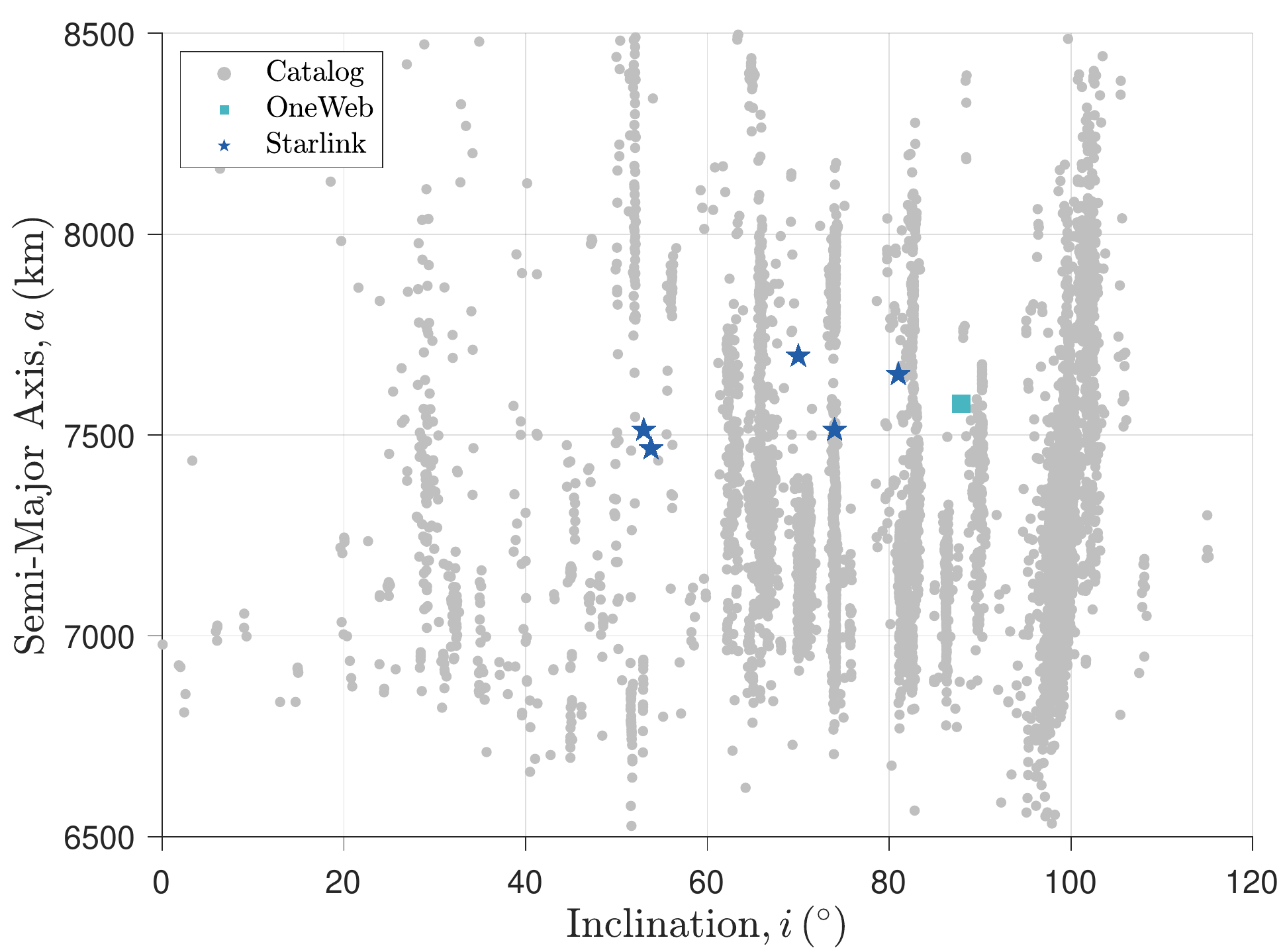}
	\caption{Altitude distribution ({\it left}) and $(i, a)$ scatter plot ({\it right}) of the OneWeb LEO and SpaceX Starlink mega-constellations, as obtained from FCC filings no. SAT-LOI-20160428-00041 and SAT-MOD-20181108-00083, respectively, against the background cataloged satellites and debris in low-Earth orbit.
	(`Norad’ Resident Space Object Catalog. www.space-track.org. Assessed 27 Nov. 2019)}
	\label{fig:leodistrib}
\end{figure}

The addition of the OneWeb LEO and SpaceX Starlink mega-constellations alone will increase the number of objects in LEO by 46 percent (Fig.~\ref{fig:leodistrib}), according to the designs described in their most recent FCC filings that actually contain constellation orbit designs (no. SAT-LOI-20160428-00041 and SAT-MOD-20181108-00083, respectively). We note that there have been additional files since, particularly for Starlink, but these do not contain sufficient orbital information to permit the type of study performed herein. 

\subsection{Physical Model}
\label{sec:physModel}
We adapt in this study a basic physical model that encompasses the gravity field of the Earth up to the 7th degree and order of the spherical harmonics, the third-body gravity of the Sun and Moon, as well as atmospheric drag using the NRLMSISE-00 model, variable F10.7 solar flux, solar radiation pressure (SRP) with a conical Earth shadow. The physical parameters of the OneWeb and Starlink satellites, needed for the computation of the non-gravitational effects, are given in Table~\ref{tab:physParams}. These physical parameters are based on common values of $C_D$ and $C_R$ as well as a brief description of the OneWeb satellite bus found on the airbus website.\footnote{\url{https://www.airbus.com/newsroom/press-releases/en/2019/07/}} For consistency and due to a lack of information, these same parameters were imposed upon the satellites of the Starlink constellation. The effects that the model fidelity has on our results is given in \ref{sec:sensitivityStudy}.

\begin{table}[h!]
\centering
\captionof{table}{Physical parameters of OneWeb LEO and Starlink constellation satellites used in their respective case studies.}
\label{tab:starlinkTarget}
\begin{tabular}{@{}lllll@{}}
\toprule
$A/m$ (m$^2$/kg) & Drag Area (m$^2$) & SRP Area (m$^2$) & $C_D$ & $C_R$ \\ 
0.0123 & 1.84 & 1.84 & 1.28  & 1.00  \\ \bottomrule
\end{tabular}
\label{tab:physParams}
\end{table}

\subsection{OneWeb LEO and Starlink Case Studies and their MiSO Variants}

\begin{figure}[h!]
	\centering  
	\captionof{table}{The initial orbital plane configuration of OneWeb LEO, as reported in FCC filings no. SAT-LOI-20160428-00041.}
    \begin{tabular}{@{}llllll@{}}
    \toprule
    $a$ (km) & $e$ & $i$ (${}^\circ$) & $\Omega$ (${}^\circ$) & $\omega$ (${}^\circ$) & Planes \\ \midrule
    7578    & 0   & 87.9  & 0 - 180& 0 & 36  \\ 
    \end{tabular}
    \label{tab:onwebConfig}
    \bigskip
    
	\includegraphics[trim = 4.6in 4.6in 3.9in 3.1in, clip,scale=0.5,width=0.35\textwidth]{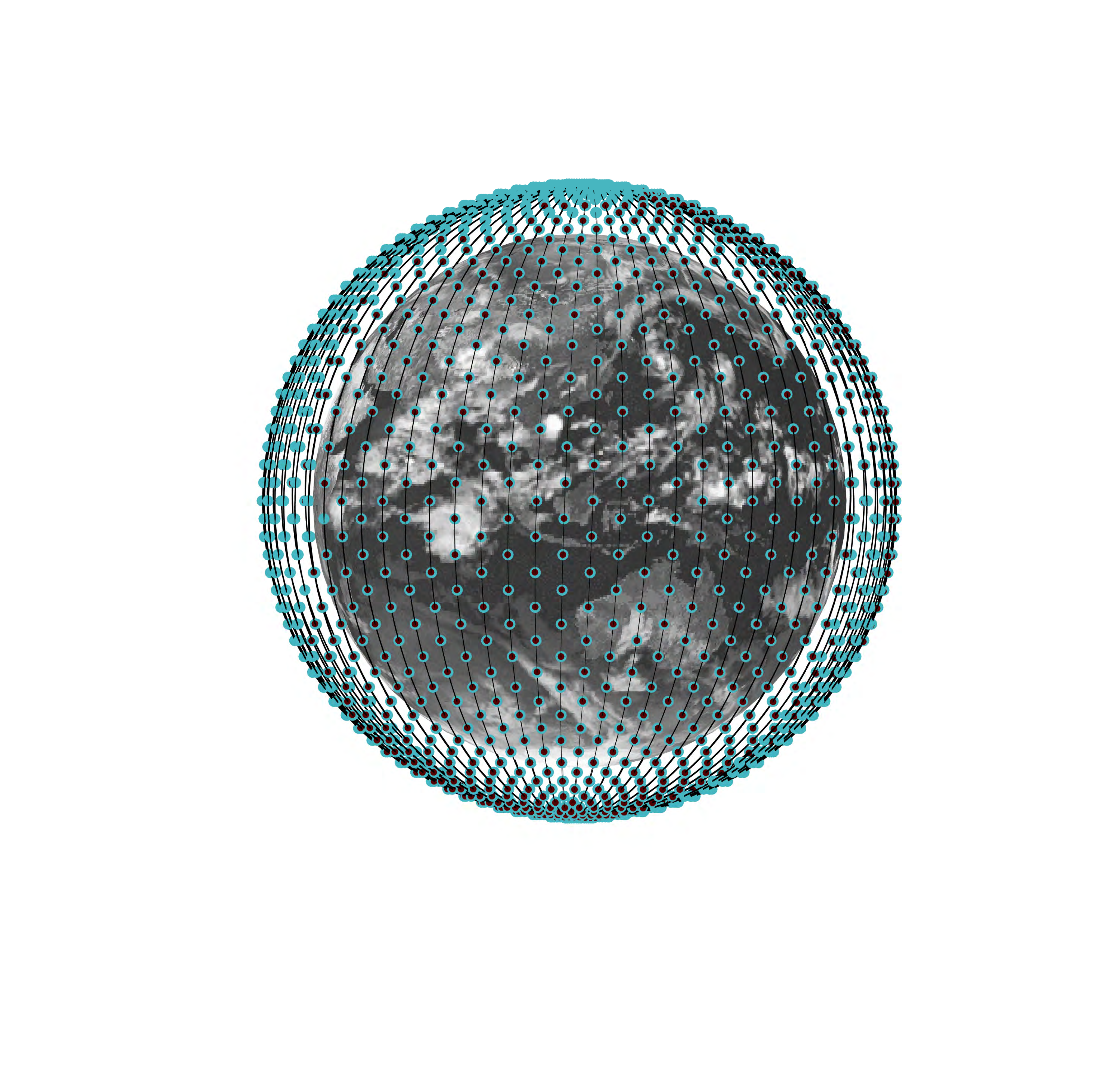}
	\includegraphics[width=0.6\textwidth]{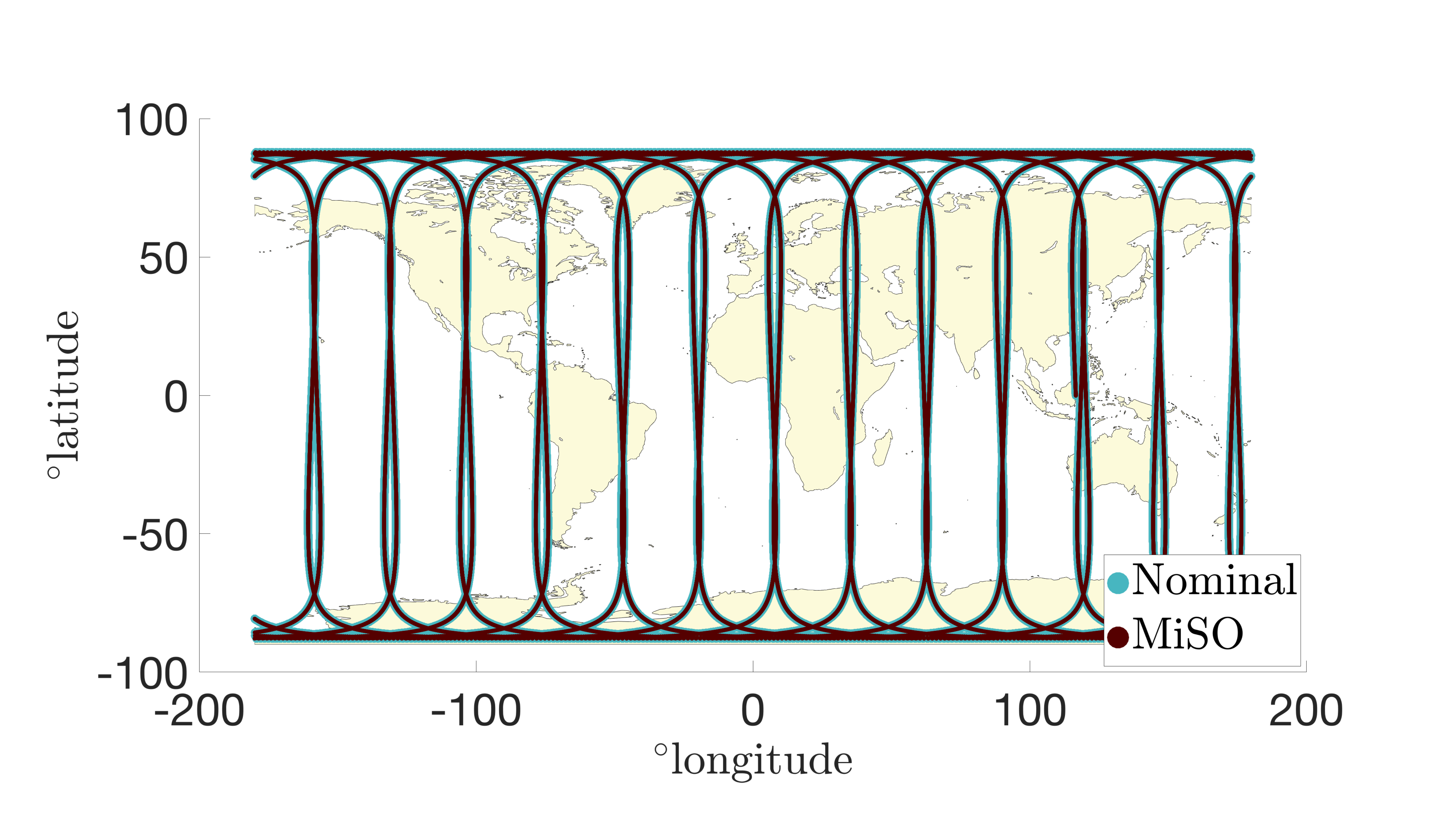}
	\vspace{0.05em}
	\captionof{figure}{The initial conditions of the nominal and MiSO variants of the OneWeb LEO constellation ({\it left}), as well as their corresponding ground tracks ({\it right}).}
	\label{fig:onewebMiso}
\end{figure}

\begin{figure}[h!]
	\centering    
	\captionof{table}{The initial orbital plane configuration of each shell of the SpaceX Starlink constellation, as reported in FCC filings no. SAT-MOD-20181108-00083.}
    \label{tab:starlinkConfig}
    \begin{tabular}{@{}llllll@{}}
    \toprule
    $a$ (km) & $e$ & $i$ (${}^\circ$) & $\Omega$ (${}^\circ$) & $\omega$ (${}^\circ$) & Planes \\ \midrule
    7512.4 & 0 & 53 & 0 - 348.8 & 0 & 32 \\ 
    7465.9 & 0 & 53.8 & 0 - 354.4 & 0 & 32 \\ 
    7696.7 & 0 & 70 & 0 - 300 & 0 & 6 \\ 
    7512.4 & 0 & 74 & 0 - 315 & 0 & 8 \\ 
    7650.8 & 0 & 53 & 0 - 288 & 0 & 5 \\ 
    \end{tabular}
    \bigskip
    
	\includegraphics[trim = 4.6in 4.6in 3.9in 3.1in, clip,scale=0.5,width=0.35\textwidth]{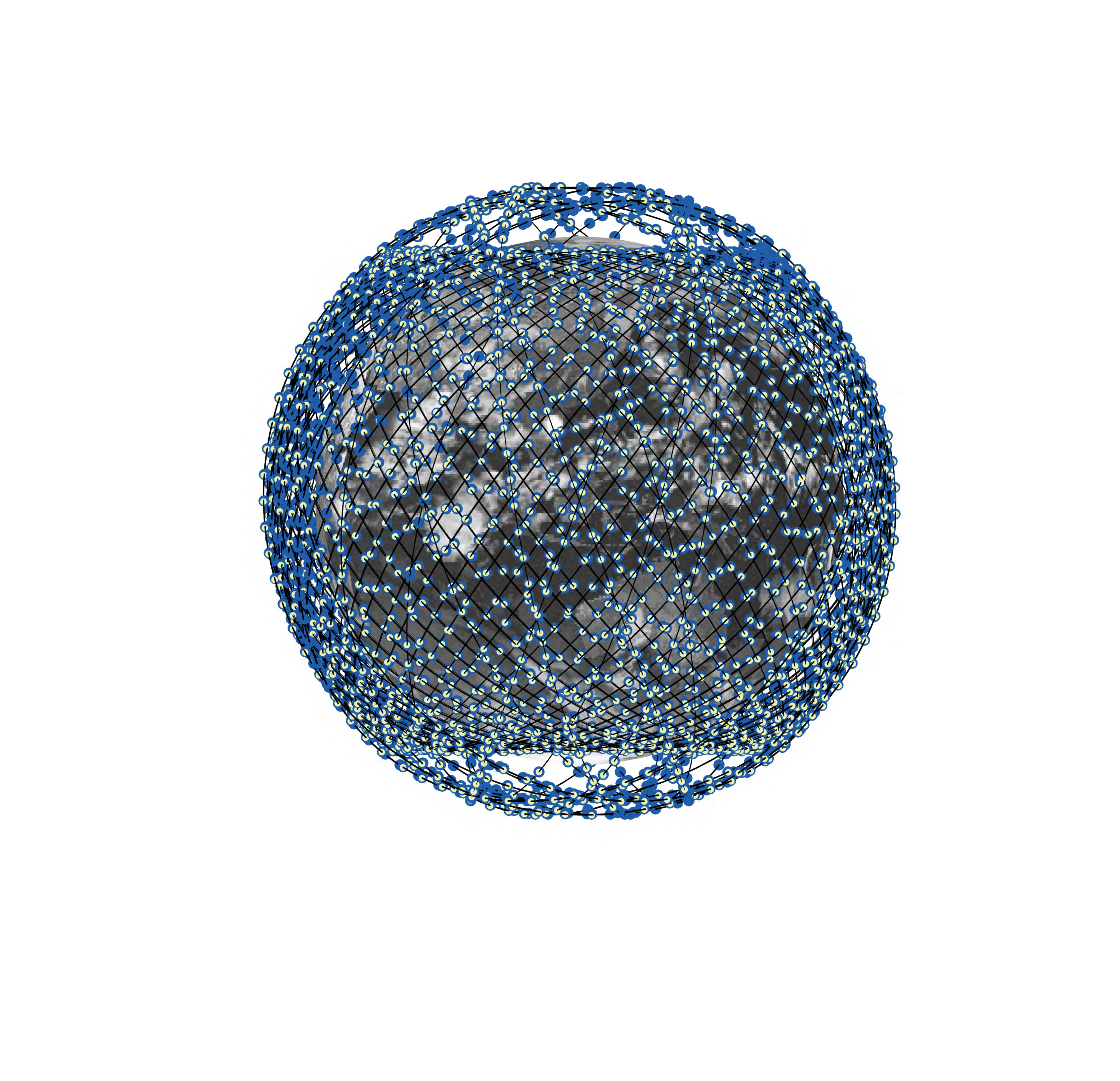}
	\includegraphics[trim = 4.6in 4.5in 3.9in 3.1in, clip,scale=0.5,width=0.35\textwidth]{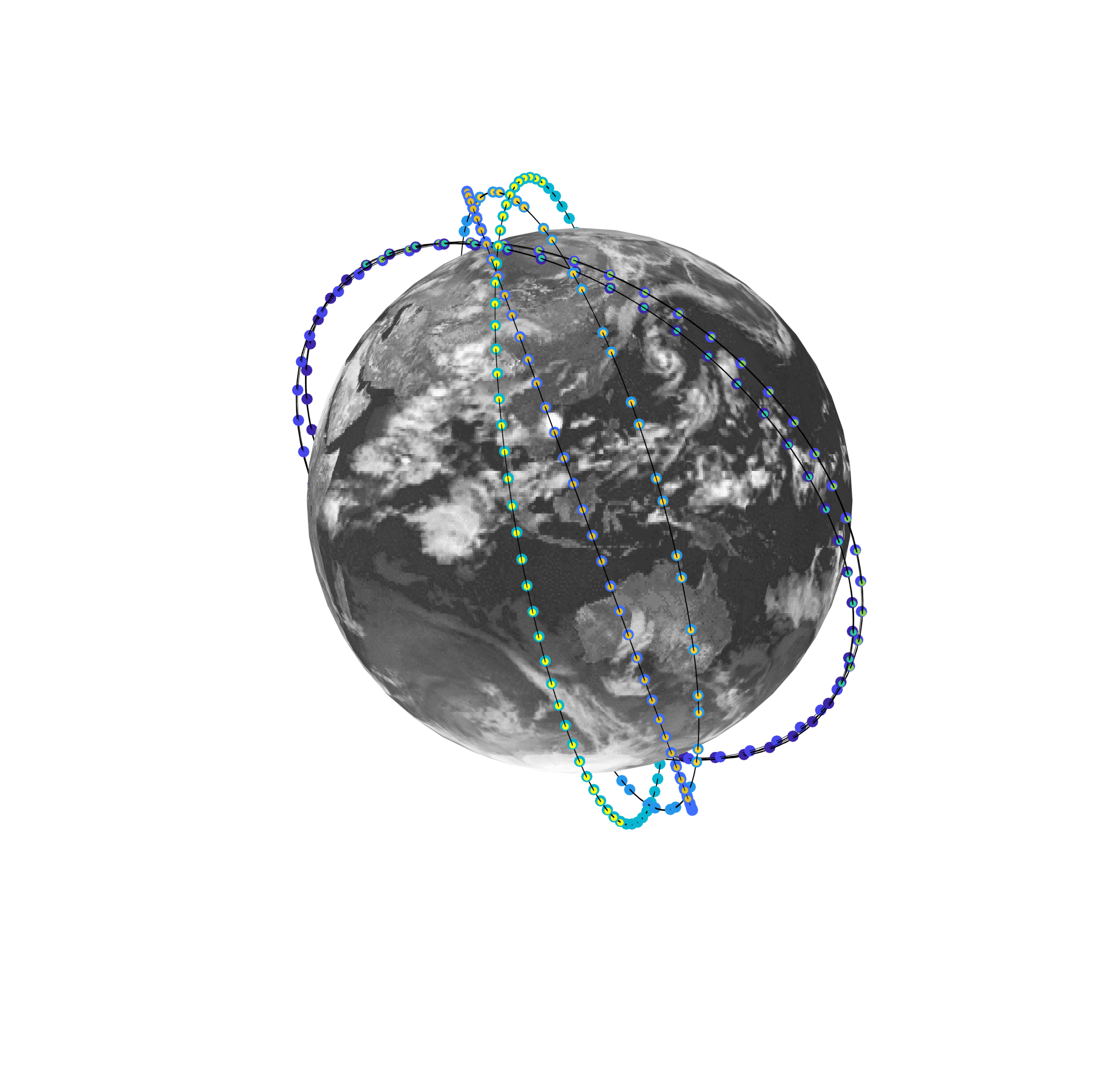}
    \includegraphics[trim = 0.7in 0.65in 1.8in 1.2in, clip,scale=0.5,width=0.60\textwidth]{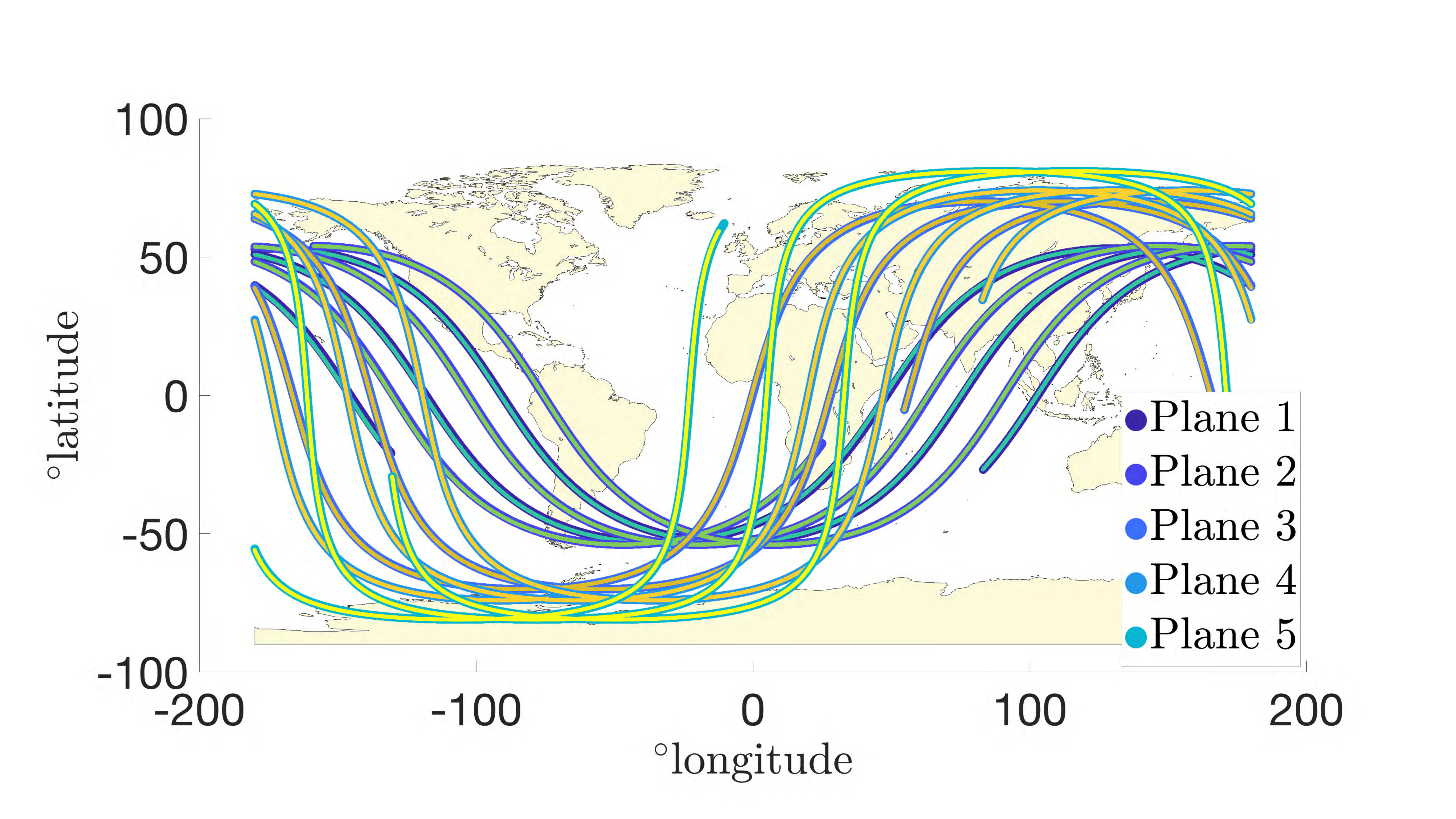}
    \vspace{0.5em}
	\captionof{figure}{The initial conditions of the nominal and MiSO variants of the SpaceX Starlink constellation ({\it top}), as well as their corresponding ground tracks ({\it bottom}). Note that only the ground tracks of one plane in each shell, as depicted in the {\it top-right} panel, is shown.}
	\label{fig:starlinkMiso}
\end{figure}

The OneWeb LEO and Starlink constellation parameters were obtained from the FCC filings no. SAT-LOI-20160428-00041 and SAT-MOD-20181108-00083, respectively. As can be seen in Tables~\ref{tab:onwebConfig} and \ref{tab:starlinkConfig}, OneWeb LEO is composed of 36 orbital planes with 1980 total satellites, while Starlink, the much larger constellation, is composed of 83 planes with 4425 satellites in total. In recent months, various other designs have been reported by researchers and reporters in the space situational awareness community. We remind that it is not our intent to critique any particular design, but rather to showcase the practical implementation of numerical techniques such as \texttt{RICA} and \texttt{JM}, as well as to demonstrate how MiSO configurations can significantly mitigate the endogenous (i.e., self inflicted) collision risks associated with mega-constellations.

The generation of the MiSO variants of the OneWeb LEO and Starlink constellations begins with determining a single optimized initial condition of one satellite in each orbital plane. These initial conditions are then propagated forwards in time using a high-accuracy force model for one orbital period. The ICs of the individual satellites in each respective orbital plane are then generated by sampling this orbital period at intervals where the mean longitude ($\varpi = \Omega + \omega + M$) of the MiSO IC matches the mean longitude of the corresponding satellite in the nominal constellation. Using this technique, the MiSO variants of the OneWeb LEO (Fig.~\ref{fig:onewebMiso}) and Starlink (Fig.~\ref{fig:starlinkMiso}) constellations are generated that are nearly indistinguishable from their nominal counterparts. Indeed, the initial conditions of the MiSO satellites are so similar to those of the nominal that the ground tracks of both sets are nearly identical. Of course they are not an exact match as shown in Fig.~\ref{fig:starlinkMiso}, where the MiSO satellite of Plane $5$ is slightly ''slower'' than the nominal satellite due to a difference in altitude. Although we are not privy to the operational requirements of Starlink, we do not expect this small discrepancy to cause much concern; especially since such a near frozen-orbit design allows the operators to avoid costly station-keeping maneuvers. 

Tables~\ref{tab:onwebTarget} and \ref{tab:starlinkTarget} list the initial osculating elements of the nominal constellations and their MiSO counterparts. Only one plane of OneWeb and one plane of each distinct shell of Starlink are shown. Note that in the nominal circular orbit case, the argument of perigee is technically undefined (only the combination of $\omega$ and $M$ is meaningful). The greatest differences between the nominal and MiSO configurations is the osculating semi-major axes. This is the result of the short-periodic fluctuations and how the MiSO ICs are generated from the optimized condition by propagating and sampling over one orbital period.    

\subsection{Endogenous vs Exogenous Conjunctions} 

We consider an ``endogenous'' conjunction/collision to be one where both parties in the event belong to the same constellation (i.e., self-induced), whereas an ``exogenous'' collision is between a constellation satellite and any other RSO that is not also a member of the constellation, including active and inactive satellites, as well as defunct man-made debris. For this study, due to limitations on the High-Performance Computing (HPC) CPU hours provided by the University, when investigating endogenous conjunctions an all-on-all style analysis was not performed, however any company capable of deploying such a constellation, or government agency required to regulate space traffic, would surely have access to the CPU hours needed to perform such an analysis. 

For the OneWeb endogenous case study, we select one plane as the set of target objects (Table~\ref{tab:onwebTarget}) and designate the remaining satellites as the field objects. Similarly, because the Starlink constellation is composed of five unique ``shells'' or plane types of distinct combinations of altitude and inclination, we designate five target planes (Table~\ref{tab:starlinkTarget}), one from each shell, where the remaining constellation satellites are considered the field objects for each corresponding target plane. The target planes for both the nominal and MiSO Starlink constellations are shown in the {\it top-right} panel of Fig.~\ref{fig:starlinkMiso}. The endogenous collision risk of these target planes with respect to the rest of the field objects is then investigated utilizing the aforementioned \texttt{RICA} and \texttt{JM} numerical algorithms as well as with the filtering method outlined in \citet{fH84} (hereafter referred to as the  \texttt{HERA} algorithm for ``Hoots Evaluation and Rapid Analysis''). Importantly, the results of the numerical investigation with \texttt{RICA} do not represent the collision risk of the entire constellation, however they can be used to compare the risk of different configurations as well as the accuracy of the \texttt{JM} and \texttt{HERA} methods.

In order to study the risk of exogenous collisions, the complete SOC was obtained from Space-Track (see, e.g., Fig.~\ref{fig:leodistrib}). The 18381 different objects were first (properly) converted from mean elements to osculating elements in the J2000 frame and then propagated using \texttt{THALASSA} to the reference epoch of the study, MJD 58849. These objects were then designated to be the ``field'' objects for the exogenous portion of the study with the same considered target planes. The collision risk of these sets of target and field objects are then computed using \texttt{RICA}. These results are relegated to \ref{sec:exogenous}.

\begin{table}[h!]
\caption{The initial osculating orbital elements at MJD $58849$ of the nominal and MiSO variants of one plane of the OneWeb LEO constellation. The mean anomalies of the satellites distributed within these planes range between 0 and $360^\circ$.}
\label{tab:onwebTarget}
\begin{tabular}{@{}llllll@{}}
\toprule
ID & $a$ (km)        & $e$               & $i$ (${}^\circ$) & $\Omega$ (${}^\circ$) & $\omega$ (${}^\circ$) \\ \midrule
Nominal       & 7578            & 0                 & 87.9             & 0                     & 0                     \\
MiSO          & 7557.6 - 7575.0 & 0.0004 - 0.0026 & 87.900 - 87.903  & 359.97 - 359.98       & 16.0 - 358.0          \\ \bottomrule
\end{tabular}
\end{table}

\begin{table}[h!]
\caption{The initial osculating orbital elements at MJD $58849$ of the nominal and MiSO variants of one plane of each shell of the SpaceX Starlink constellation. The mean anomalies of the satellites distributed within these planes range between 0 and $360^\circ$.}
\label{tab:starlinkTarget}
\begin{tabular}{@{}llllll@{}}
\toprule
ID        & $a$ (km)        & $e$             & $i$ (${}^\circ$) & $\Omega$ (${}^\circ$) & $\omega$ (${}^\circ$) \\ \midrule
Nominal 1 & 7512.4          & 0               & 53               & 348.8                 & 0                     \\
MiSO 1    & 7539.7- 7550.8  & 0.0003- 0.0014  & 53.00 - 53.03    & 348.50 - 348.74       & 49.9 - 131.6          \\ 
Nominal 2 & 7465.9          & 0               & 53.8             & 354.4                 & 0                     \\
MiSO 2    & 7501.8 - 7513.2 & 0.0003 - 0.0015 & 53.80 - 53.83    & 354.13- 354.37        & 51.2 - 130.1          \\
Nominal 3 & 7696.7          & 0               & 70               & 300                   & 0                     \\
MiSO 3    & 7704.6 - 7719.8 & 0.0003 - 0.0018 & 70.00 - 70.02    & 299.86 - 299.99       & 9.0 - 353.3           \\
Nominal 4 & 7512.4          & 0               & 74               & 315                   & 0                     \\
MiSO 4    & 7507.0 - 7523.3 & 0.0004 - 0.0022 & 74.00 - 74.02    & 314.88 - 314.99       & 19.3 - 348.6          \\
Nominal 5 & 7650.8          & 0               & 81               & 288                   & 0                     \\
MiSO 5    & 7654.3 - 7671.2 & 0.0004 - 0.0024 & 81.00 - 81.01    & 287.93 - 287.99       & 8.5 - 354.1           \\ \bottomrule
\end{tabular}
\end{table}

\section{Endogenous Assessment of the OneWeb LEO Constellation}

\begin{figure}[t!]
	\centering    
	\includegraphics[trim = 1.1in 0.1in 2.2in 0.8in, clip,scale=0.5,width=0.49\textwidth]{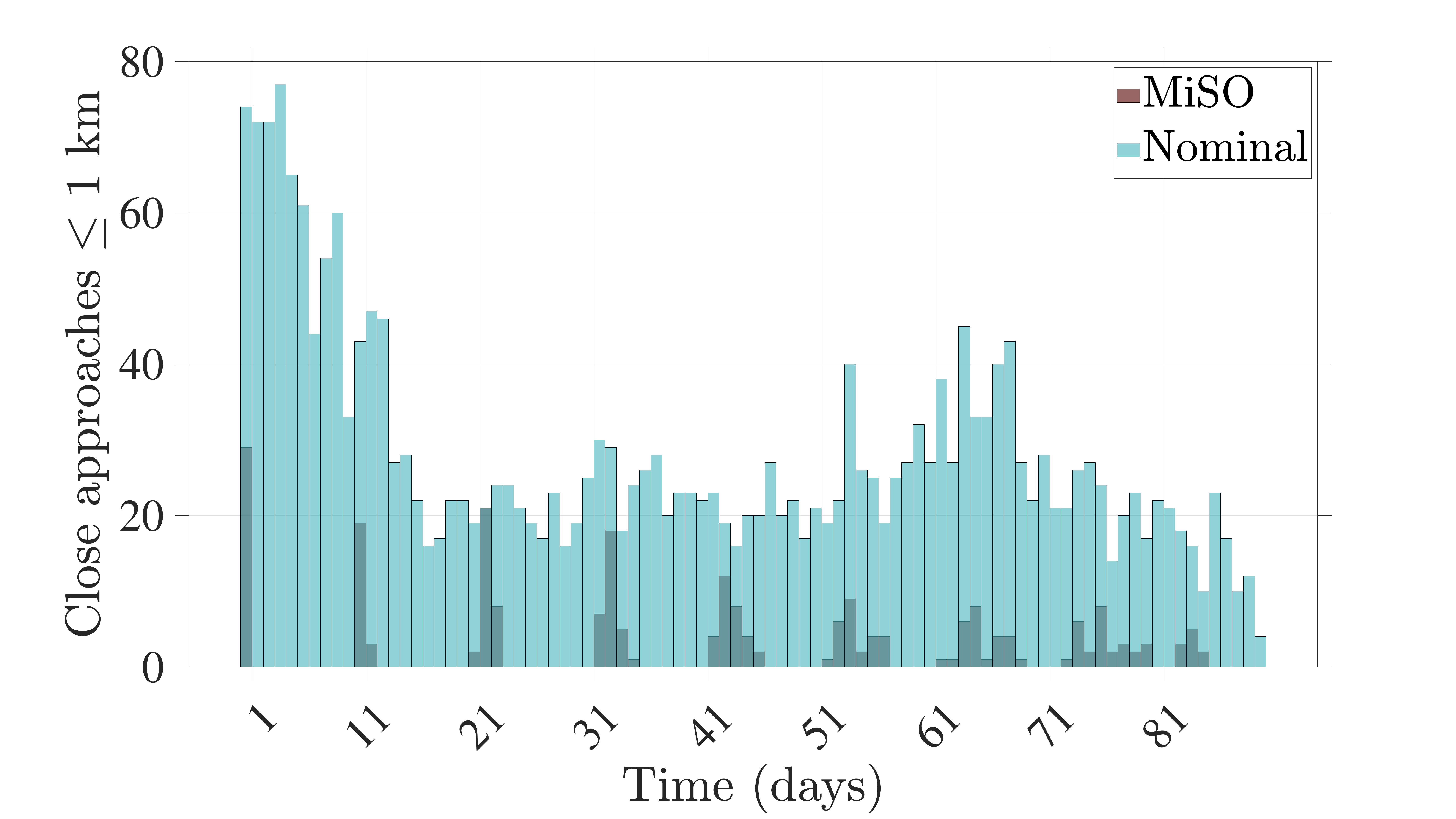}
	\includegraphics[trim = 1.1in 0.1in 2.2in 0.8in, clip,scale=0.5,width=0.49\textwidth]{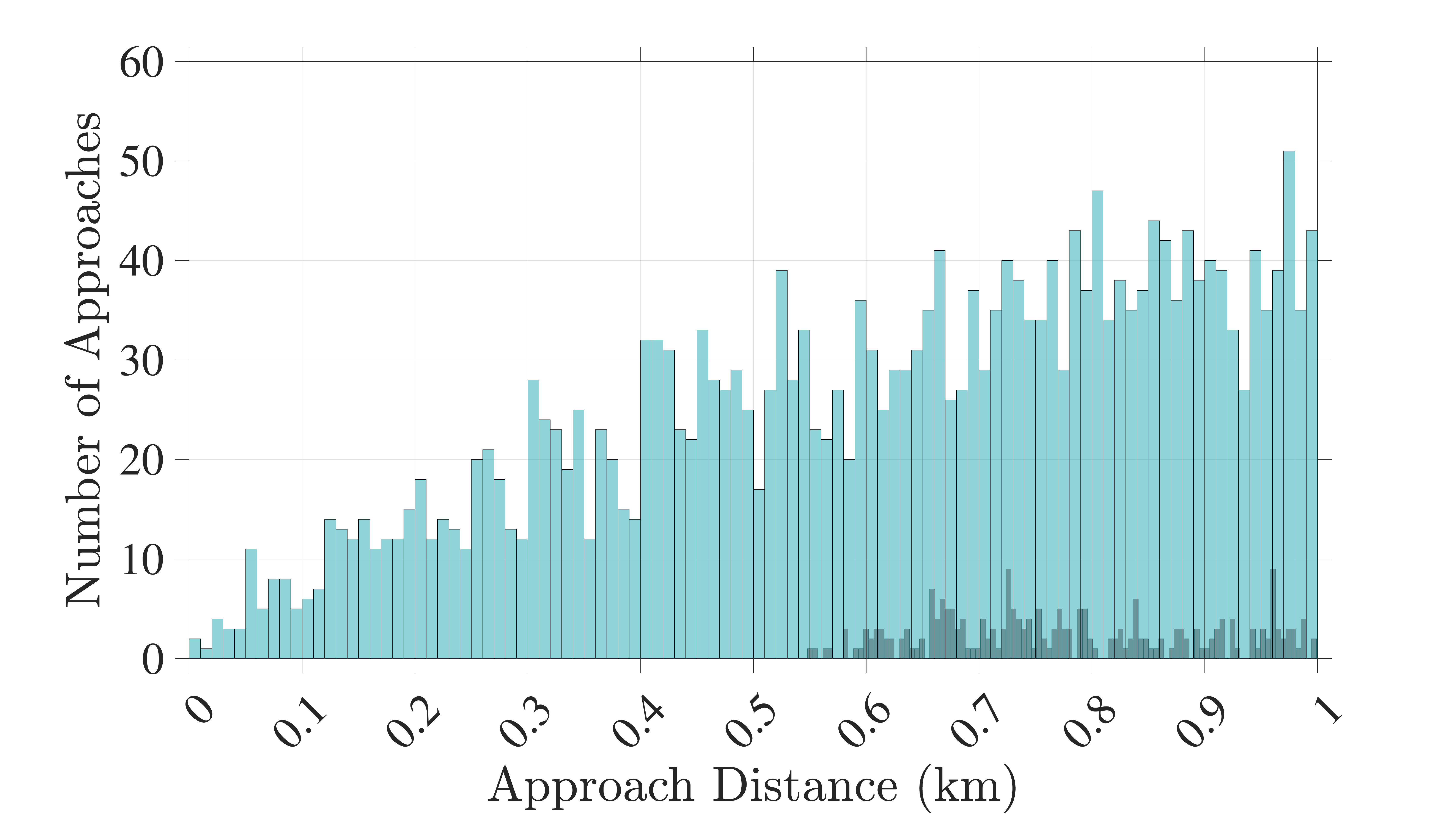}
	\includegraphics[trim = 1.1in 0.1in 2.2in 0.8in, clip,scale=0.5,width=0.49\textwidth]{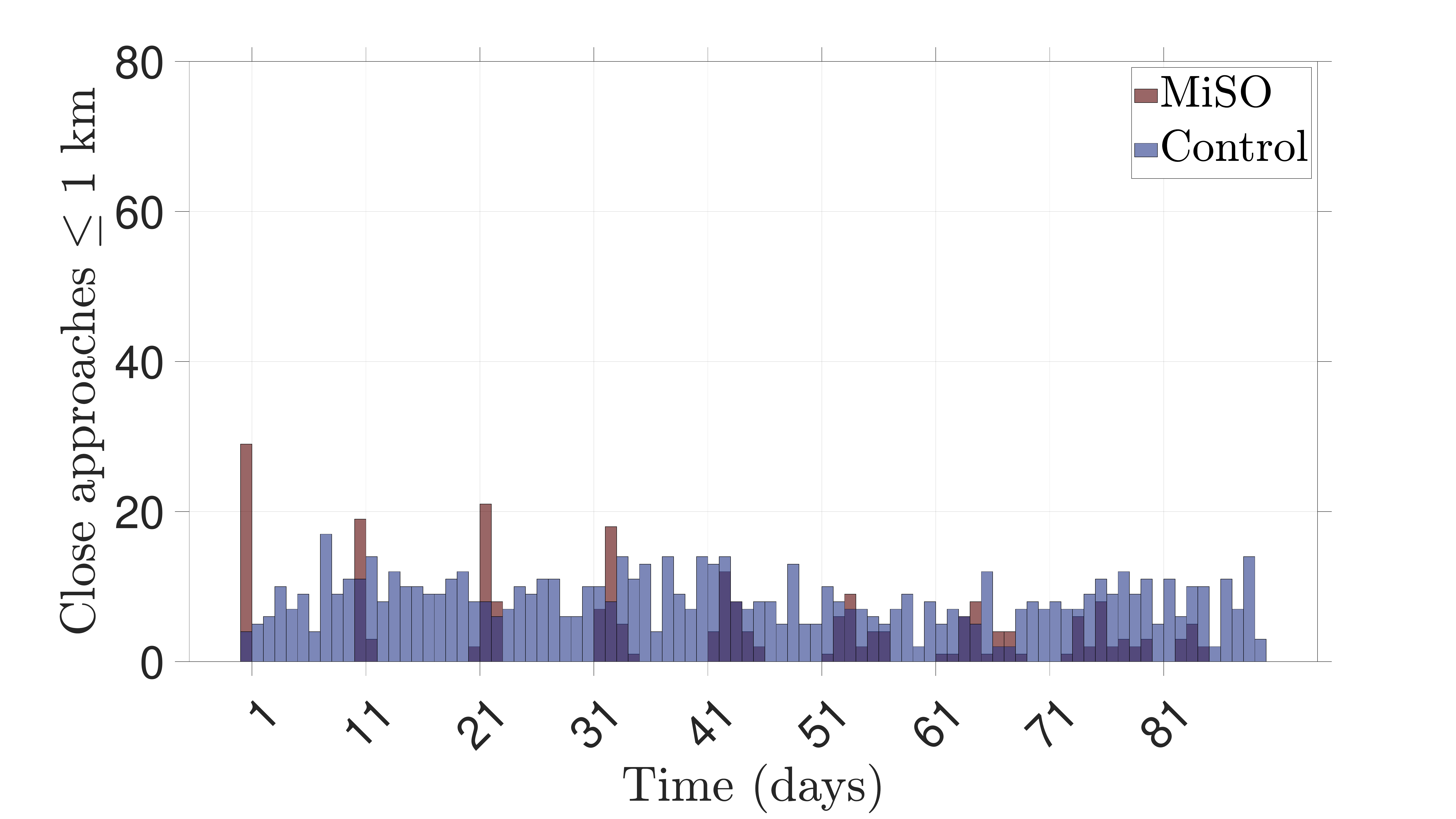}
	\includegraphics[trim = 1.1in 0.1in 2.2in 0.8in, clip,scale=0.5,width=0.49\textwidth]{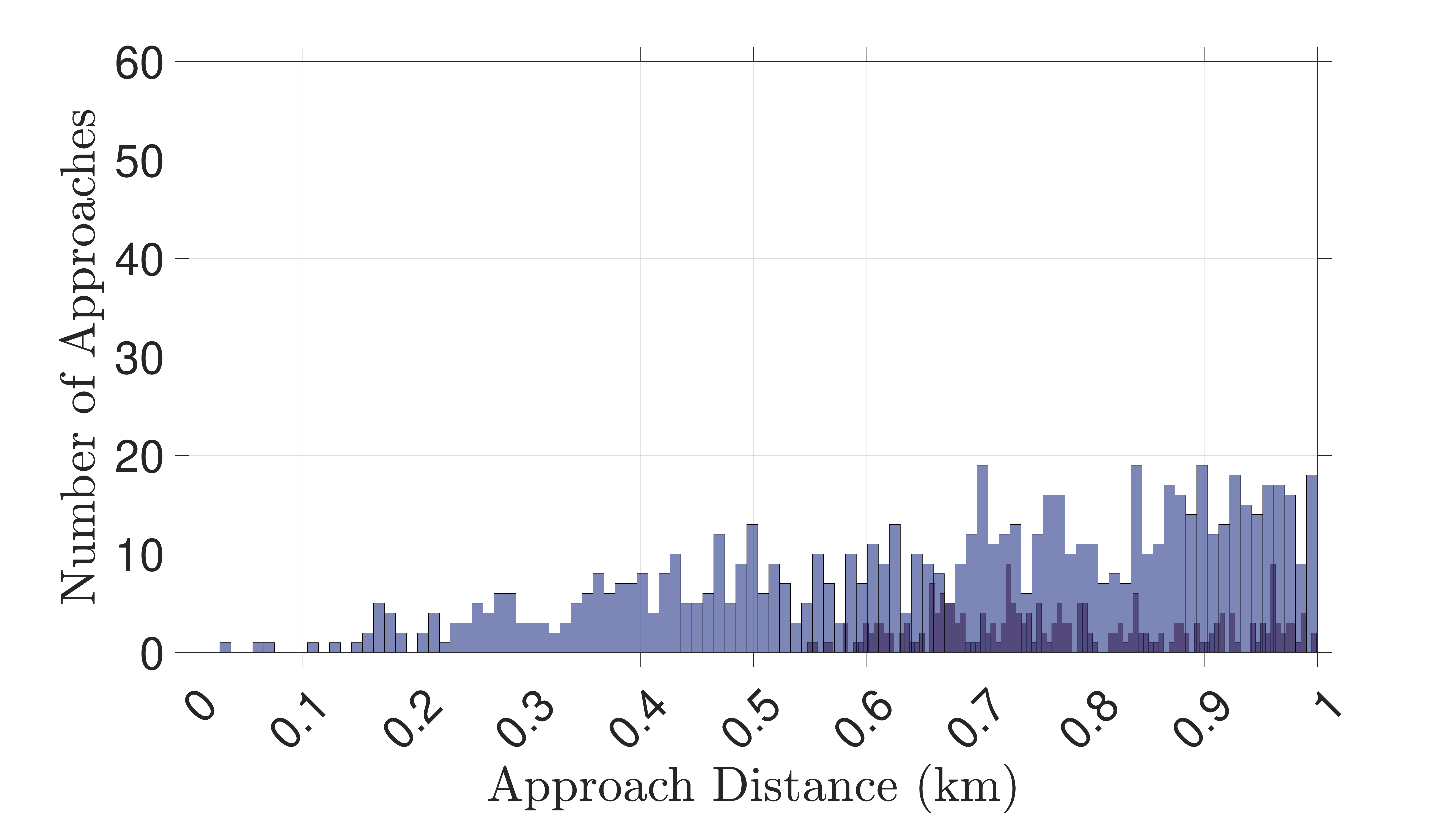}
	\caption{Frequency of close approaches within 1 km ({\it left}) and close approach distance ({\it right}) for the nominal ({\it top}) and control ({\it bottom}) configurations of the OneWeb LEO constellation compared against the MiSO counterpart.}
	\label{fig:onewebMisoResults}
\end{figure}

\begin{figure}[h!]
	\centering    
	\includegraphics[trim = 0.6in 0.1in 1.8in 0.8in, clip,width=0.575\textwidth]{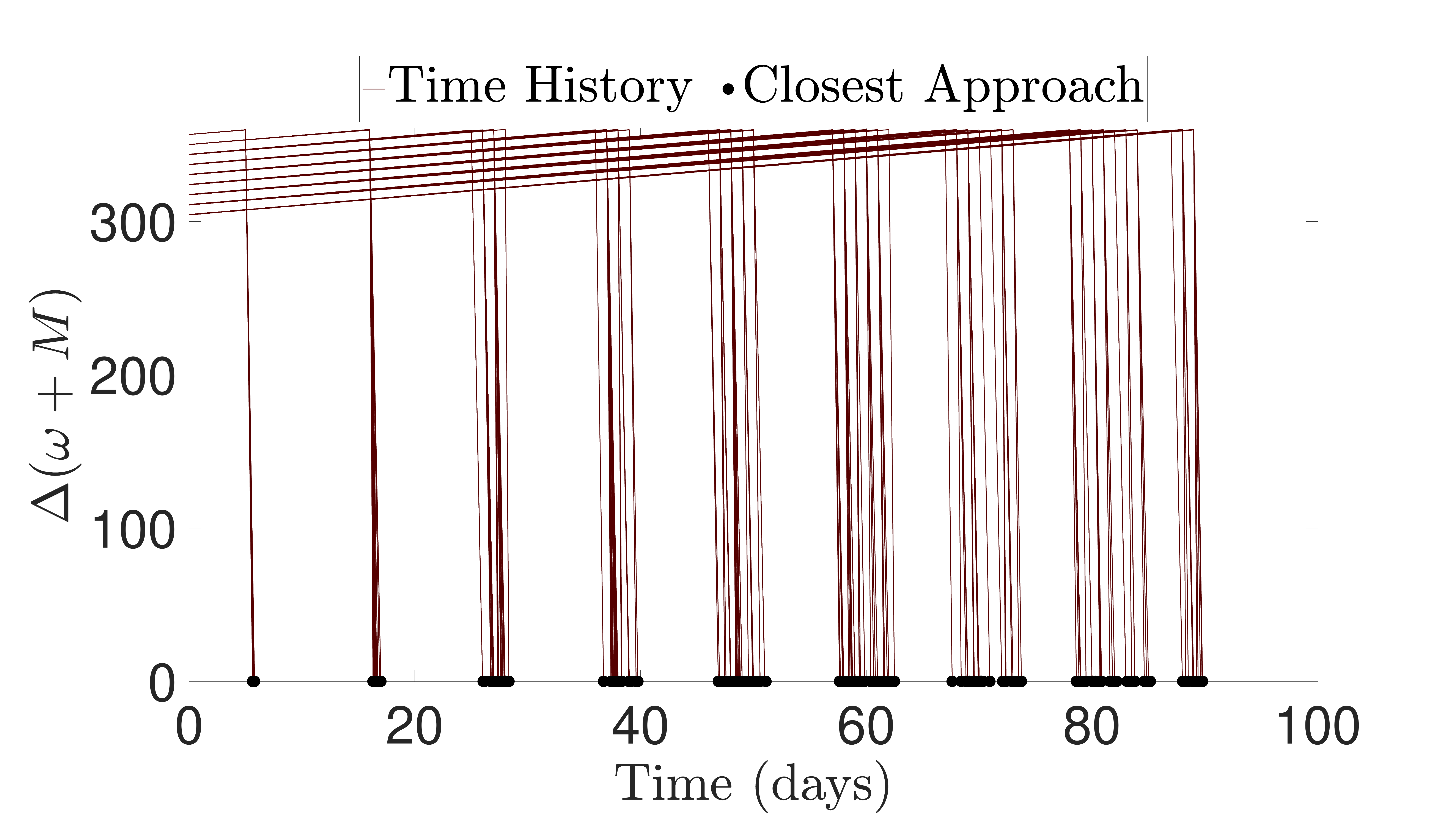}
	\caption{Close approaches within the OneWeb MiSO satellites only occur when the difference in $\omega + M$ of the target and field satellites is zero. As the orbits become more perturbed from their initial configuration, the frequency with which this occurs becomes increasingly less consistent.}
	\label{fig:onewebPeriodicity}
\end{figure}

In the endogenous assessment of the OneWeb LEO constellation, a time span of interest of 90 days is considered with close-approach distances of $\tau_1 = 20$ km, $\tau_2 = 5$ km, and $\tau_3 = 1$ km, respectively, for the first, second, and third filters. Running on the UA’s HPC cluster with over 200 CPUs for 13 hours registered a staggering $2,522$ unique close approaches of less than 1 km within the nominal case, as well as a minimum approach distance of 6.4 m. As seen in Fig.~\ref{fig:onewebMisoResults}, as many as 75 close approaches of 1 km or less occur on a single day within the nominal configuration. The periodic spikes in the frequency of close approaches experienced by the constellation is equivalent to the period of $\Delta(\omega + M)$ of the approaching target and field objects, which changes as the satellites become increasingly perturbed (Fig.~\ref{fig:onewebPeriodicity}).

\begin{table}[h!]
\centering
\caption{The number of endogenous close approaches within 1 km and the minimum approach distance experienced by the OneWeb LEO constellation after 90 days of operation, compared against the control case and MiSO variant.}
\label{tab:onewebComparison}
\begin{tabular}{@{}lcc@{}}
\toprule
ID & Number of Approaches & Minimum Approach Distance (km) \\ \midrule
Nominal & 2522 & 0.0064 \\
Control & 762 & 0.0353 \\
MiSO & 232 & 0.5502 \\ \bottomrule
\end{tabular}
\end{table}

The MiSO variant of the OneWeb LEO constellation is then evaluated following the same procedure. In stark contrast to the nominal configuration, after 90 days of operation, the MiSO variant only experiences 232 close approaches within 1 km with a minimum approach distance of 550 m (Table~\ref{tab:onewebComparison}). Although MiSO significantly outperforms the nominal configuration, it is not without its flaws. The periodic spikes in the close approach frequencies could be cause for concern, fortunately they are quite predictable since they occur when $\Delta(M + \omega) = 0$ and $\lambda = i$, which is the maximum latitude $\lambda$ that the satellites can experience. Furthermore, the approaches are limited to about half a kilometer and would likely not pose a significant threat as compared to those in the nominal case. 

\begin{figure}[h!]
	\centering    
	\includegraphics[trim = 1.8in 0.2in 2.5in 1.4in, clip,width=0.65\textwidth]{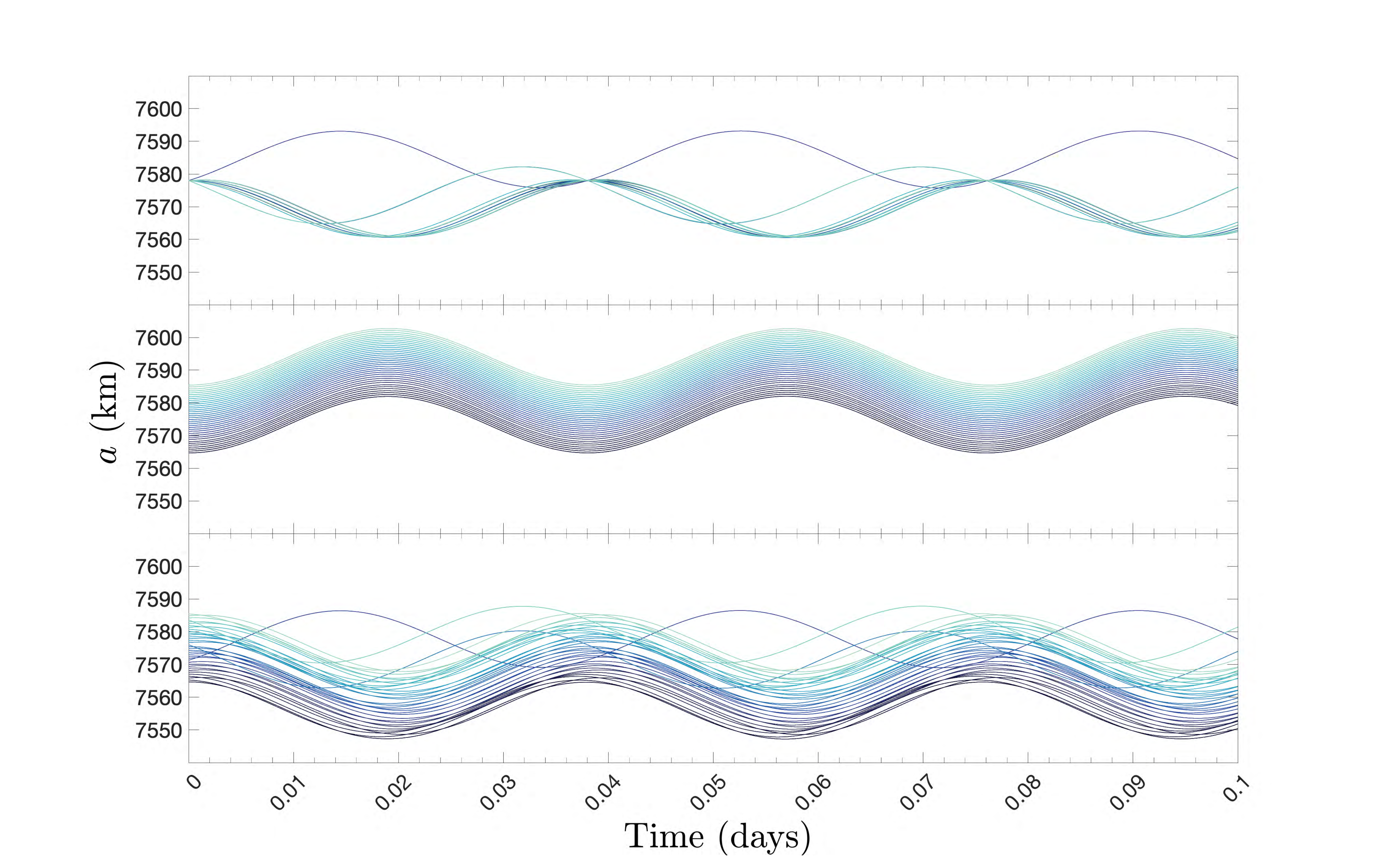}
	\caption{Comparison of the evolution of the osculating semi-major axis of a satellite in each plane of the nominal ({\it top}), MiSO ({\it middle}), and control case ({\it bottom}) configurations of the OneWeb LEO constellation.}
	\label{fig:misoVsControlSMA}
\end{figure}

Recall that the MiSO variant of the OneWeb constellation is generated using one optimized initial condition for each orbital plane, and such that the difference in initial altitude of the adjacent planes is 600 m. In order to test whether or not the reduced frequency and magnitude of close approaches experienced by the MiSO configuration was caused by the ``frozen'' nature of the orbits, or merely due to the difference in altitude between the planes, a control case was designed. The orbital elements for the control case are identical to the nominal configuration with the exception that the altitude of the planes was modified to match the altitudes of the initial conditions used to generate the MiSO constellation. As indicated by Fig.~\ref{fig:onewebMisoResults}, the MiSO variant significantly outperforms the control constellation, showing that simply spacing out the planes is not sufficient to minimize risk. The special nature of our design is shown Fig.~\ref{fig:misoVsControlSMA}, where the adjacent orbital plane spacing is conserved in the MiSO configuration, but not in the nominal or control cases.

\begin{table}[h!]
\centering
\caption{Close approaches less than 1 km predicted by HERA within the target planes of the nominal and MiSO OneWeb constellations.}
\label{tab:onewebHERA}
\begin{tabular}{@{}lll@{}}
\toprule
ID      & Predicted Approaches & False-Positives \\ \midrule
Nominal & 1                    & 1               \\ 
MiSO    & 0                    & 0               \\ \bottomrule
\end{tabular}
\end{table}

\begin{figure}[h!]
	\centering    
	\includegraphics[width=0.45\textwidth]{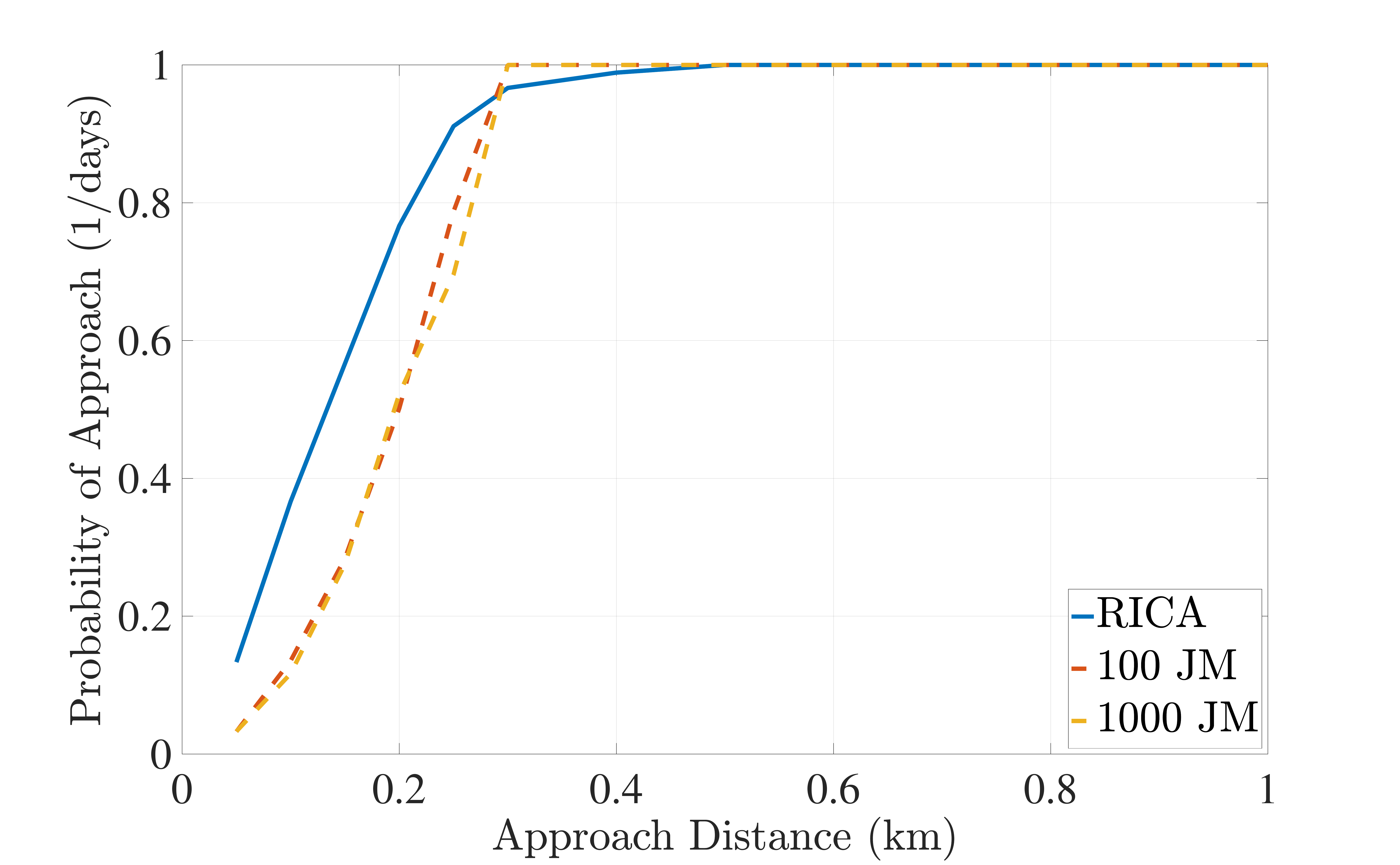} 
	\includegraphics[width=0.45\textwidth]{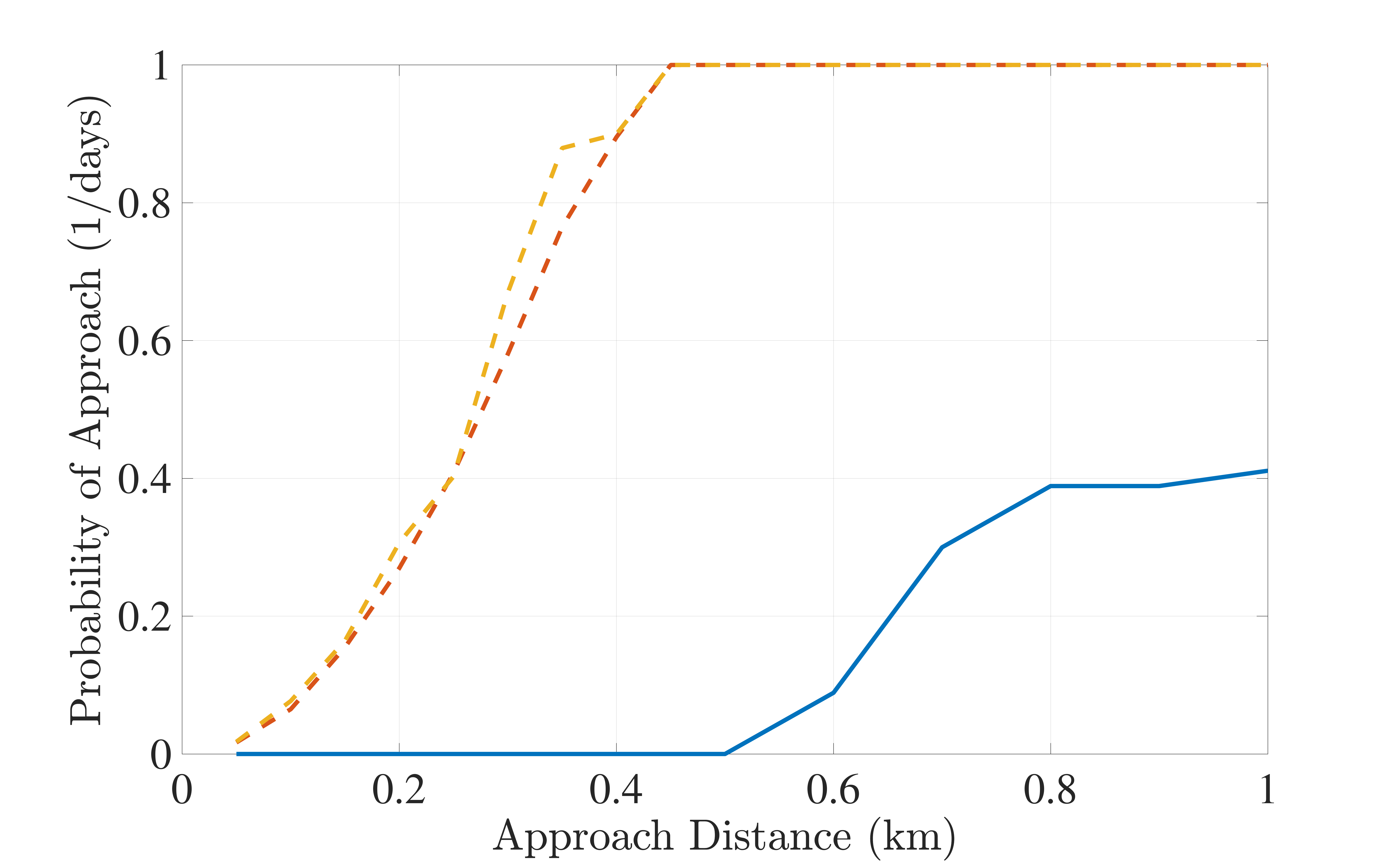}
	\caption{Comparison of the collision probability of the OneWeb nominal ({\it left}) and MiSO ({\it right} configurations.}	
	\label{fig:onewebJM}
\end{figure}

The performance of \texttt{JM} can be seen in Fig.~\ref{fig:onewebJM}. In both the nominal and MiSO cases, the probabilities of both the 100 and 1000 clone runs agree quite well. Furthermore, for the nominal case, \texttt{JM} does a reasonably good job at predicting the probability of approaches that occur between 0 and 0.25 km and extremely well at predicting approaches beyond this range; however it performs quite poorly for the MiSO configuration, for approaches of all distances other than 0.05 km (where it accurately captures the low probability of approach). The inability of \texttt{JM} to predict the probability of approaches greater than 0.05 km unfortunately indicates that the method is not sensitive enough to ``respond'' to subtle differences between the nominal and MiSO configurations. Despite its inability to accurately predict the close approach probability for the MiSO configuration, \texttt{JM} vastly outperforms the purely MOID-based \texttt{HERA} algorithm (Table~\ref{tab:onewebHERA}), which only predicted one close approach that upon post processing was found to be a false-positive.

\section{Endogenous Assessment of the SpaceX Starlink Constellation}

\begin{figure}[htp!]
	\centering    
	\includegraphics[trim = 0.7in 0.1in 2.2in 0.8in, clip,scale=0.5,width=0.4\textwidth]{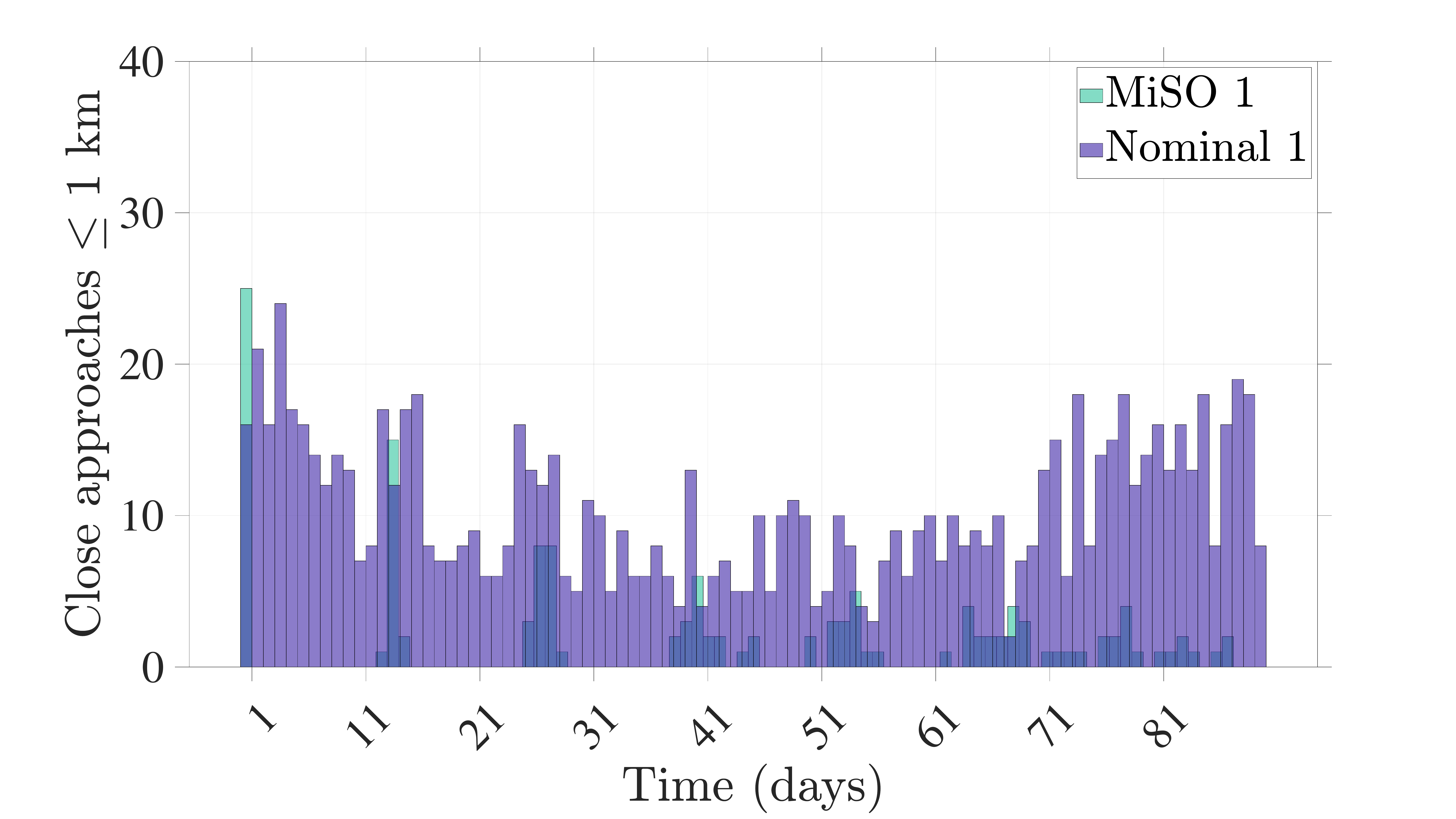}
	\includegraphics[trim = 1.1in 0.1in 2.2in 0.8in, clip,scale=0.5,width=0.4\textwidth]{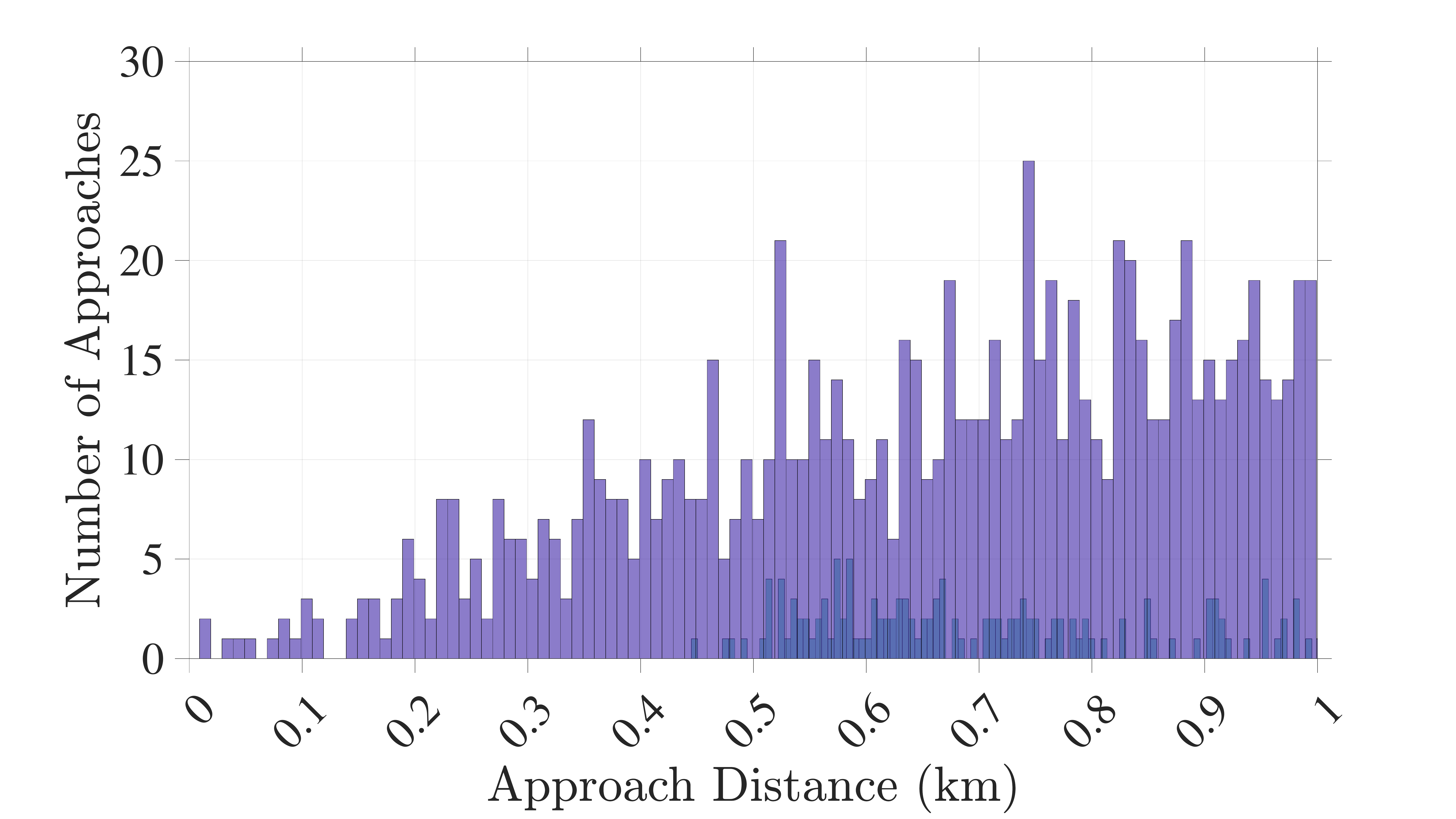}
	\includegraphics[trim = 0.7in 0.1in 2.2in 0.8in, clip,scale=0.5,width=0.4\textwidth]{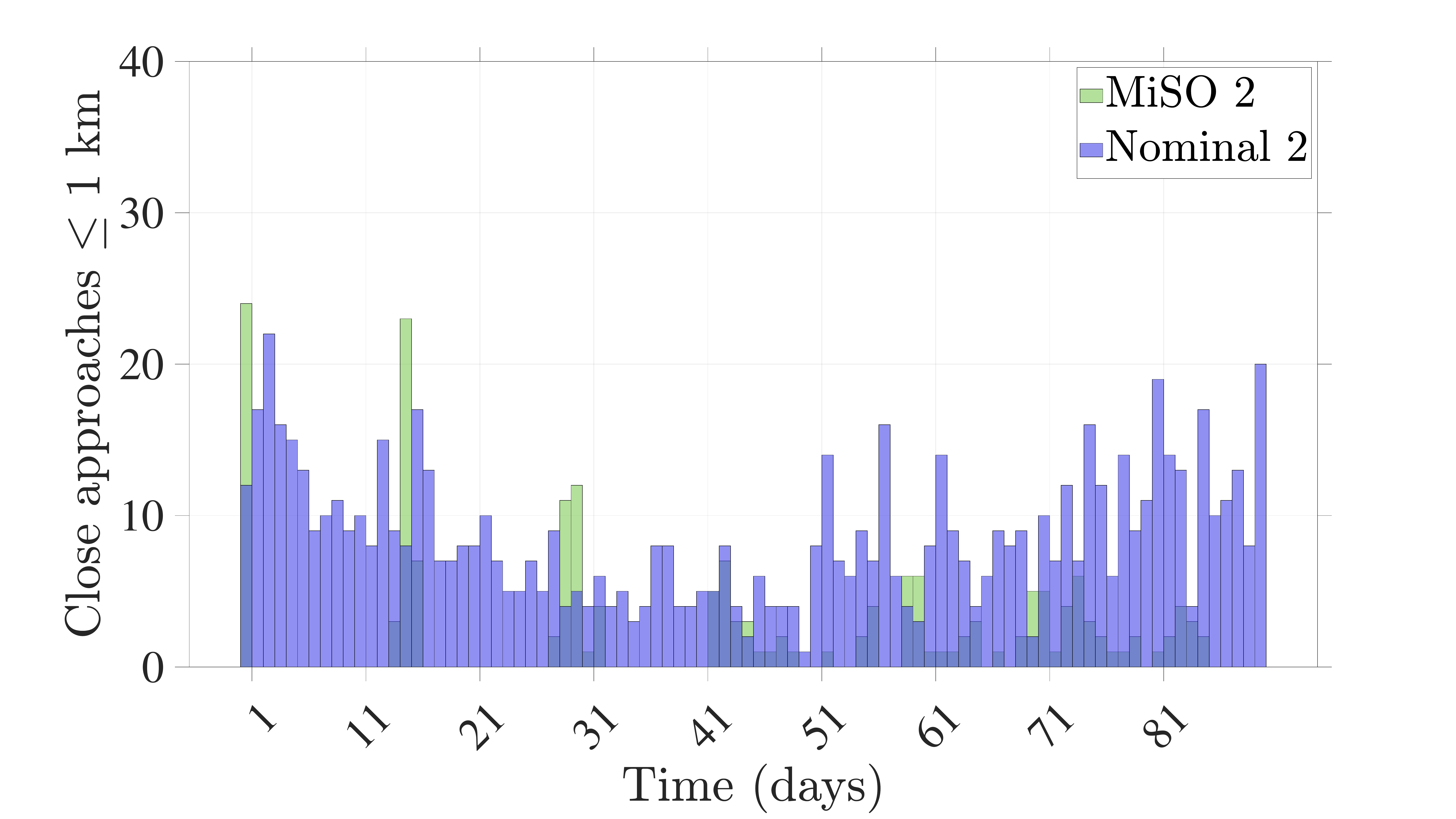}
	\includegraphics[trim = 1.1in 0.1in 2.2in 0.8in, clip,scale=0.5,width=0.4\textwidth]{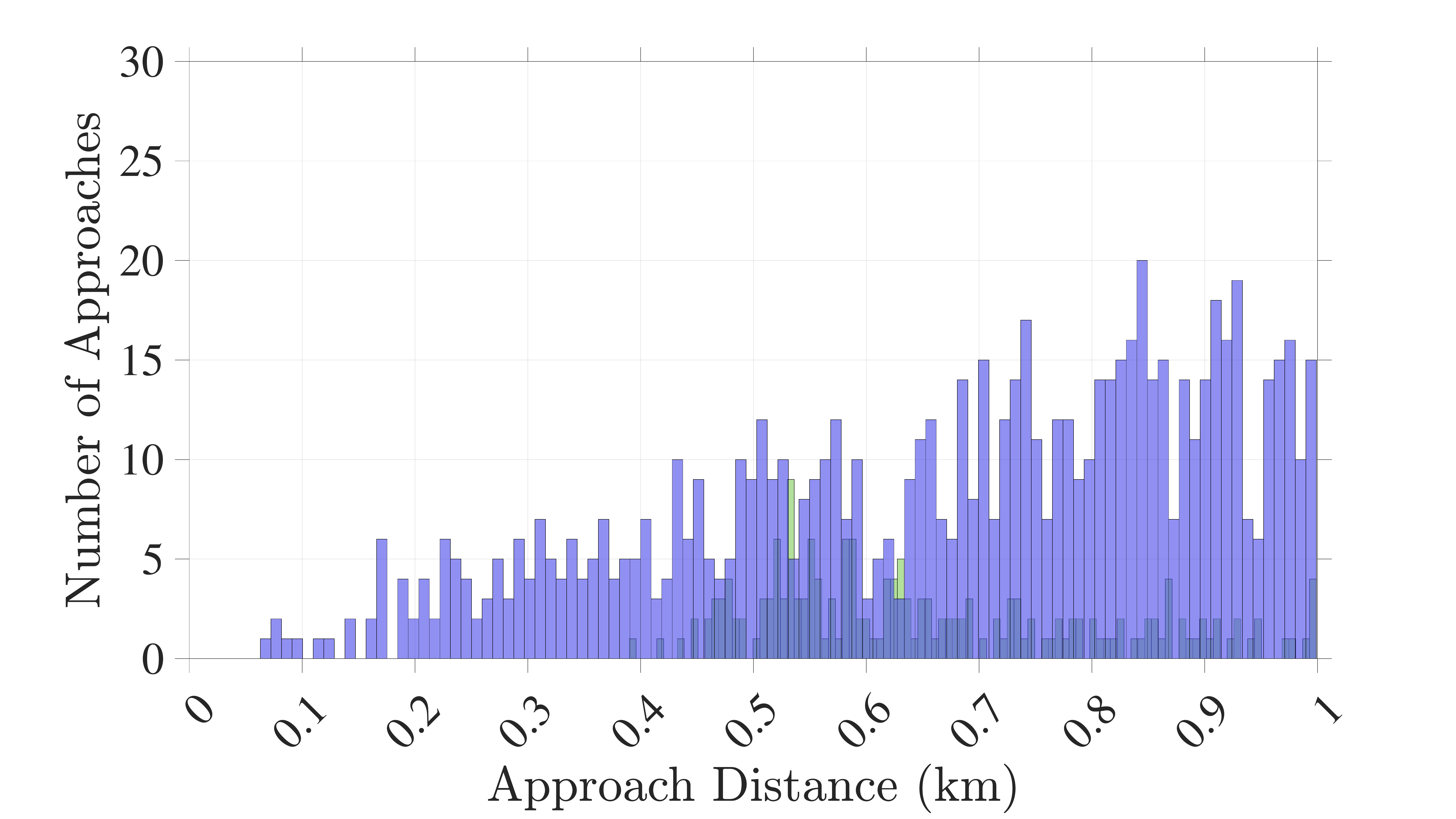}
	\includegraphics[trim = 0.7in 0.1in 2.2in 0.8in, clip,scale=0.5,width=0.4\textwidth]{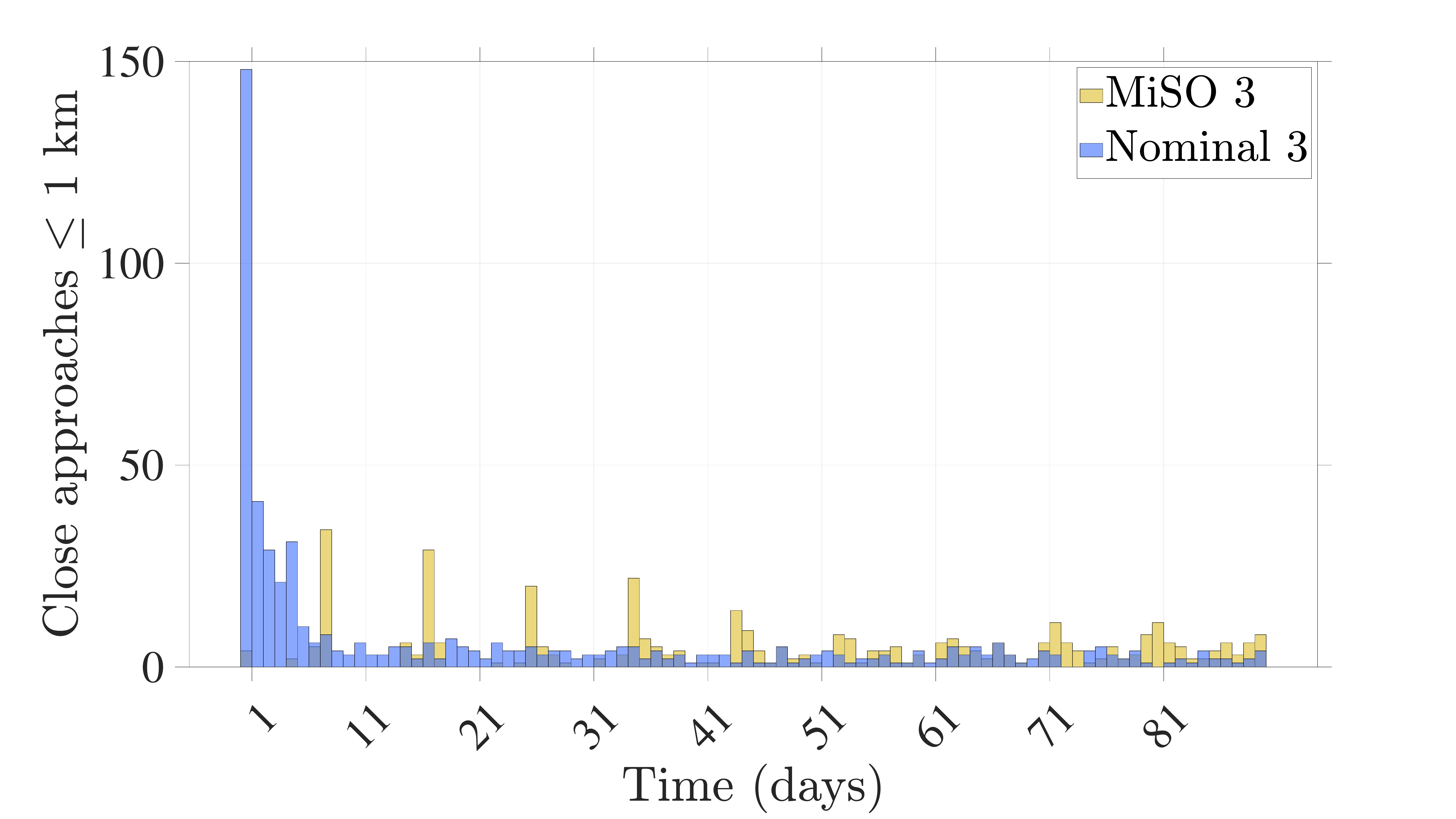}
	\includegraphics[trim = 1.1in 0.1in 2.2in 0.8in, clip,scale=0.5,width=0.4\textwidth]{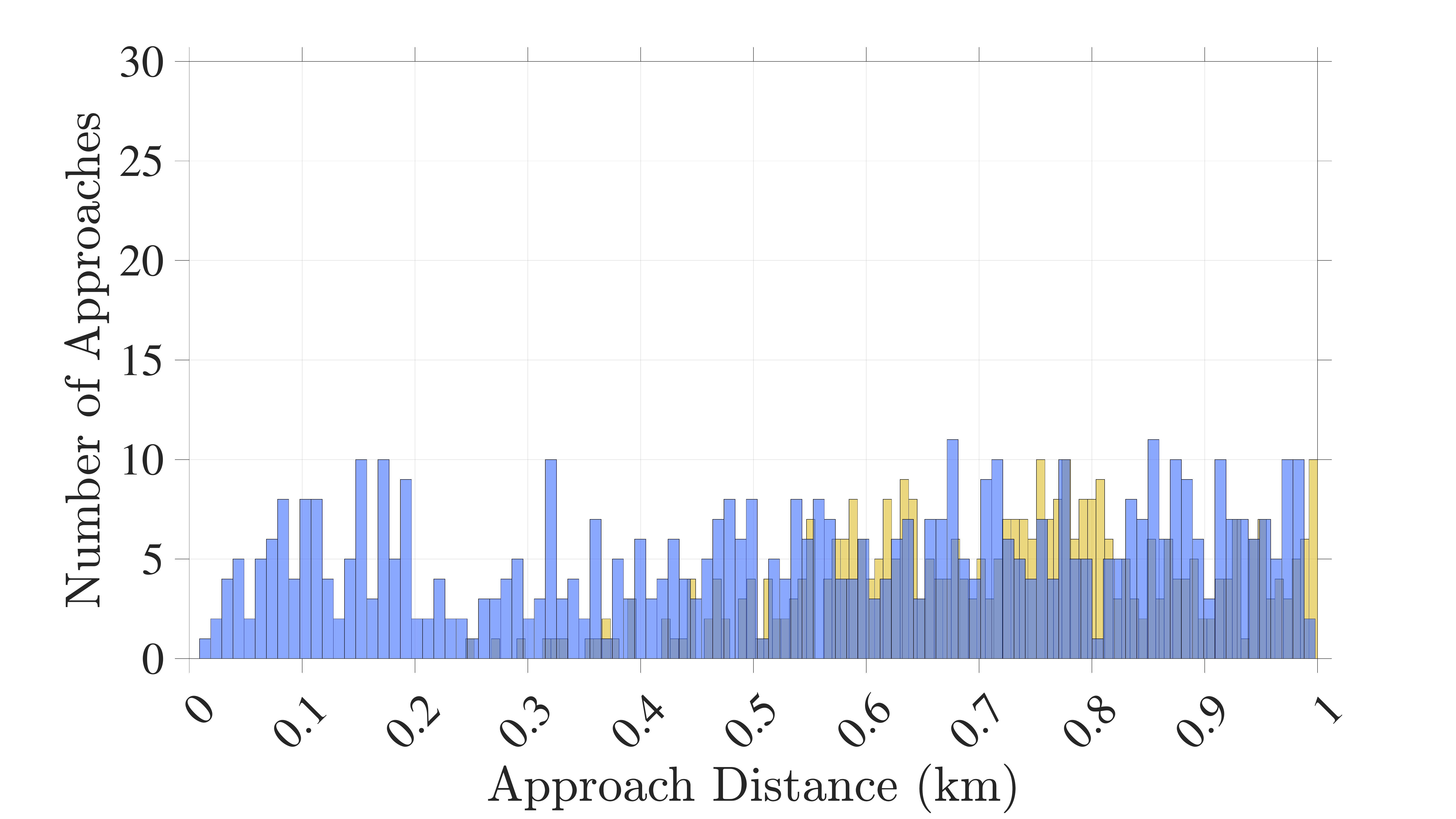}
	\includegraphics[trim = 0.7in 0.1in 2.2in 0.8in, clip,scale=0.5,width=0.4\textwidth]{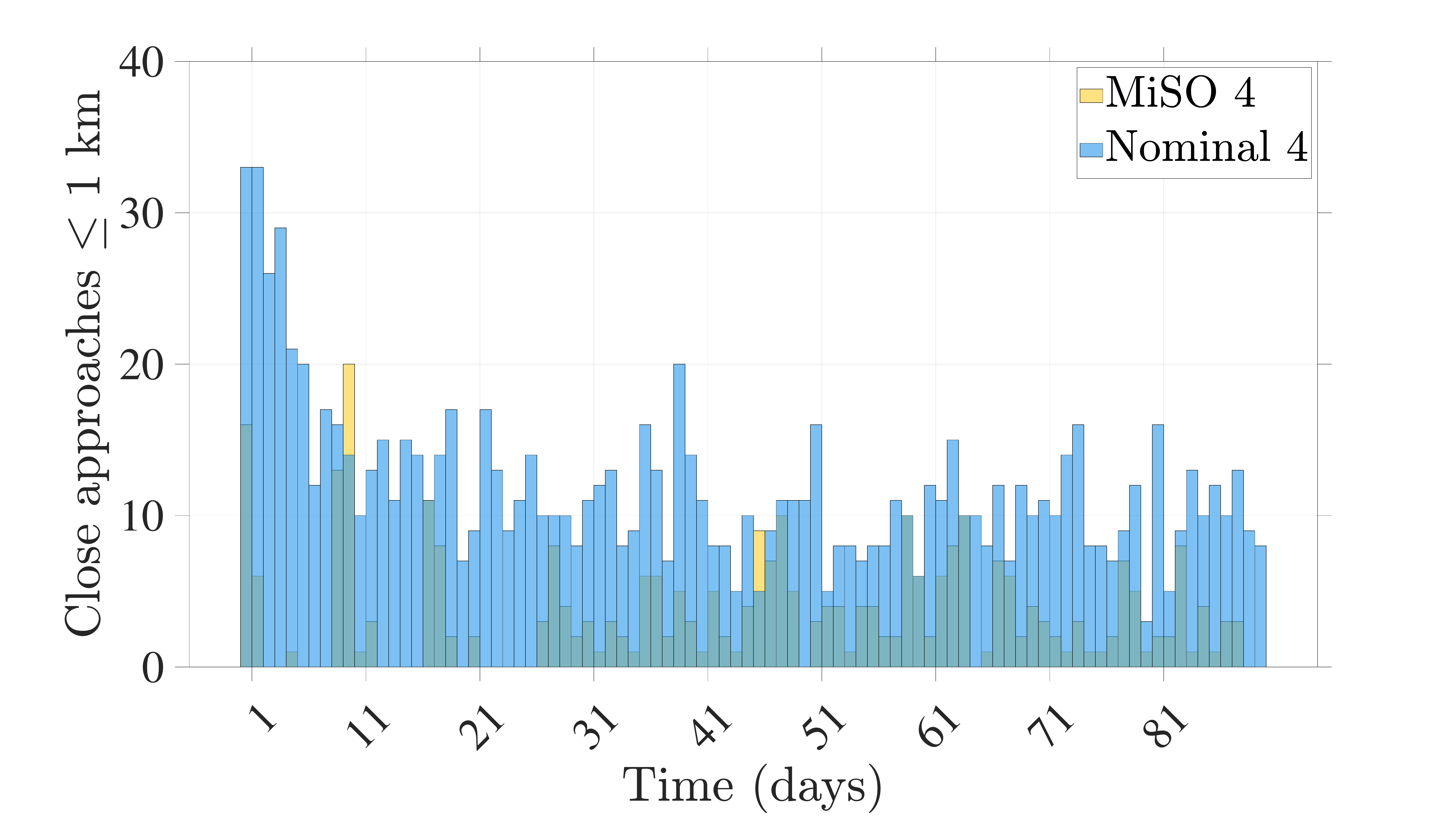}
	\includegraphics[trim = 1.1in 0.1in 2.2in 0.8in, clip,scale=0.5,width=0.4\textwidth]{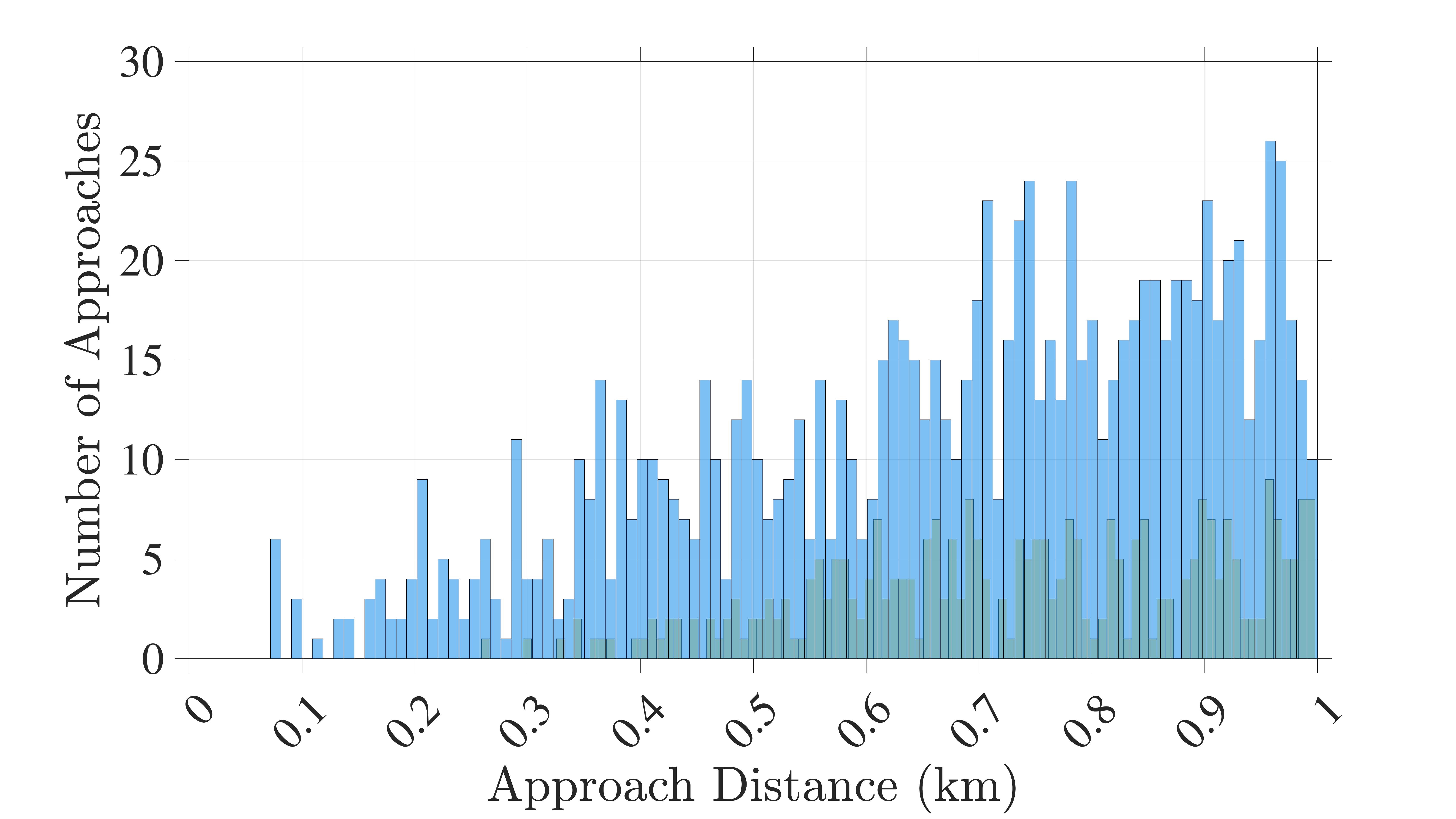}
	\includegraphics[trim = 0.7in 0.1in 2.2in 0.8in, clip,scale=0.5,width=0.4\textwidth]{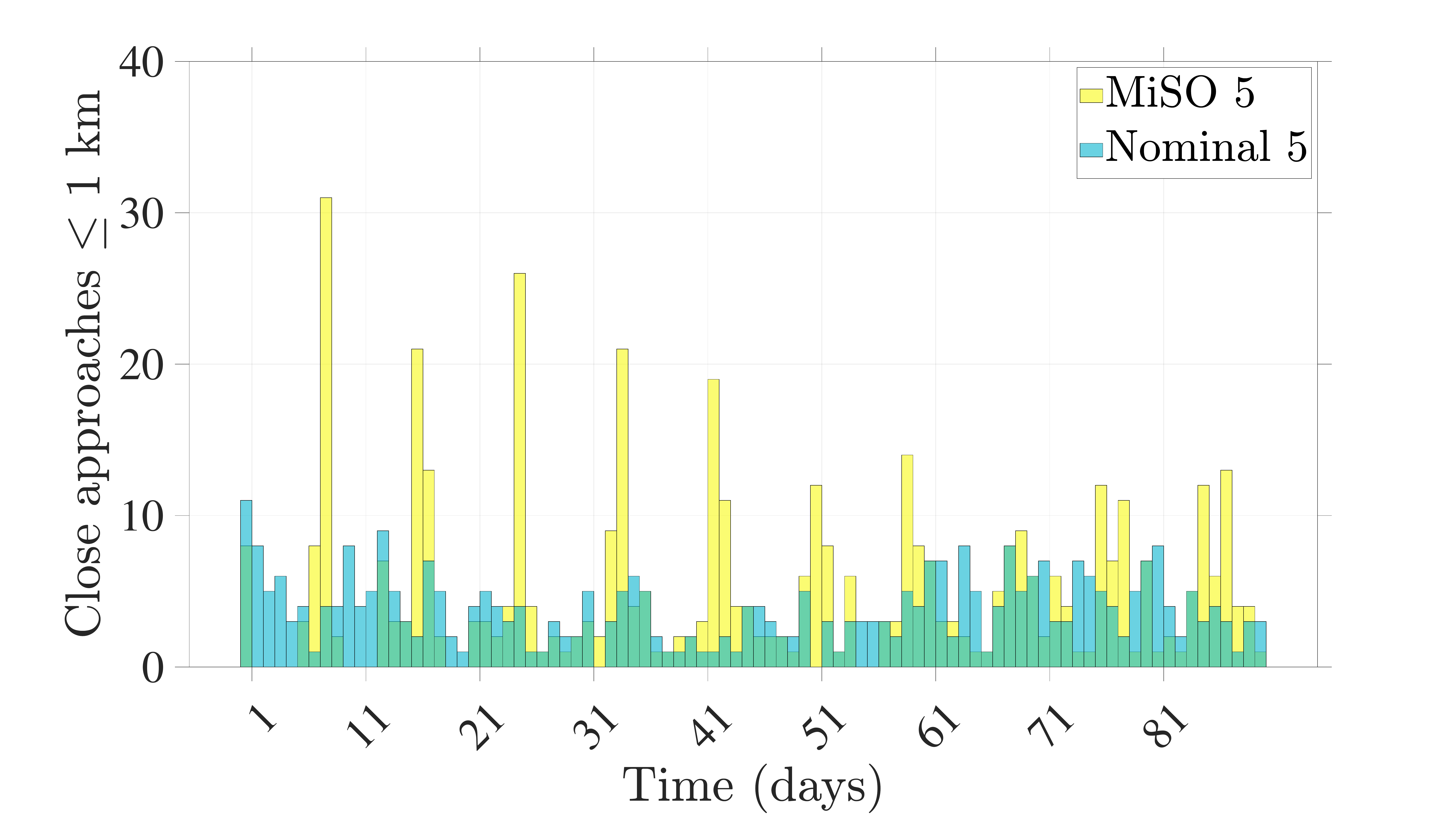}
	\includegraphics[trim = 1.1in 0.1in 2.2in 0.8in, clip,scale=0.5,width=0.4\textwidth]{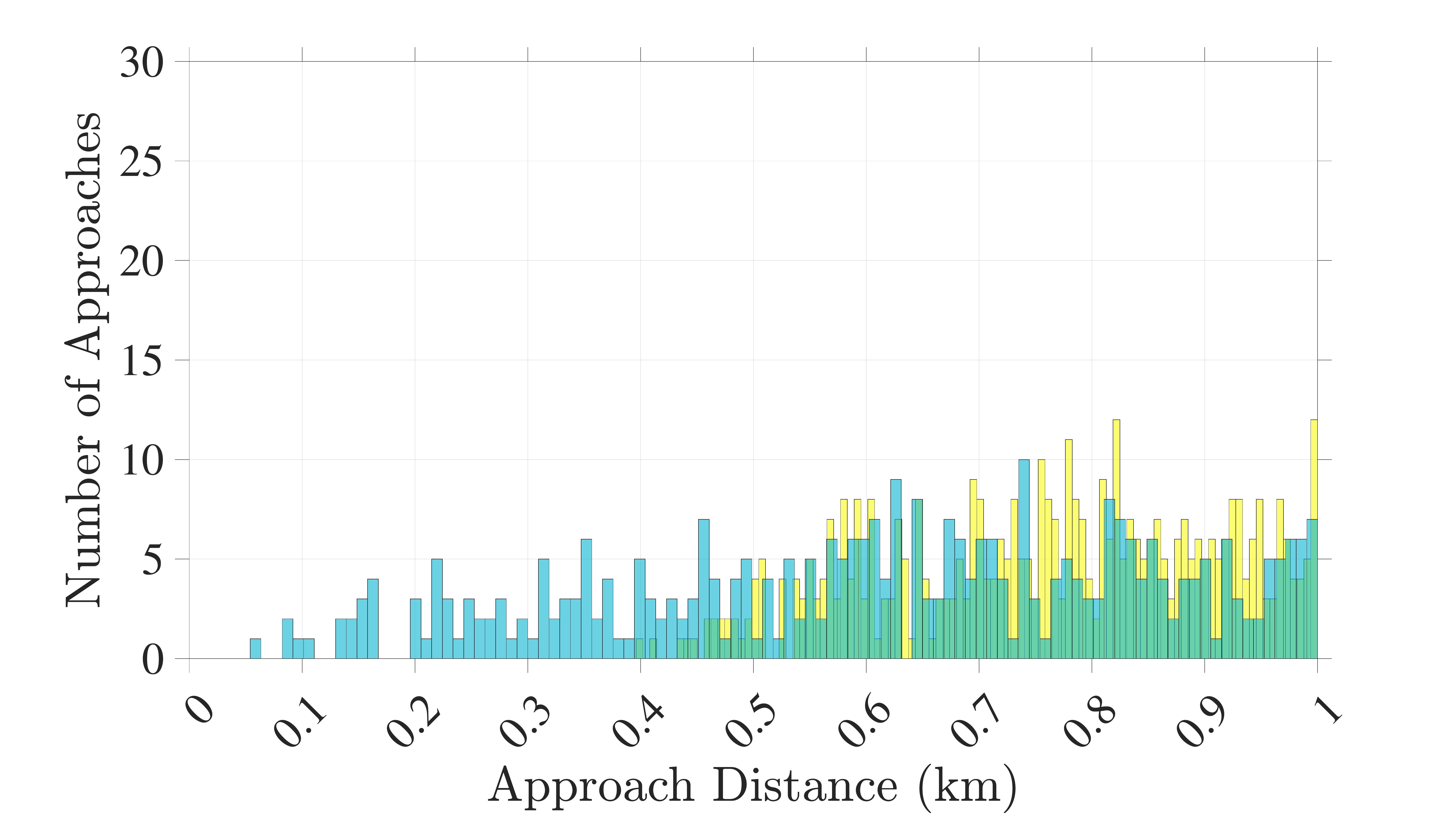}
	\caption{Frequency of close approaches within 1 km ({\it left}) and close approach distance ({\it right}) for each shell of the SpaceX Starlink constellation compared against the MiSO counterpart.}
	\label{fig:starlinkMisoResults}
\end{figure}

To investigate the collision risk of the Starlink constellation, each set of target and field satellites for both the nominal and MiSO variants (Table~\ref{tab:starlinkTarget}) were run with \texttt{RICA}, \texttt{JM}, and \texttt{HERA} for a time span of 90 days. Fig.~\ref{fig:starlinkMisoResults} shows that every nominal Starlink target plane has a significantly smaller minimum approach distance than their MiSO counterparts, with the statistics provided in Table~\ref{tab:starlinkComparison}. Additionally, with the exception of Plane~5, every MiSO plane experiences less total close approaches than the nominal versions of the same planes.

\begin{table}[h!]
\centering
\caption{The number of endogenous close approaches within 1 km calculated by \texttt{RICA} and the minimum approach distance experienced by the SpaceX Starlink constellation after 90 days of operation, compared against the MiSO variant.}
\label{tab:starlinkComparison}
\begin{tabular}{@{}lcc@{}}
\toprule
ID & Number of Approaches & Minimum Approach Distance (km) \\ \midrule
Nominal 1 & 940 & 0.0167 \\
MiSO 1 & 135 & 0.4482 \\
Nominal 2 & 783 & 0.0631 \\
MiSO 2 & 182 & 0.3938 \\
Nominal 3 & 539 & 0.0172 \\
MiSO 3 & 375 & 0.2515 \\
Nominal 4 & 1068 & 0.0728 \\
MiSO 4 & 312 & 0.2612 \\
Nominal 5 & 346 & 0.0584 \\
MiSO 5 & 454 & 0.3968 \\ \bottomrule
\end{tabular}
\end{table}

\begin{figure}[htp!]
	\centering    
	\includegraphics[trim = 0.4in 0.6in 0.05in 1.2in, clip,scale=0.5,width=0.45\textwidth]{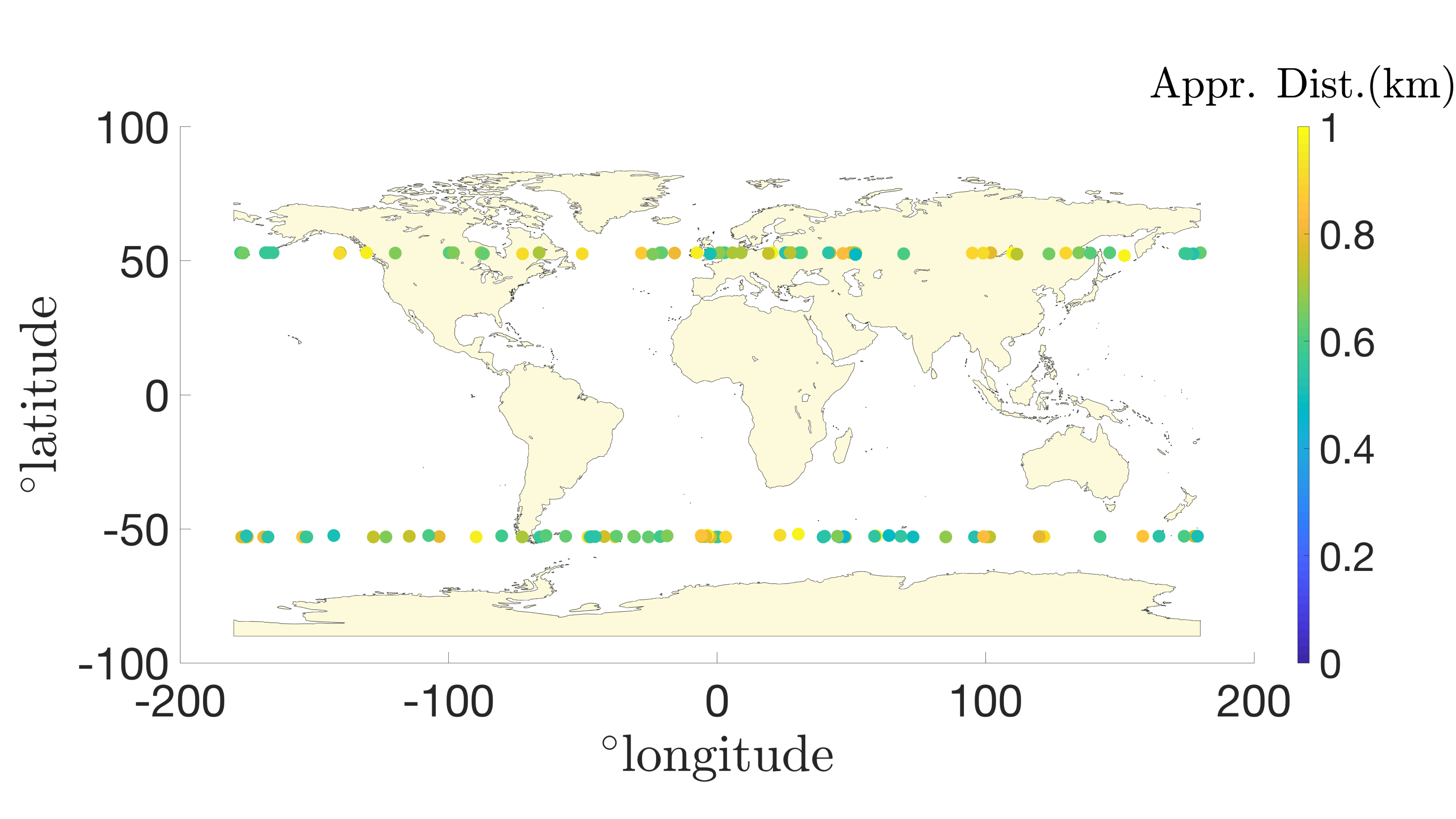}
	\includegraphics[trim = 0.5in 0.4in 1.9in 1.7in, clip,scale=0.5,width=0.45\textwidth]{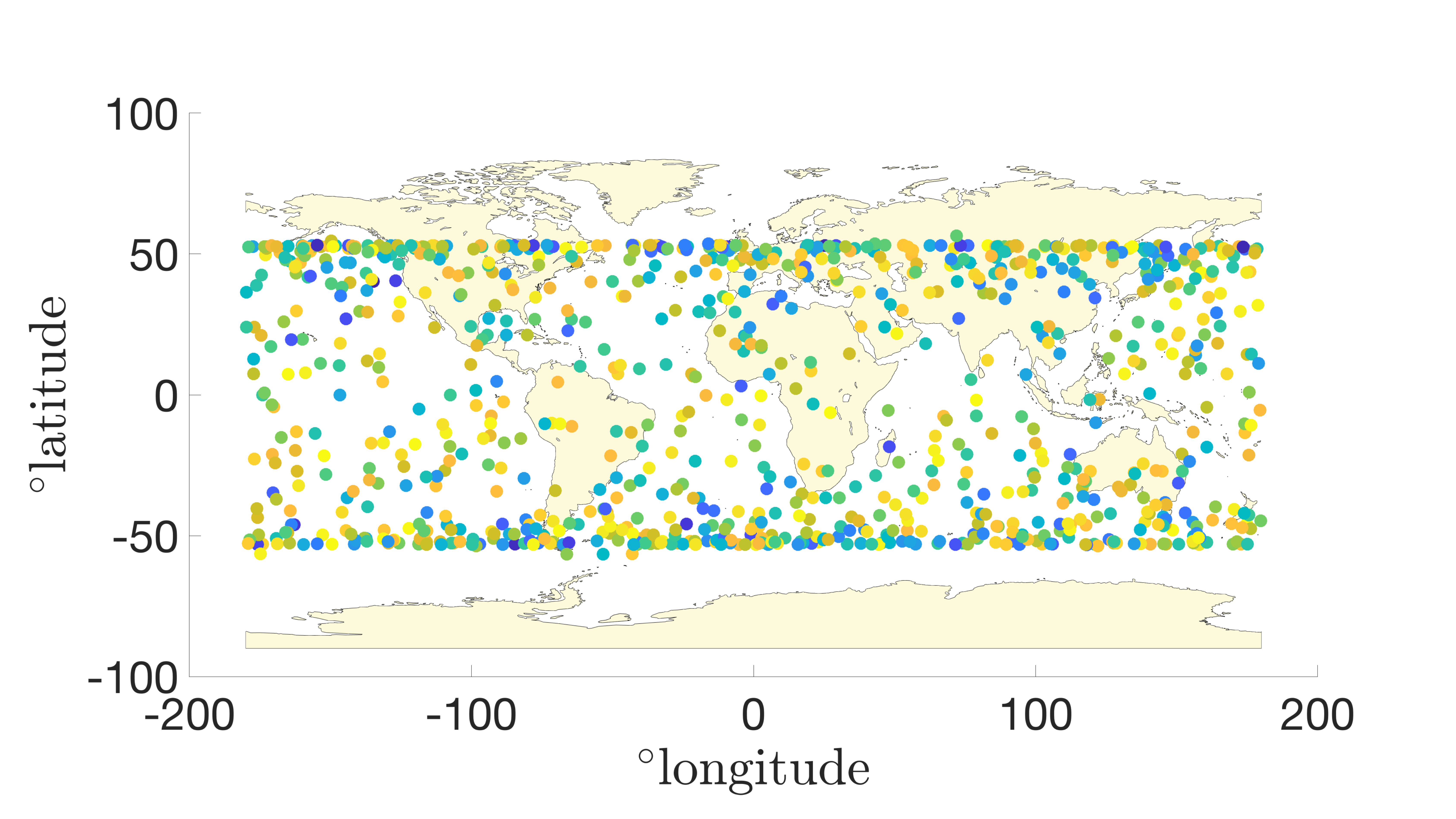}
	\caption{Latitude and longitude coordinates of the target satellites during close approach for the MiSO ({\it left}) and nominal ({\it right}) configurations.}
	\label{fig:spacexMAplusARGP}
\end{figure}

\begin{figure}[h!]
	\centering    
	\includegraphics[trim = 0.1in 0.1in 1.7in 1in, clip,scale=0.5,width=0.5\textwidth]{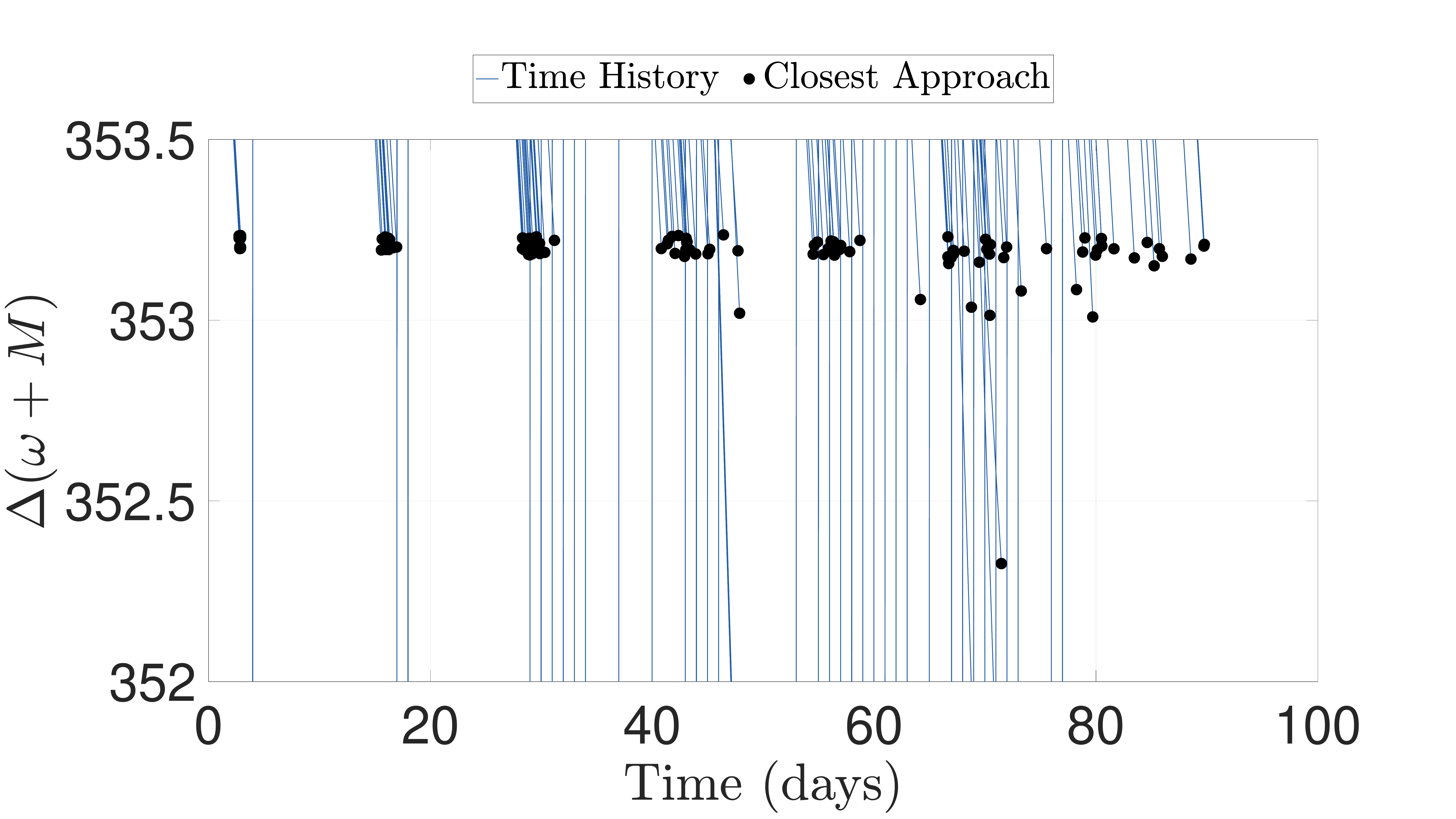}
	\includegraphics[trim = 0.6in 0.1in 1.8in 0.8in, clip,scale=0.5,width=0.45\textwidth]{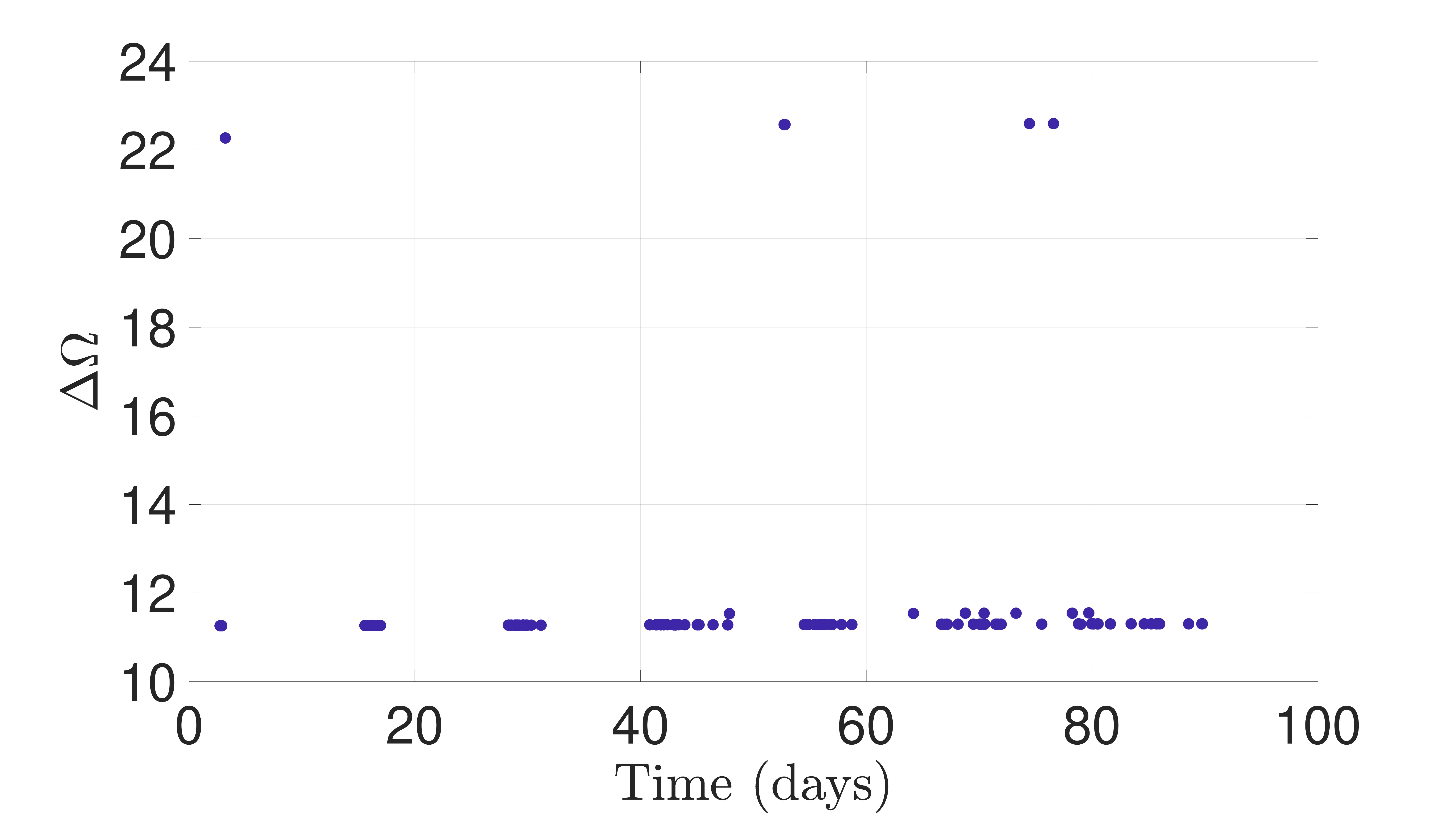}
	\caption{Close approaches within the Starlink MiSO satellites only occur when the difference in $\omega + M$ of the target and field satellites is near zero. As the orbits become more perturbed from their initial configuration, the frequency with which this occurs becomes increasingly less consistent.}
	\label{fig:spacexApprLoc}
\end{figure}

In Fig.~\ref{fig:starlinkMisoResults}, we see the same periodicity in the frequency of close approaches in the MiSO variants of the Starlink target planes that was observed in OneWeb (Fig.~\ref{fig:onewebMisoResults}). Fig.~\ref{fig:spacexApprLoc} shows that every close approach experienced by the satellites of the MiSO target Plane~1 occur near the maximum latitudes of the target satellites ($\pm {50}^{\circ}$). Although there is certainly an abundance of approaches at these same latitudes in the nominal target plane, there are also a significant number of approaches in the range between $\pm {50}^{\circ}$. Investigating further, in Fig.~\ref{fig:spacexMAplusARGP}, we notice that close approaches only occur when the difference in $\omega + M$ between the approaching target and field objects is around $355^{\circ}$. Initially, when the orbits of the target and field objects have not been significantly perturbed, the target and field objects approach one another at latitudes of $\pm {50}^{\circ}$ within $\Delta(\omega + M) = 355^{\circ}$ at a regular interval. In contrast to the OneWeb MiSO constellation (Fig.~\ref{fig:onewebPeriodicity}), we see in Fig.~\ref{fig:spacexMAplusARGP} that approaches occur at other values $\Delta(\omega + M)$, not just $360^{\circ}$. This same phenomenon occurs in each target plane of the Starlink MiSO constellation and is a result of approaches occurring between satellites of non-adjacent orbital planes or adjacent planes which have a relatively large difference in $\Omega$. Despite this difference, the periodicity of close approaches also decays in the Starlink MiSO target planes as the orbits are increasingly perturbed. The change in $\dot{M}$ and $\dot{\omega}$ of each satellite is distinct and therefore the moments when the approach conditions ($\Delta(\omega + M)$ and $\lambda$) are met no longer occur at the same interval and this previously observed periodicity of close approaches is slowly destroyed.

\begin{figure}[t!]
	\centering    
	\includegraphics[trim = 1.3in 0.2in 2.5in 0.9in, clip,scale=0.5,width=0.24\textwidth]{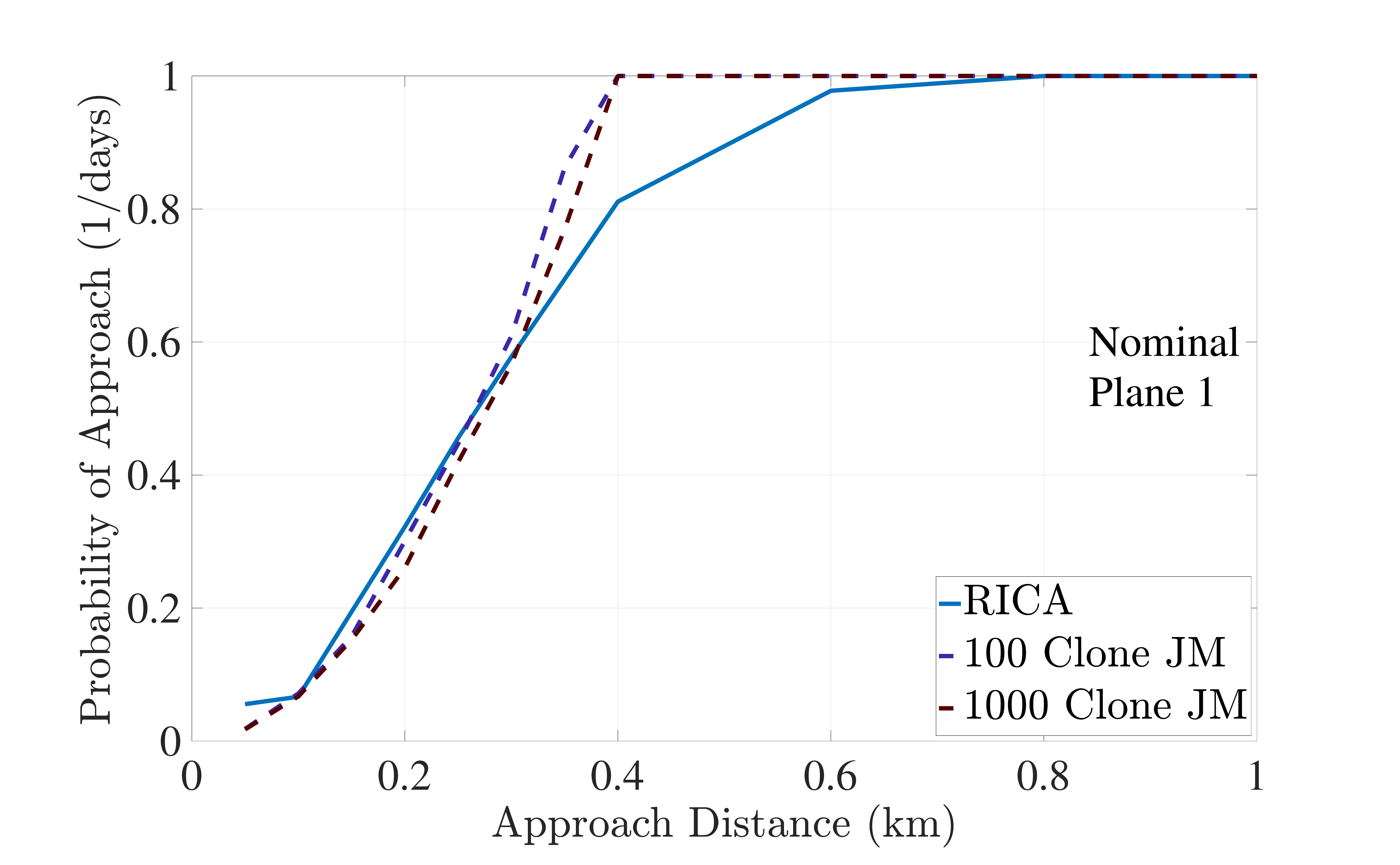}
	\includegraphics[trim = 1.3in 0.2in 2.5in 0.9in, clip,scale=0.5,width=0.24\textwidth]{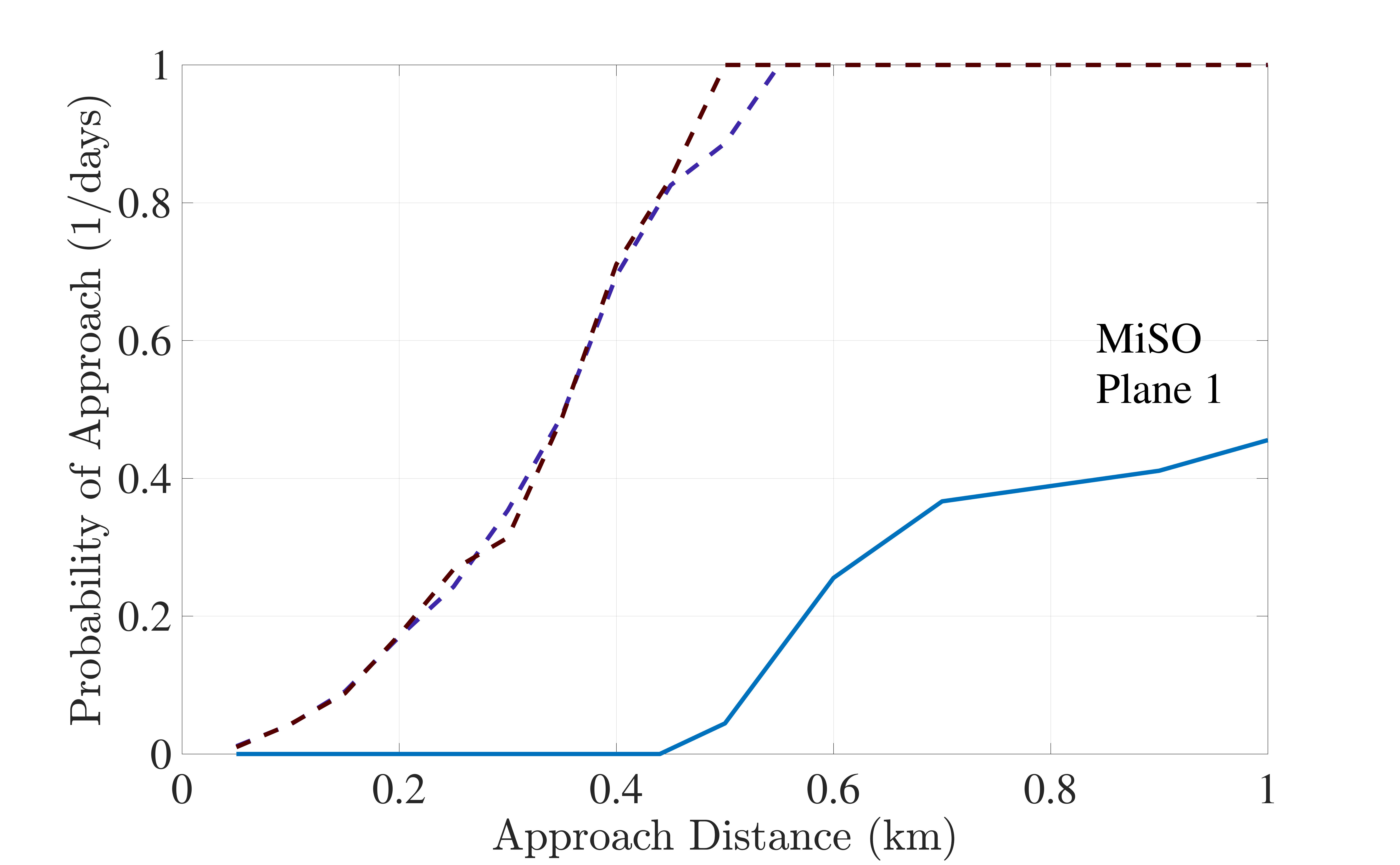}
	\includegraphics[trim = 1.3in 0.2in 2.5in 0.9in, clip,scale=0.5,width=0.24\textwidth]{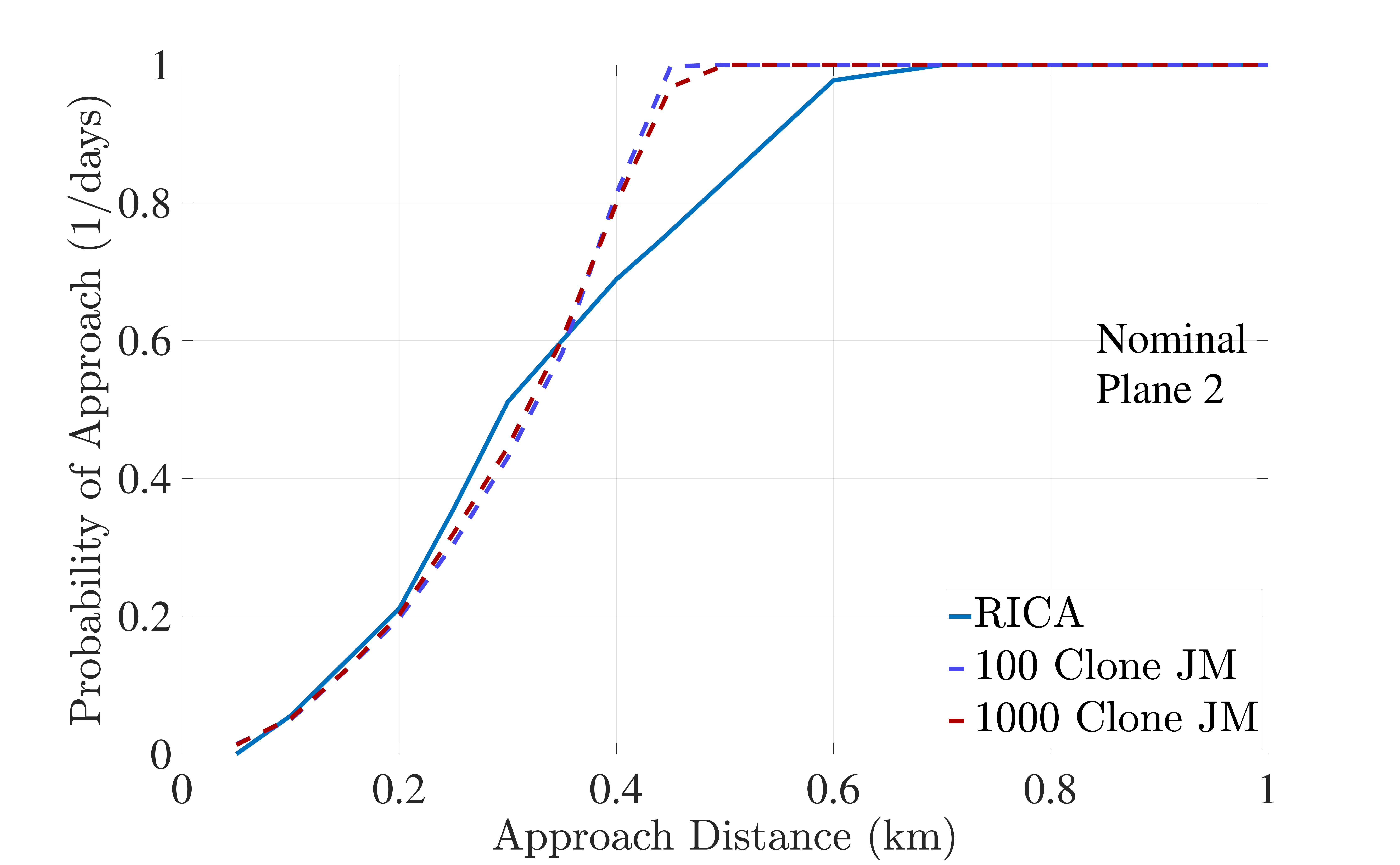}
	\includegraphics[trim = 1.3in 0.2in 2.5in 0.9in, clip,scale=0.5,width=0.24\textwidth]{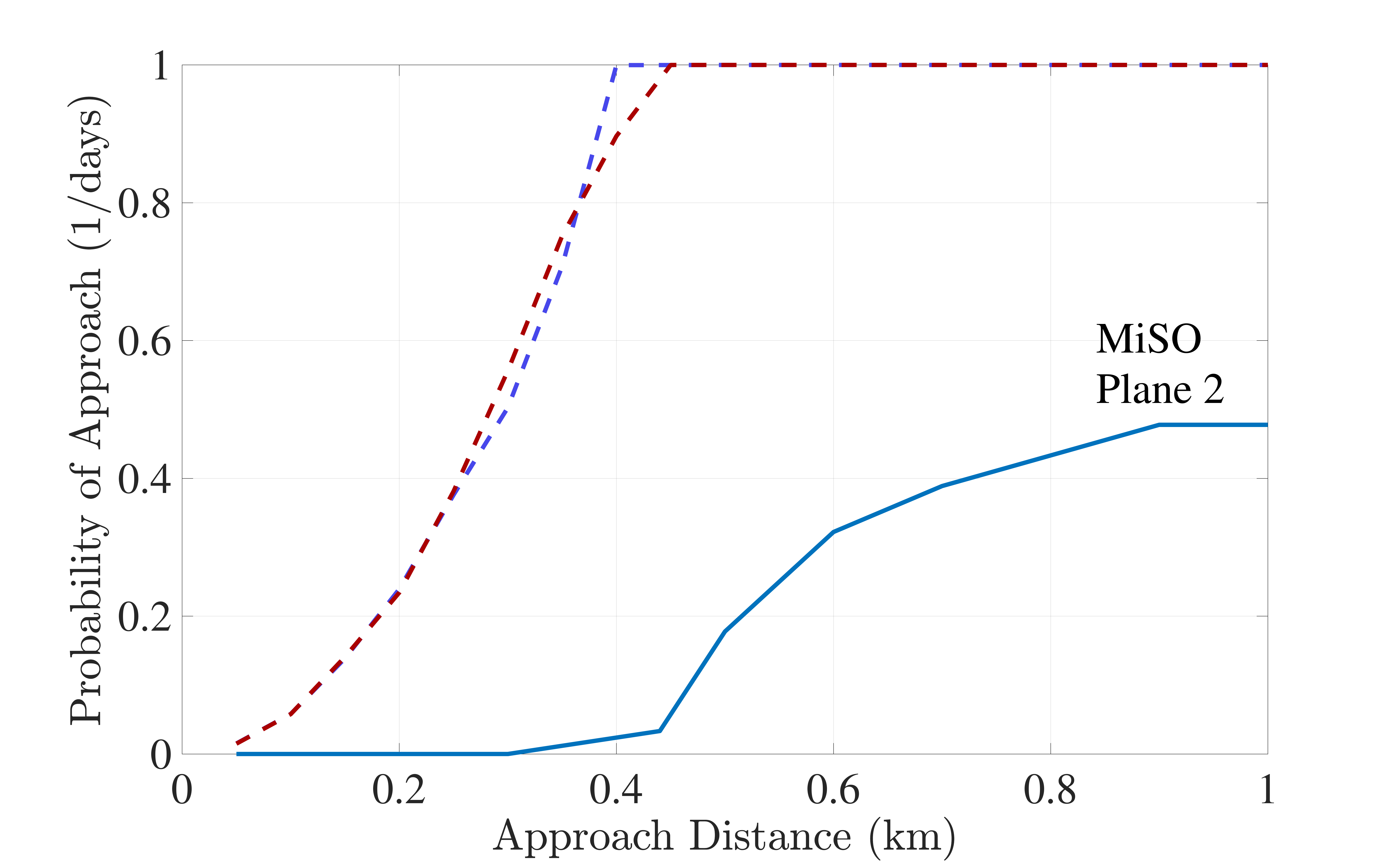}
	\includegraphics[trim = 1.3in 0.2in 2.5in 0.9in, clip,scale=0.5,width=0.24\textwidth]{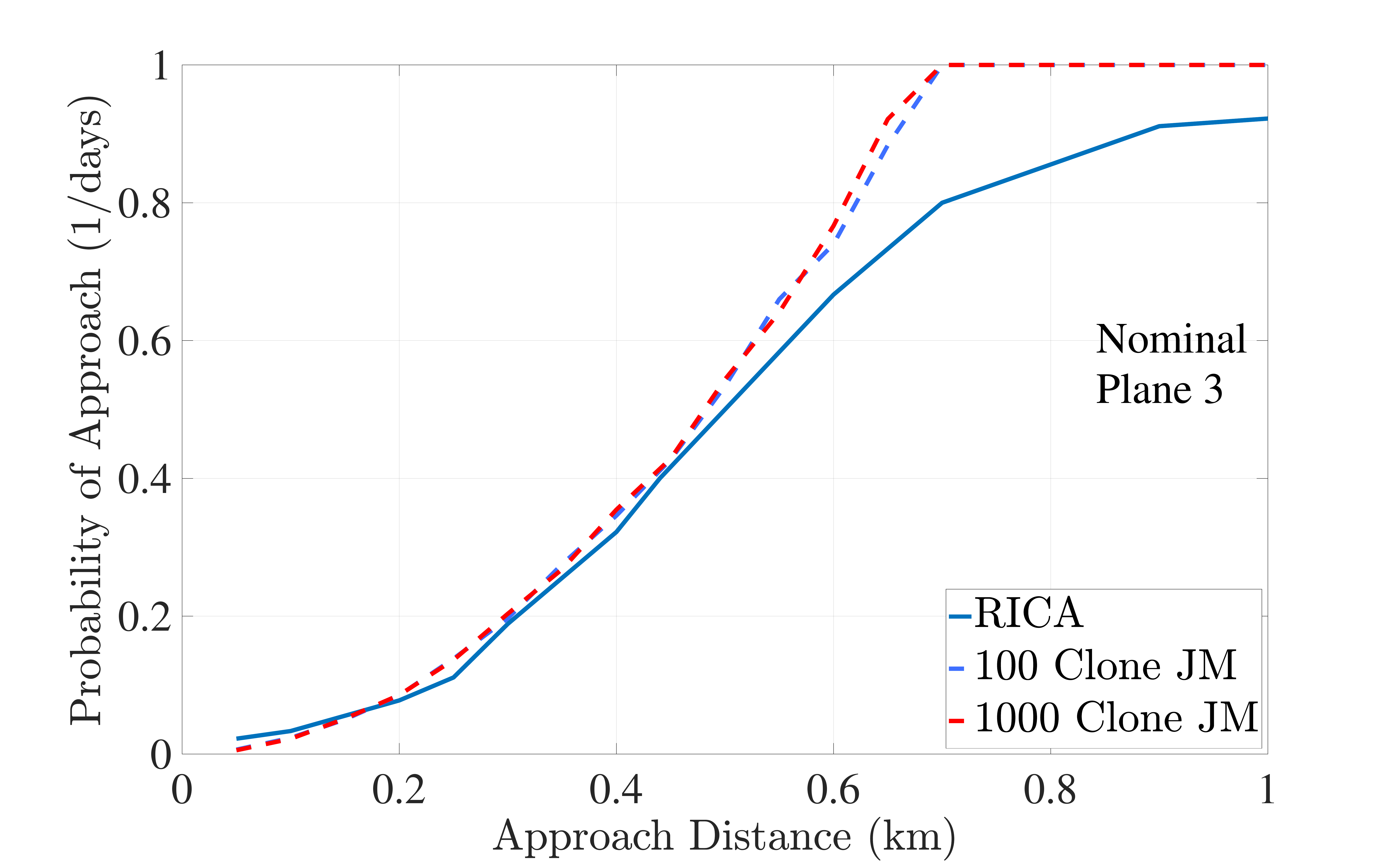}
	\includegraphics[trim = 1.3in 0.2in 2.5in 0.9in, clip,scale=0.5,width=0.24\textwidth]{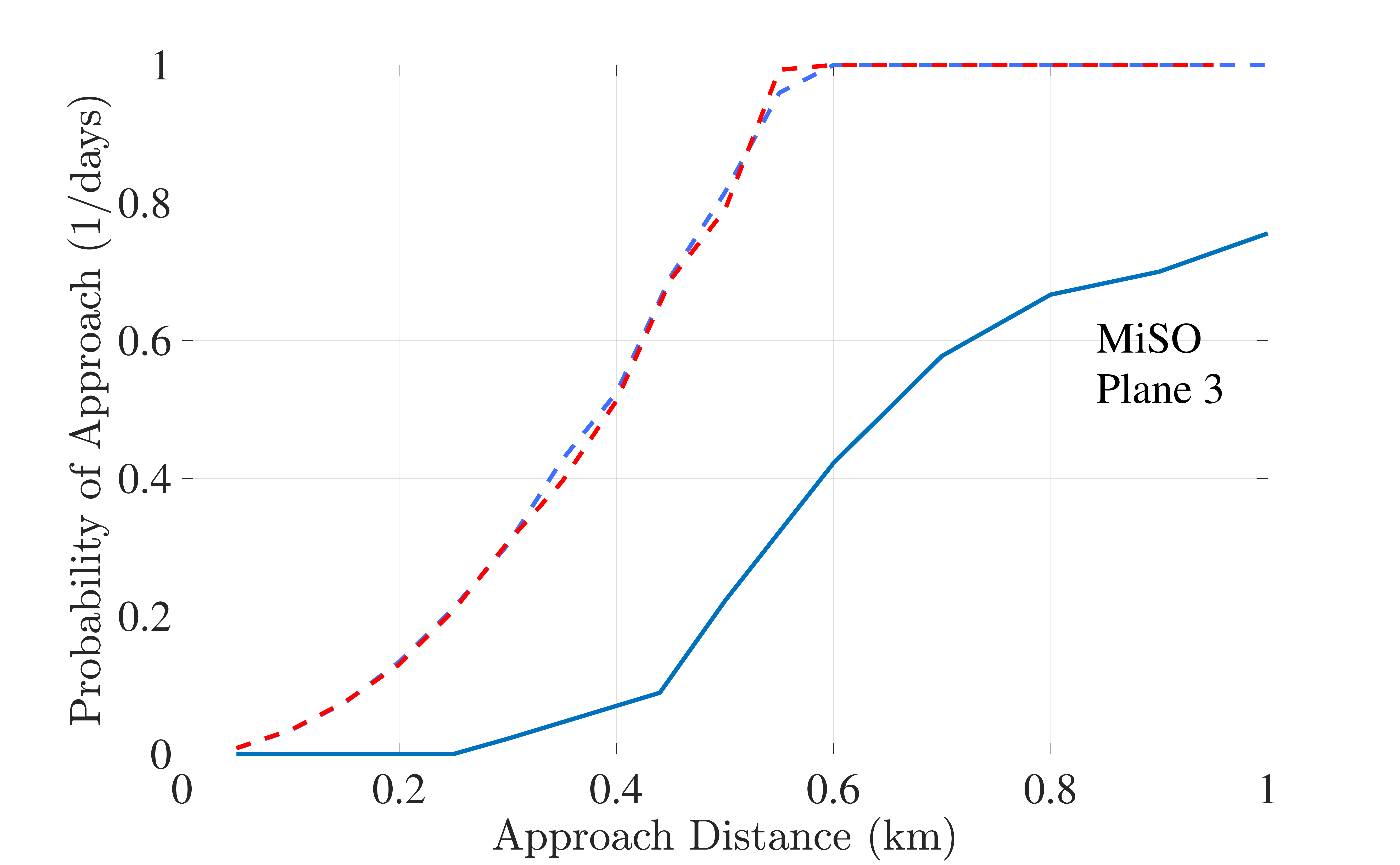}
	\includegraphics[trim = 1.3in 0.2in 2.5in 0.9in, clip,scale=0.5,width=0.24\textwidth]{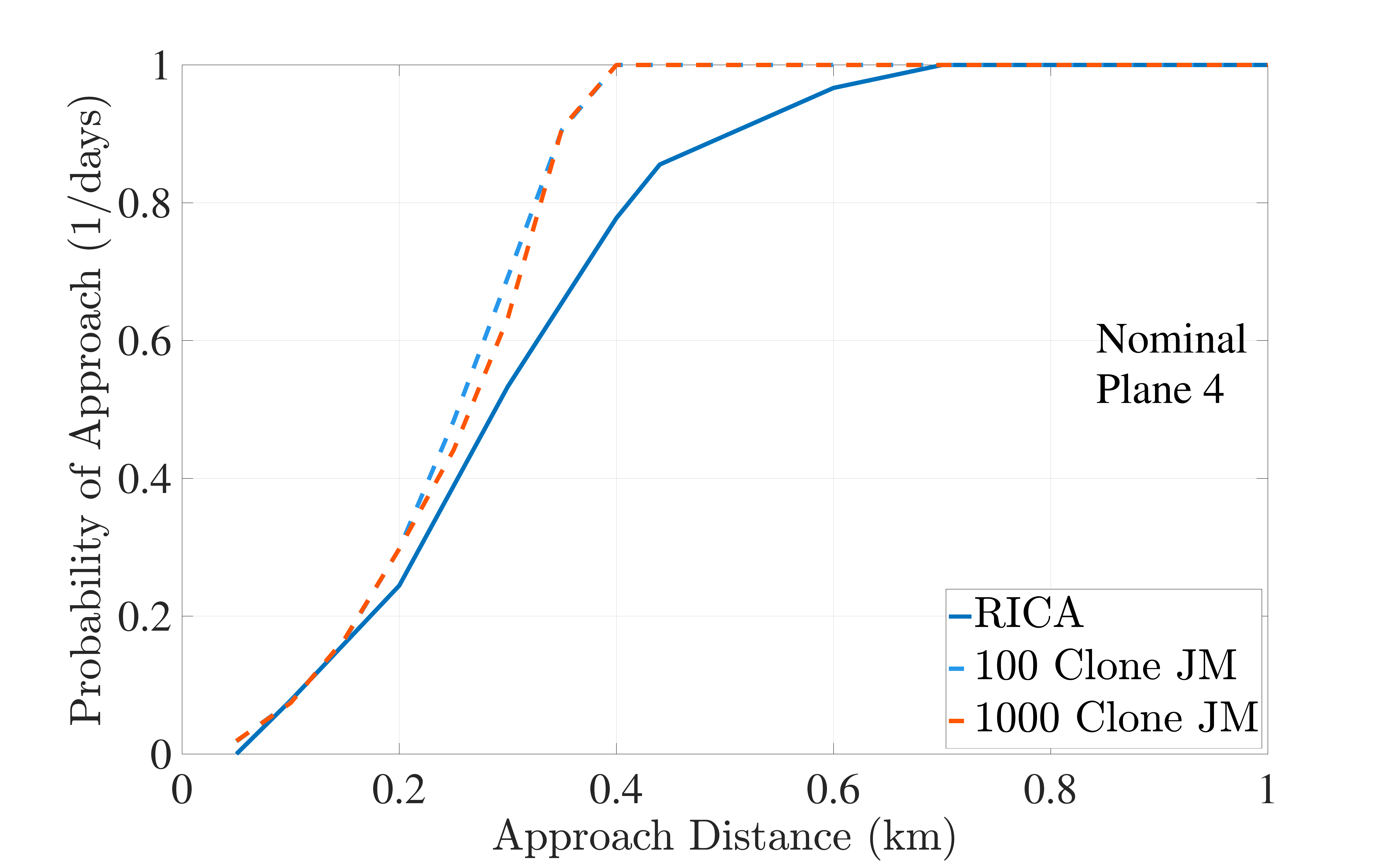}
	\includegraphics[trim = 1.3in 0.2in 2.5in 0.9in, clip,scale=0.5,width=0.24\textwidth]{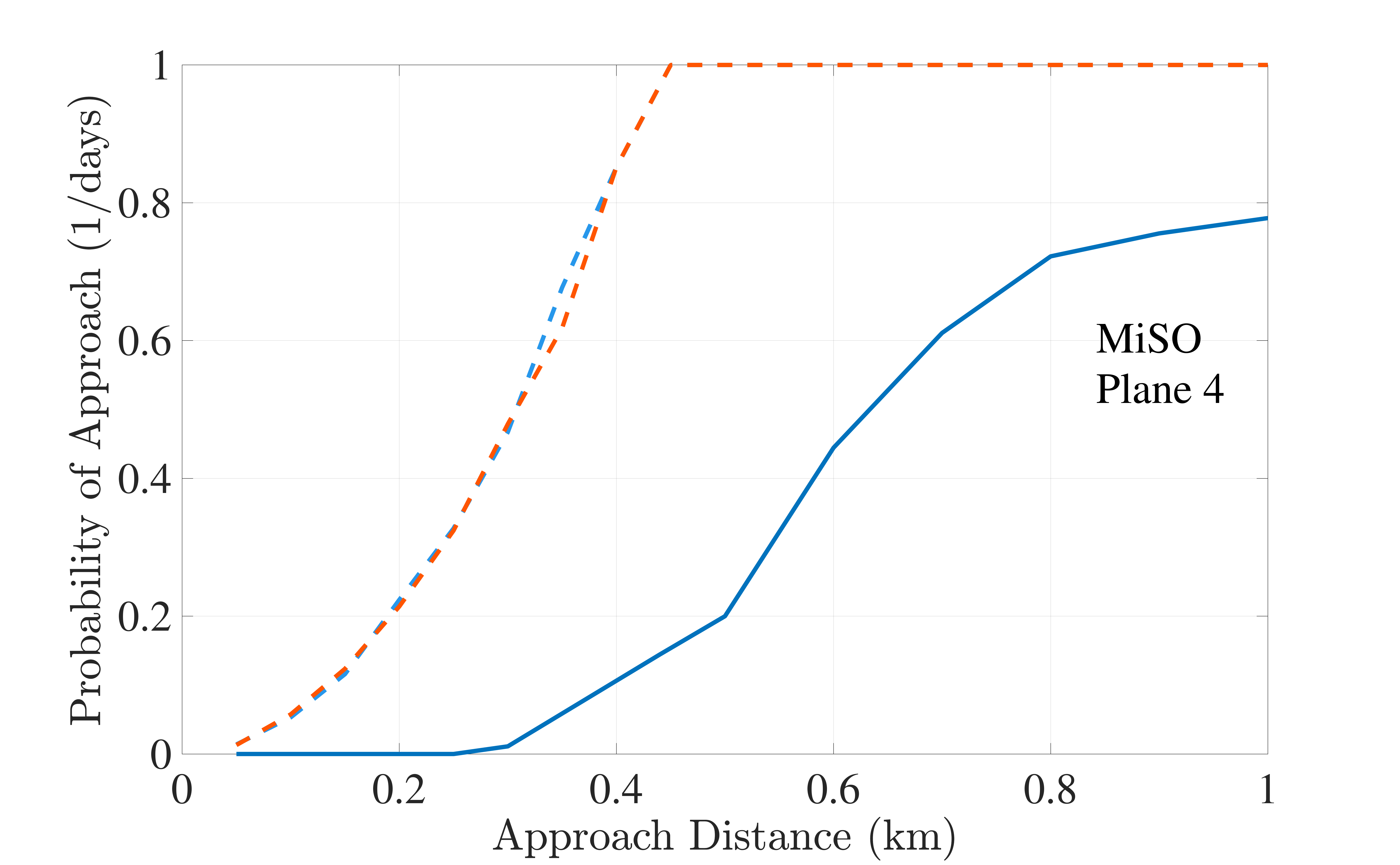}
	\includegraphics[trim = 1.3in 0.2in 2.5in 0.9in, clip,scale=0.5,width=0.24\textwidth]{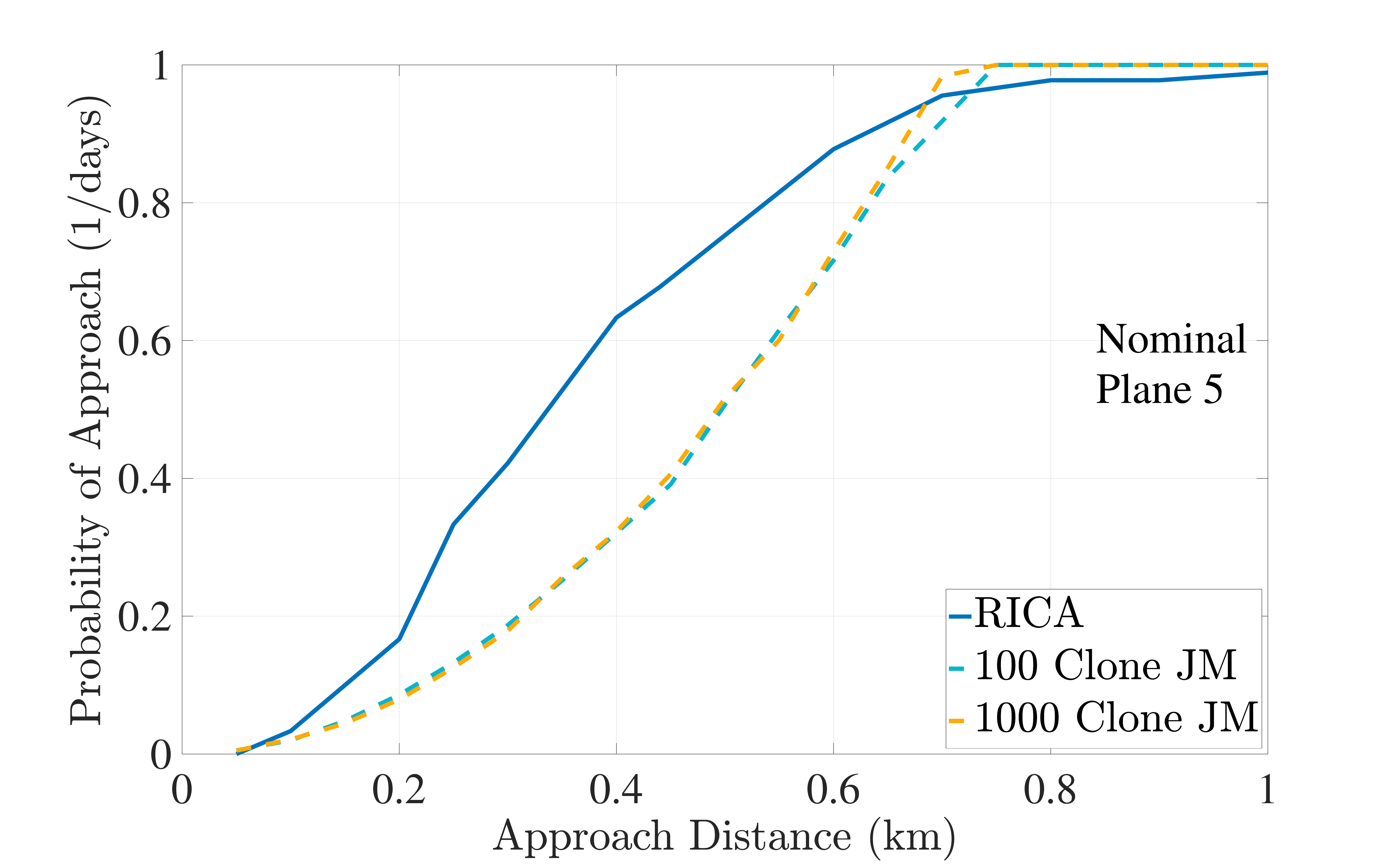}
	\includegraphics[trim = 1.3in 0.2in 2.5in 0.9in, clip,scale=0.5,width=0.24\textwidth]{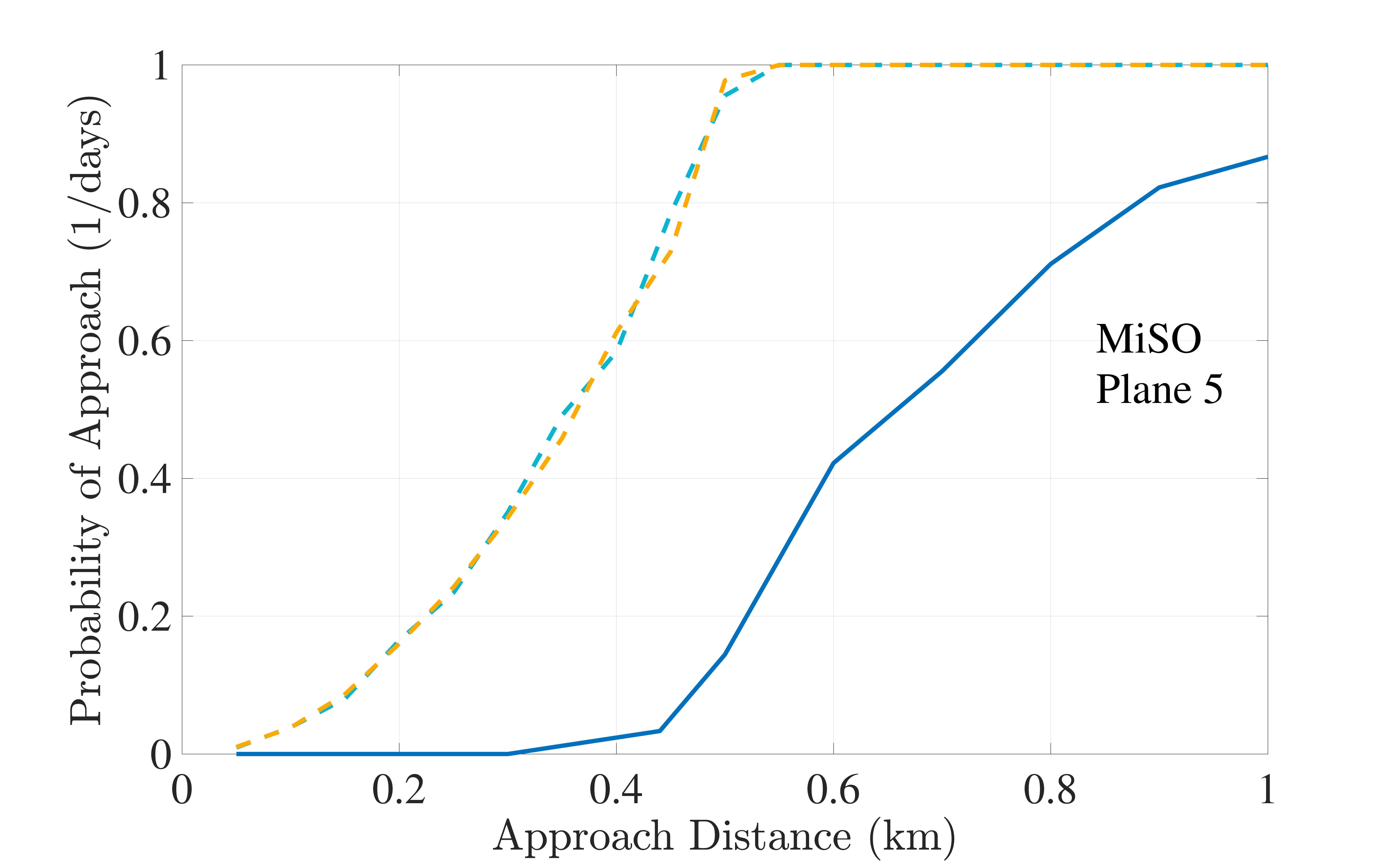}
	\caption{Comparison of the collision probability of the Starlink nominal ({\it left}) and MiSO ({\it right}) target planes with their respective fields.}
	\label{fig:starlinkJM}
\end{figure}

Overall, the performance of \texttt{JM} for the nominal Starlink constellation is similar to its performance for OneWeb, in both nominal and MiSO cases. In Fig.~\ref{fig:starlinkJM}, we see that for Planes 1 and 2 and for approach distances from 0 to 0.35 km and from 0.6 to 1 km, \texttt{JM} performs extremely well, but for the approach distances in between, the probabilities predicted by \texttt{RICA} and the \texttt{JM} diverge quite significantly. For Plane 3, the performance is excellent between 0 and 0.45 km, but for larger distances \texttt{JM} suffers. In Plane 4, we see good performance until about 0.2 km and finally in Plane 5 the performance is quite poor for the vast majority of approach distances and for every MiSO target plane, \texttt{JM} is fairly inaccurate. Here, similarly to the OneWeb case study, \texttt{HERA} is shown to be inadequate with respect to predicting conjunctions between the nominal Starlink and MiSO target planes and their respective fields. It is able to predict that more close approaches within 1 km are are experienced by the MiSO 5 than nominal 5 target plane, however all of these predicted approaches are false-positives.

\begin{table}[h!]
\centering
\caption{Close approaches less than 1 km predicted by HERA for the target planes of the nominal and MiSO Starlink constellations.}
\label{tab:starlinkHERA}
\begin{tabular}{@{}llllll@{}}
\toprule
Plane Number            & Plane 1 & Plane 2 & Plane 3 & Plane 4 & Plane 5 \\ \midrule
Nominal Approaches      & 0       & 0       & 0       & 0       & 21      \\
Nominal False Positives & 0       & 0       & 0       & 0       & 21      \\
MiSO Approaches         & 0       & 0       & 0       & 0       & 79      \\
MiSO False Positives    & 0       & 0       & 0       & 0       & 79      \\ \bottomrule
\end{tabular}
\end{table}

\section{Discussion}

The deployment and management of mega-constellations requires new standards within the fields of SSA and STM. As we have seen in the endogenous case studies of both the OneWeb and Starlink constellations, techniques that are not purely deterministic are not sensitive enough to be solely relied upon. While \texttt{JM} is significantly more accurate and fast than \texttt{HERA} and faster than \texttt{RICA}, it is not robust enough to capture the nuanced differences in dynamics of the nominal versus MiSO constellation variants. The accuracy of the \texttt{JM} technique could be improved by no longer treating time as a stochastic variable, however this would sacrifice the main advantage of the algorithm, its speed, due to the enormous amount of propagations that would be required. Furthermore, the Hoot's method and other MOID-based approaches to conjunction assessment are better suited to mean elements and are unreliable when osculating elements are considered. Short-periodic variation in the elements cause not only missed conjunctions, but all false positives in \texttt{HERA}. This problem is significantly exacerbated when dealing with mega-constellations in the highly perturbed LEO environment, where the magnitude of the short-period variations in $a$ alone is on the order of 15 kilometers. Thus, we argue that in order to account for these challenges, the orbits of the considered objects would have to be sampled at a rate similar to what is used in \texttt{RICA} and accordingly, the additional computations need to calculate a parameter such as the MOID would become an unnecessary hindrance. Accordingly, it seems that the only rigorous method to design constellations that prevent endogenous conjunctions and to determine the real need for avoidance-maneuvers is a brute-force approach similar to \texttt{RICA}. Furthermore, with the use of regularization and parrallelization the problem in now not only tractable, but practical.

Although accurate close-approach prediction software and collision avoidance maneuvers will be central in preventing endogenous collisions in satellite mega-constellations, the results of the OneWeb and SpaceX case studies indicate that if constellations are designed smartly, such as our MiSO configurations, collision avoidance maneuvers may not even be necessary, or at least minimized. This is especially important considering the predicted failure rate of these relatively inexpensively manufactured satellites, where they can potentially become debris hazards within the constellations.

\section{Conclusion}

 The brute-force approach offers detailed insight into the collision risk of operational satellites, and provides a very useful tool for constellation designers as well as the SSA and STM communities. Specifically, although other approaches are faster, it provides a great service vis-a-vis the ability to determine, which approaches are suited to what scenarios. For example, \texttt{HERA}, which is based on \cite{fH84}, has been shown to be far too conservative for the new heavily congested orbital environment, and the far more modern \texttt{JM} approach, although able to better capture the dynamics of the problem, is not reliable enough to be used as a stand-alone method for evaluating collision risk. Furthermore, in addition to using the brute-force approach to evaluate other methods for conjunction prediction, it has been successfully utilized to evaluate subtly different constellation designs. Using our developed algorithm, \texttt{RICA}, we have demonstrated that the frozen-orbit-based MiSO optimization \citep{cB18} of the Nominal OneWeb and Starlink constellations is incredibly effective and could likely remove the need for collision avoidance maneuvers.
 
 Although we do not know the true designs of the planned mega-constellations, the lack of any currently available astrodynamics tools such as \texttt{RICA} and MiSO (to our knowledge), leads us to assume that the operators of mega-constellations currently have a sub-optimal design in place. Such dynamical assessments reported herein not only have a profound and tangible influence on satellite constellation design, perhaps attacking the debris problem at its source, but could also provide crucial insight into new satellite and space mission concepts that are not simply predicated by Keplerian motion.
 
 In addition to improving the efficiency of \texttt{RICA} and \texttt{JM}, an interesting future route is to take the historical high-risk conjunction assessments, reported in the literature and other outlets, and evaluate them under these tools. In the past, accurate orbit propagation has not played a major role in predicting the evolution of the orbital debris environment, however its place in solving this challenging problem can no longer be ignored.

\section*{Acknowledgements}

Aspects of this work were presented at the 20th {\it Advanced Maui Optical and Space Surveillance Technologies Conference}, 2018, Maui, Hawaii. This research is funded in part by the Universities Space Research Association (USRA) under Grant Agreement SUBK-19-0020 (Subagreement No. 90006.004/08102). N.R. gratefully acknowledges support from the National Science Foundation Bridge to Doctorate Fellowship (NSF 1809591). We especially thank D. Amato, now of the University of Colorado, for his technical contributions to this paper through \texttt{THALASSA} and for inspiring \texttt{RICA}.

\bibliographystyle{elsarticle-harv}
\bibliography{references}

\clearpage
\begin{appendix}

\section{Algorithms}
\label{sec:algos}

\vspace{-\baselineskip}

\begin{algorithm}[h!]
\begin{spacing}{1.1}
\footnotesize
\caption{MAX VELOCITY({\it objects})}\label{euclid}
\label{alg:maxVelocity_pseudo}
\begin{algorithmic}[1]

\For{$i \gets 1,$ length({\it objects})} 
    \vspace{0.25\baselineskip}
    \State $\ds v(i) \gets \mu \left( \frac{2}{a(i)(1 - e(i))} - \frac{1}{a(i)} \right)$
    \vspace{0.25\baselineskip}
\EndFor
\Return max($v$)
\end{algorithmic}
\end{spacing}
\end{algorithm}

\vspace{-\baselineskip}

\begin{algorithm}[h!]
\begin{spacing}{1.1}
\footnotesize
\caption{PROPAGATE({\it objects, tspan})}\label{euclid}
\label{alg:propagate_pseudo}
\begin{algorithmic}[1]

\For{$i \gets 1,$ length({\it objects})}
    \State $ {\bm r}(t)_i \gets \text{Propagate \textit{objects(i)} for duration {\it tspan} with \textit{THALASSA}} $
\EndFor

\Statex \Return $\{{\bm r}(t)\}$
\end{algorithmic}
\end{spacing}
\end{algorithm}

\vspace{-\baselineskip}

\begin{algorithm}[h!]
\begin{spacing}{1.1}
\footnotesize
\caption{FILTER STAGE({\it targets, fields, $\tau$, tspan})}\label{euclid}
\label{alg:filterStage_pseudo}
\begin{algorithmic}[1]

\State $v_\text{max} \gets$ max velocity({\it targets}) $+$ max velocity({\it fields})
\vspace{0.25\baselineskip}
\State $\ds t_\text{step} \gets \frac{1}{f_s} \frac{\tau}{v_\text{max}}$
\vspace{0.25\baselineskip}
\State $\ds n \gets \frac{tspan}{t_\text{step}}$
\vspace{0.25\baselineskip}
\State $\{ {\bm r}(t)_{targets} \} \gets$ PROPAGATE({\it targets}, {\it tspan}) 
\State $\{ {\bm r}(t)_{fields} \} \gets$ PROPAGATE({\it fields}, {\it tspan})
\For{$k \gets 1,\, n$}
    \State $t \gets k \cdot t_\text{step}$
    \State Calculate $|{\bm r}(t)_{target(i)} - {\bm r} (t)_{field(j)}|_k$ of $\mathcal{S} \gets {{targets} \choose {fields}}$
\EndFor
\State $\{ \textit{reduced targets} \}  \gets {targets(i)},$ which satisfy $|{\bm r}_\textit{{targets(i)}} - {\bm r}_\textit{{fields(j)}}|_k \leq \tau$
\State $\{ \textit{reduced fields} \} \gets {fields(i)},$ which satisfy $|{\bm r}_\textit{{targets(i)}} - {\bm r}_\textit{{fields(j)}}|_k \leq \tau$
\Statex
\Return $\textit{\{reduced targets\}},\textit{\{reduced fields\}}$

\end{algorithmic}
\end{spacing}
\end{algorithm}

\vspace{-\baselineskip}

\begin{algorithm}[h!]
\begin{spacing}{1.1}
\footnotesize
\caption{RICA}\label{euclid}
\label{alg:fica_pseudo}
\begin{algorithmic}[1]

\State $\textit{$targets_1$} \gets \text{Set of target objects}$
\State $\textit{$fields_1$} \gets \text{Set of field objects}$

\Statex

\State $\textit{$\tau_1$} \gets \text{Stage 1 close approach distance}$
\State $\textit{$\tau_2$} \gets \text{Stage 2 close approach distance}$
\State $\textit{$\tau_3$} \gets \text{Stage 3 close approach distance}$

\Statex

\State $\textit{$tspan$} \gets \text{Time span of interest}$

\Statex

\State $\textit{$targets_2$, $fields_2$} \gets \text{FILTER STAGE(\textit{$targets_1$, $fields_1$, $\tau_1$, tspan})}$
\State $\textit{$targets_3$, $fields_3$} \gets \text{FILTER STAGE(\textit{$targets_2$, $fields_2$, $\tau_2$, tspan})}$
\State $\textit{$targets_4$, $fields_4$} \gets \text{FILTER STAGE(\textit{$targets_3$, $fields_3$, $\tau_3$, tspan})}$

\end{algorithmic}
\end{spacing}
\end{algorithm}

\vspace{-\baselineskip}

\begin{algorithm}[h!]
\begin{spacing}{1.1}
\footnotesize
\caption{CALCULATE PROBABILITY($\textbf{\oe}_a,\textbf{\oe}_b$)}\label{euclid}
\label{alg:probability_pseudo}
\begin{algorithmic}[1]
\Procedure{Calculate Collision Probability}{}
\State $ {\bm v}_a \gets \text{Velocity of object $a$ at time of close approach} $
\State $ {\bm v}_b \gets \text{Velocity of object $b$ at time of close approach}$

\Statex

\If {$|{\bm v}_a| > |{\bm v}_b|$}
\State $\textbf{\oe}_1 \gets \textbf{\oe}_a$
\State $\textbf{\oe}_2 \gets \textbf{\oe}_b$
\Else
\State $\textbf{\oe}_1 \gets \textbf{\oe}_b$
\State $\textbf{\oe}_2 \gets \textbf{\oe}_a$
\EndIf

\Statex

\If {$i_1 > 90 \And i_2 < 90 $}
\State $\ds k \gets -\frac{|{\bm v}_2|}{|{\bm v}_1|}$
\ElsIf{$i_1 < 90 \And i_2 > 90 $}
\State $\ds k \gets -\frac{|{\bm v}_2|}{|{\bm v}_1|}$
\Else
\State $\ds k \gets \frac{|{\bm v}_2|}{|{\bm v}_1|}$
\EndIf

\Statex

\State $\ds \sin \alpha \gets \frac{1 + e_1 \sin{f_1}}{\sqrt{1 + 2 e_1 \cos{f_1} + {e_1}^2}}$
\State $\ds \theta \gets \arccos{(\frac{{\bm v}_1 \cdot {\bm v}_2}{|{\bm v}_1| \cdot |{\bm v}_2|})}$
\State $\ds \theta_c \gets \frac{0.9}{k |{\bm v}_1|} \sqrt{(1 - k^2) \tau \cdot \sin{\alpha}}$
\Statex
\If {$ \textit{$\theta$} < \textit{$\theta c$} $} 
\State $\ds P \gets \frac{1.7}{T_1 \cdot T_2} \sqrt{\frac{(1 - k)\tau}{(1 + k)g \cdot \sin \alpha}}$
\Else
\State $\ds P \gets \frac{\pi \tau |{\bm v}_1 - {\bm v}_2|}{2 |{\bm v}_1 \times {\bm v}_2 | T_1 \cdot T_2}$
\EndIf

\Return $\textit{P}$

\EndProcedure
\end{algorithmic}
\end{spacing}
\end{algorithm}

\vspace{-\baselineskip}

\begin{algorithm}[h!]
\begin{spacing}{1.1}
\footnotesize
\caption{GENERATE CLONES($N_c,\textbf{\oe}(t))$}\label{euclid}
\label{alg:generateClones_pseudo}
\begin{algorithmic}[1]
\State $i \gets 1$
\State $j \gets 1$
\State $k \gets 1$
\State $nobjs \gets$ length($\textbf{\oe}(t)$)

\While{$j \leq nobjs$} 
    \vspace{0.25\baselineskip}
    \State $npts \gets$ length($\textbf{\oe}(t)_i$) 
    \vspace{0.25\baselineskip}
    \While{$i\leq N_c$}
        \vspace{0.25\baselineskip}
        \State $l \gets 1 \leq  \text{random integer} \leq npts $
        \State $ {\bm {clones}}_k \gets \textbf{\oe}(l)_j $
        \State $i \gets i + 1$
        \State $k \gets k + 1$
    \EndWhile
    
    \State $i \gets 1$
    \State $j \gets j + 1$
    
\EndWhile
\end{algorithmic}
\end{spacing}
\end{algorithm}

\vspace{-\baselineskip}

\begin{algorithm}[h!]
\begin{spacing}{1.1}
\footnotesize
\caption{JM APPROACH}\label{euclid}
\label{alg:jmColProb_pseudo}
\begin{algorithmic}[1]

\State $\textit{targets} \gets \text{Set of target objects}$
\State $\textit{fields} \gets \text{Set of field objects}$

\Statex

\State $ \textbf{\oe}(t)_\textit{targets} \gets \text{PROPAGATE(\textit{targets}, tspan)} $
\State $ \textbf{\oe}(t)_\textit{fields} \gets \text{PROPAGATE(\textit{fields}, tspan)} $

\Statex

\State $ {\bm {clones}}_\textit{target} \gets \text{GENERATE CLONES}(1, \textbf{\oe}(t)_\textit{targets}) $
\State $ {\bm {clones}}_\textit{field} \gets \text{GENERATE CLONES}(N_c, \textbf{\oe}(t)_\textit{fields}) $

\Statex

\State $\ds \mathcal{S} \gets {{\bm {clones}}_\textit{target} \choose {\bm {clones}}_\textit{field}} $ 

\Statex

\For{$i \gets 1,$ length($\mathcal{S}$)}
    \State $\textbf{target}_i \gets \text{target object of }\mathcal{S}_i  $
    \State $\textbf{field}_i \gets \text{field object of }\mathcal{S}_i  $
    \State $ s \gets \text{GRONCHI MOID}(\textbf{target}_i, \textbf{field}_i) $
    \If {$s \leq \tau $}
        \State $ P_i \gets \text{CALCULATE PROBABILITY}(\textbf{target}_i, \textbf{field}_i) $
    \Else
        \State $ P_i \gets 0 $
    \EndIf
    
\EndFor

\Statex

\State $\ds P_{total} \gets \frac{1}{N_c} \sum P_i$

\end{algorithmic}
\end{spacing}
\end{algorithm}

\clearpage
\section{Exogenous Assessment}
\label{sec:exogenous}

\begin{figure}[h!]
	\centering    
	\includegraphics[trim = 0.8in 0.1in 3in 0.7in, clip,width=0.45\textwidth]{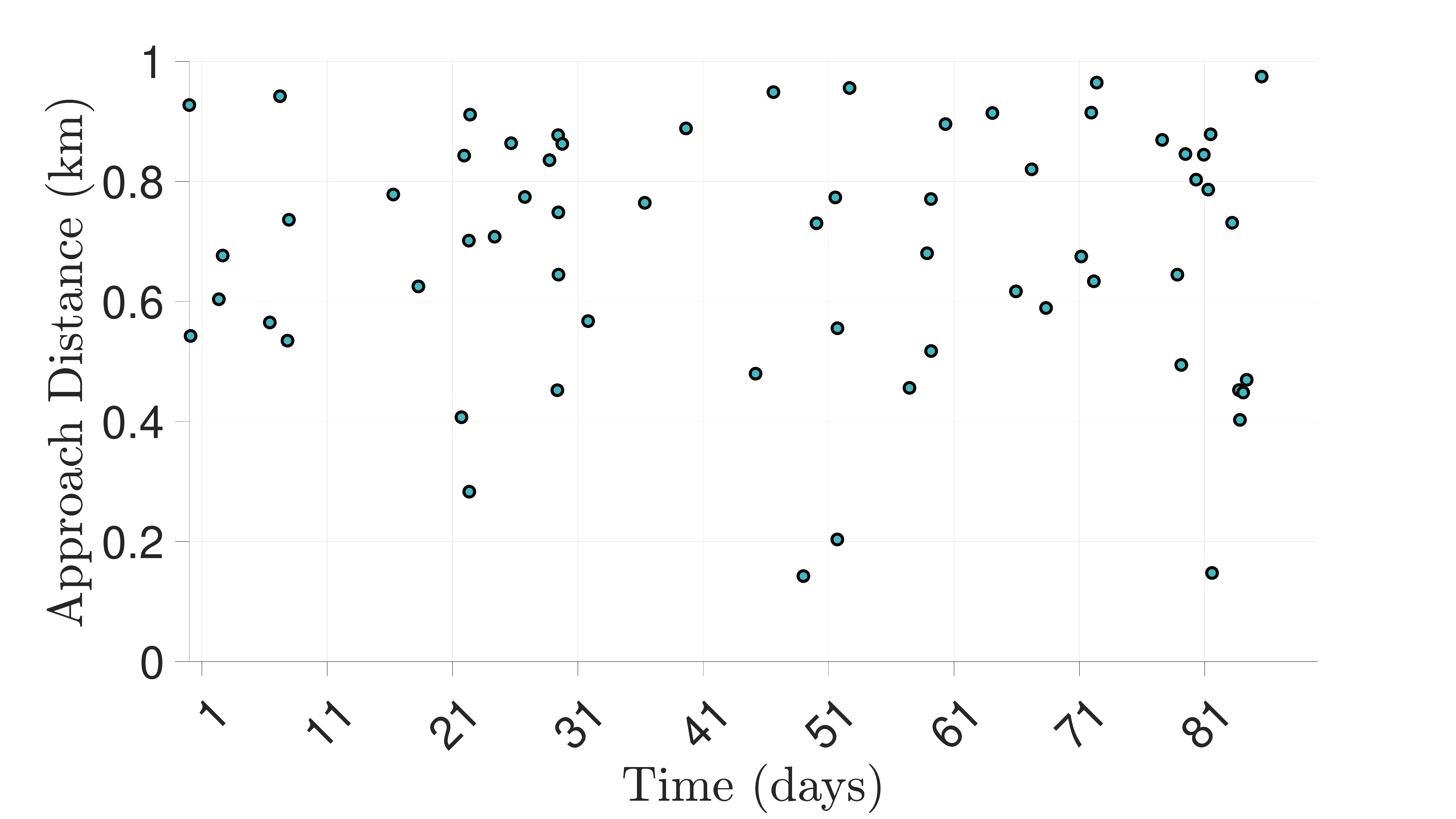}
	\caption{Unique exogenous conjunctions predicted by \texttt{RICA} between the nominal OneWeb target plane and all 18381 Space-Track-provided RSOs.}
	\label{fig:onewebExo}
\end{figure}

\begin{figure}[h!]
	\centering
	\includegraphics[trim = 0.7in 0.1in 2.2in 0.8in, clip,scale=0.5,width=0.32\textwidth]{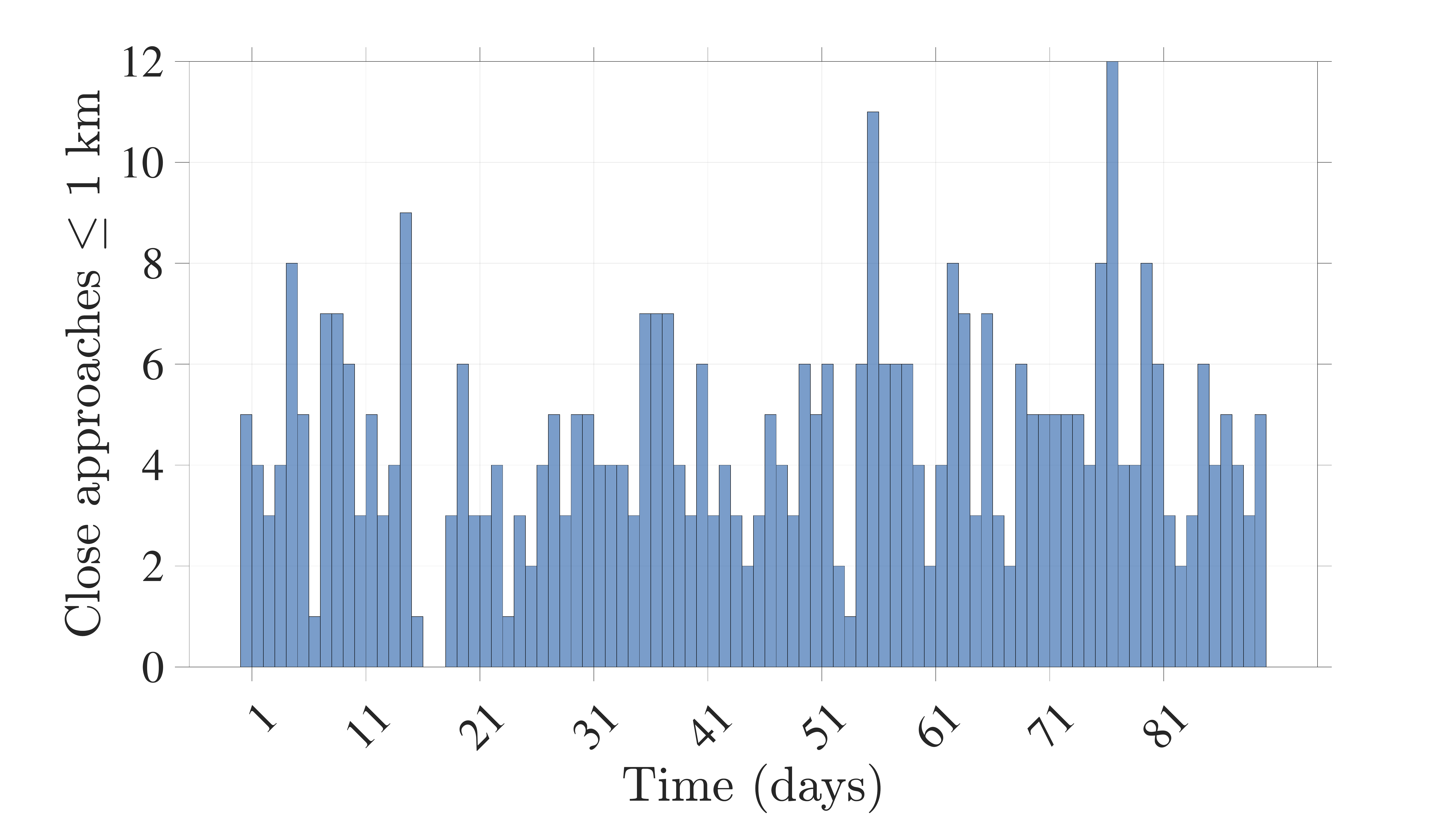}
	\includegraphics[trim = 1.1in 0.1in 2.2in 0.8in, clip,scale=0.5,width=0.32\textwidth]{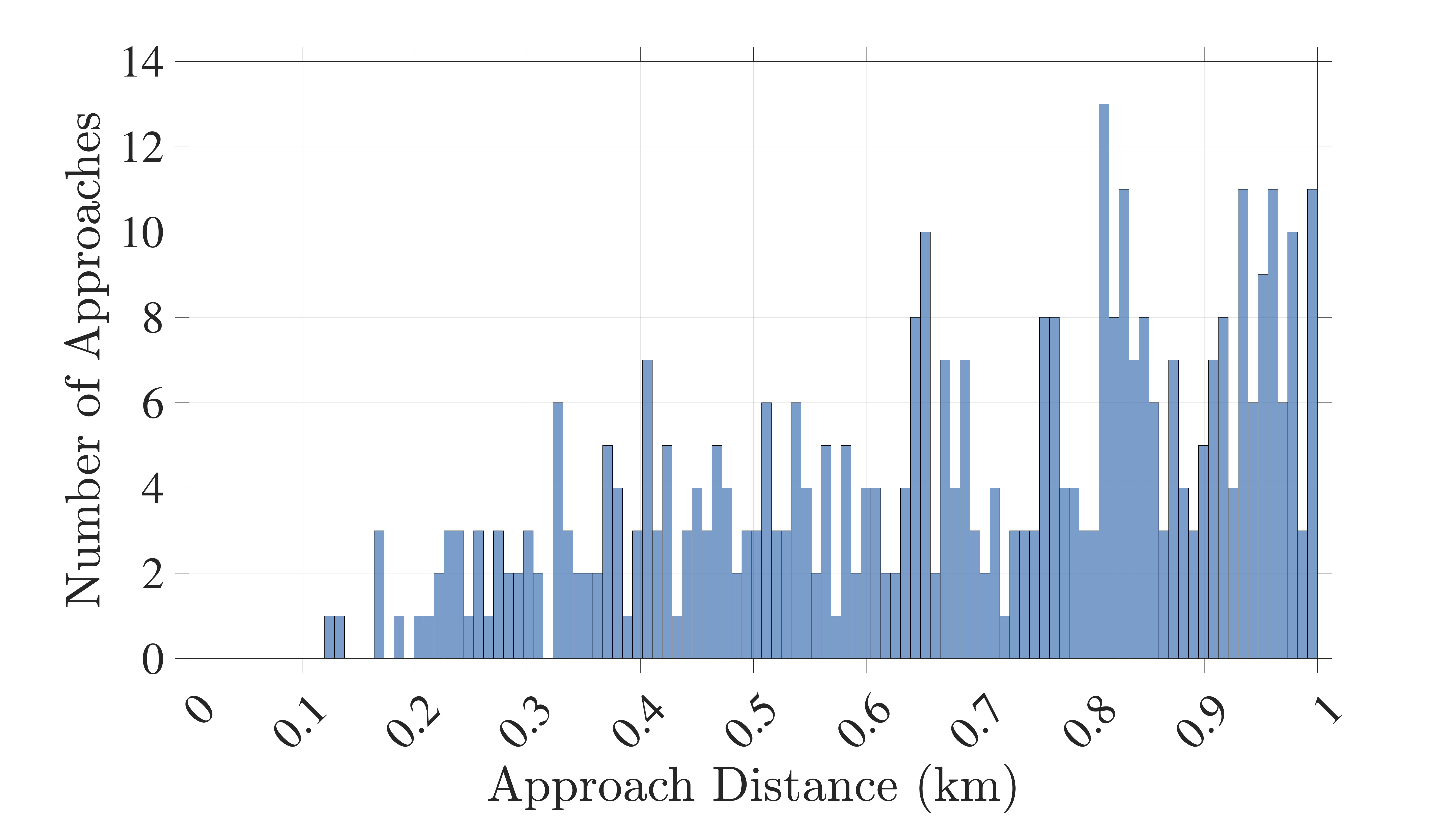}
	\includegraphics[trim = 1.1in 0.4in 2.6in 1.2in, clip,scale=0.5,width=0.32\textwidth]{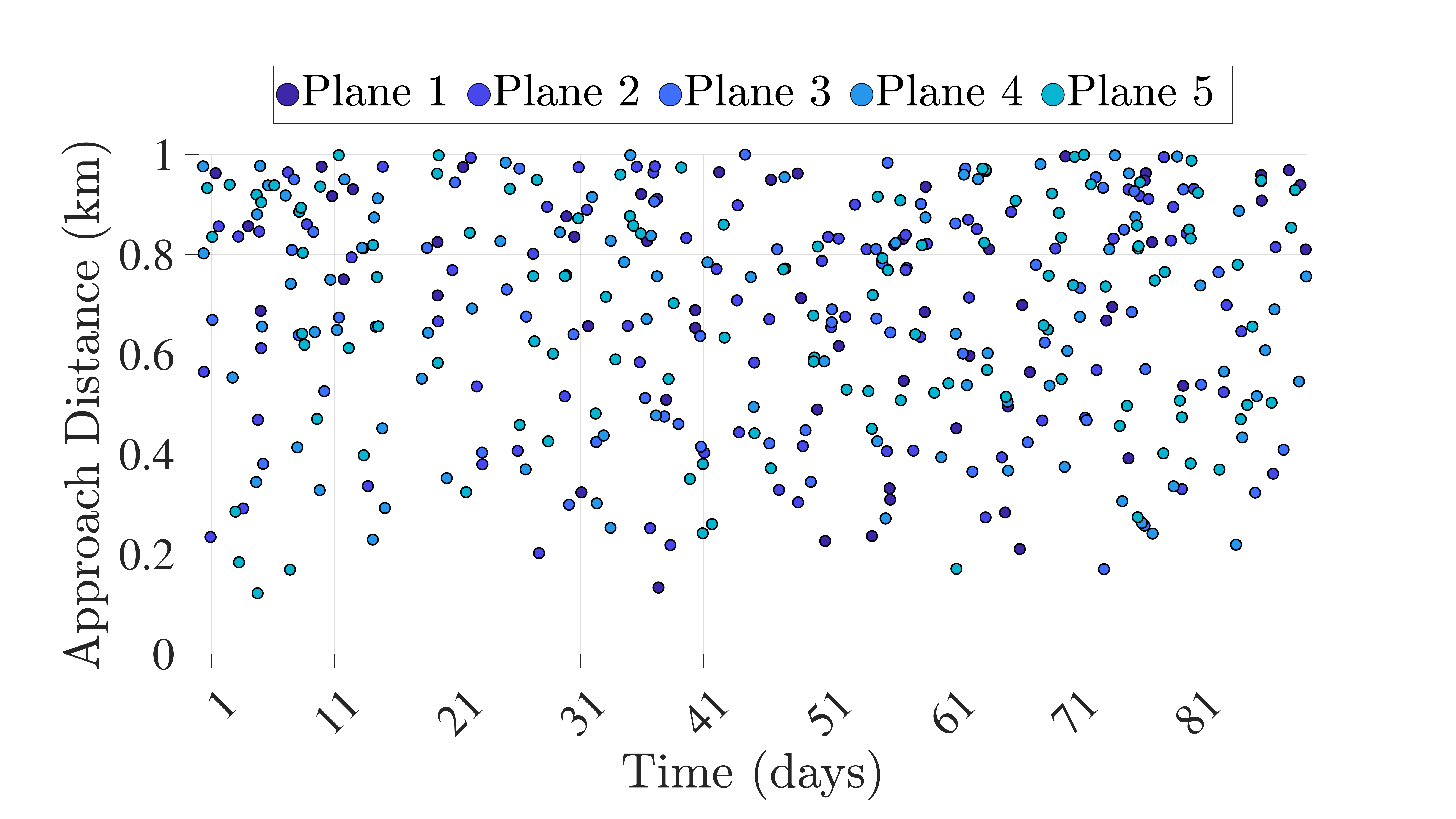}
	\caption{Frequency ({\it left}) and distance of ({\it middle}) \texttt{RICA}-predicted close approaches experienced between all five nominal Starlink target planes and the SOC obtained from Space-Track as well as approaches color-coded based on target plane ({\it right}). }
	\label{fig:starlinkExo}
\end{figure}

For the exogenous assessment, only the nominal target planes of the OneWeb LEO and Starlink constellations were considered. The reason for this is the similarity between the orbits of the satellites of the nominal and MiSO constellations with respect to TLE catalog. As in the endogenous assessment, the target plane of the OneWeb constellation and target planes 1 through 5 of the Starlink constellation were run with \texttt{RICA} against the entire TLE catalog for a duration of 90 days with $\tau_1 = 20$ km, $\tau_2 = 5$ km, and $\tau_3 = 1$ km. Fig.~\ref{fig:onewebExo} and Table~\ref{tab:onewebExo} show that the OneWeb target plane only experiences 63 close approaches with a minimum approach distance of 0.1426 km. Because this is only one plane of the constellation, we could estimate that the remaining 35 planes would likely experience a similar number of close approaches. Likewise in Fig.~\ref{fig:starlinkExo} and Table~\ref{tab:starlinkExo} we see that each Starlink target plane experienced a similar number of close approaches to the OneWeb target plane with similar minimum approach distances. The single exception is the Starlink target Plane 5, which experienced 118 close approaches, which could be due to its location in the semi-major axis and inclination phase-space (see, e.g., Fig.~\ref{fig:leodistrib}). 

\begin{table}[h!]
\centering
\caption{Close approaches with exogenous objects experienced by the target plane of the nominal OneWeb Constellation.}
\label{tab:onewebExo}
\begin{tabular}{@{}lcc@{}}
\toprule
ID & Number of Approaches & Minimum Approach Distance (km) \\ \midrule
Nominal & 63 & 0.1426 \\ \bottomrule
\end{tabular}
\end{table}

\begin{table}[h!]
\centering
\caption{Close approaches with exogenous objects experienced by the target planes of the nominal Starlink Constellation.}
\label{tab:starlinkExo}
\begin{tabular}{@{}lcc@{}}
\toprule
ID & Number of Approaches & Minimum Approach Distance (km) \\ \midrule
Nominal 1 & 60 & 0.1330 \\
Nominal 2 & 86 & 0.2020 \\
Nominal 3 & 59 & 0.1698 \\
Nominal 4 & 82 & 0.2188 \\
Nominal 5 & 118 & 0.1215 \\ \bottomrule
\end{tabular}
\end{table}

\section{Sensitivity Study}
\label{sec:sensitivityStudy}

\begin{figure}[h!]
	\centering    
	\includegraphics[trim = 0.7in 0.1in 2.2in 0.8in, clip,scale=0.5,width=0.4\textwidth]{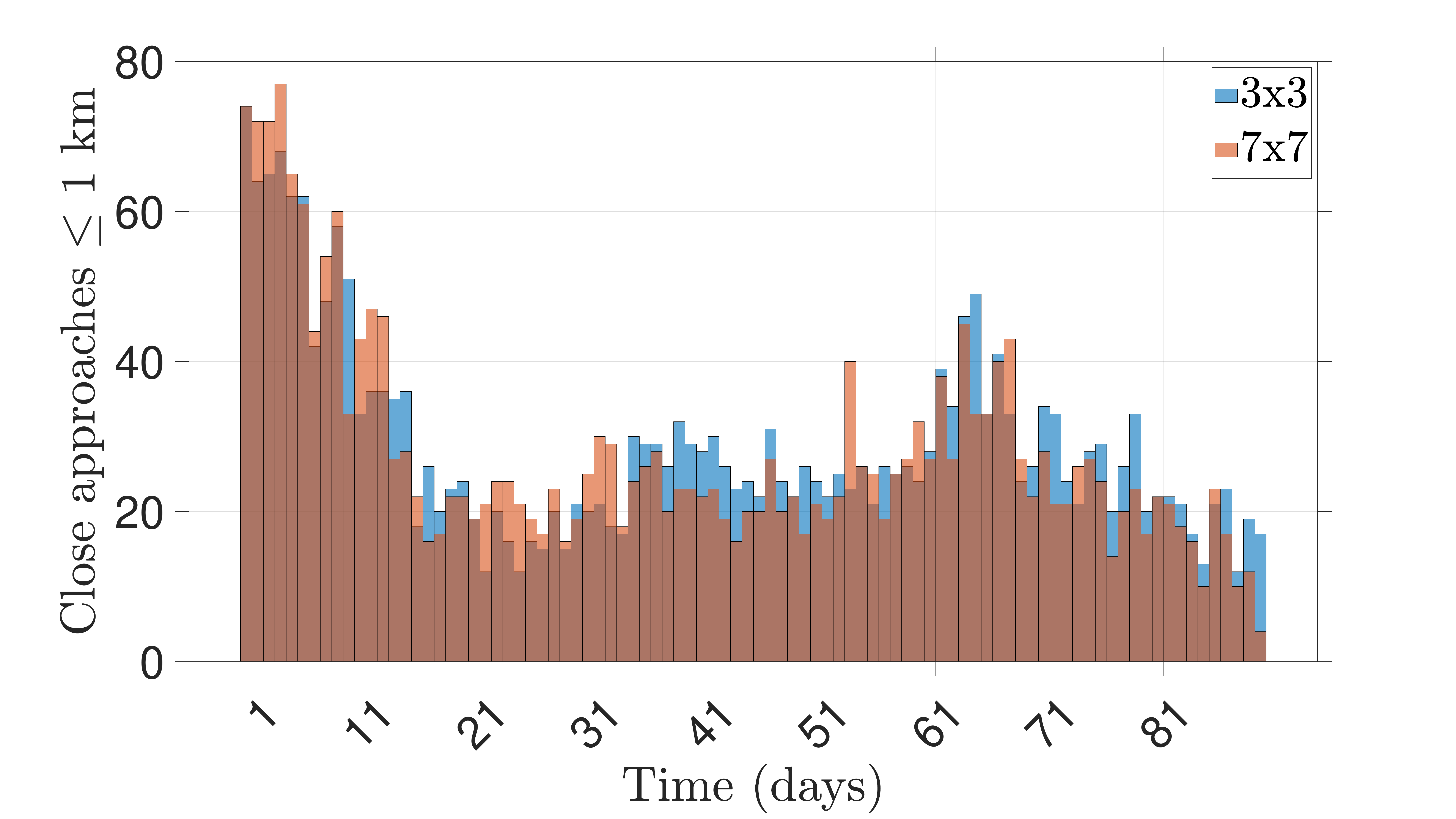}
	\includegraphics[trim = 1.1in 0.1in 2.2in 0.8in, clip,scale=0.5,width=0.4\textwidth]{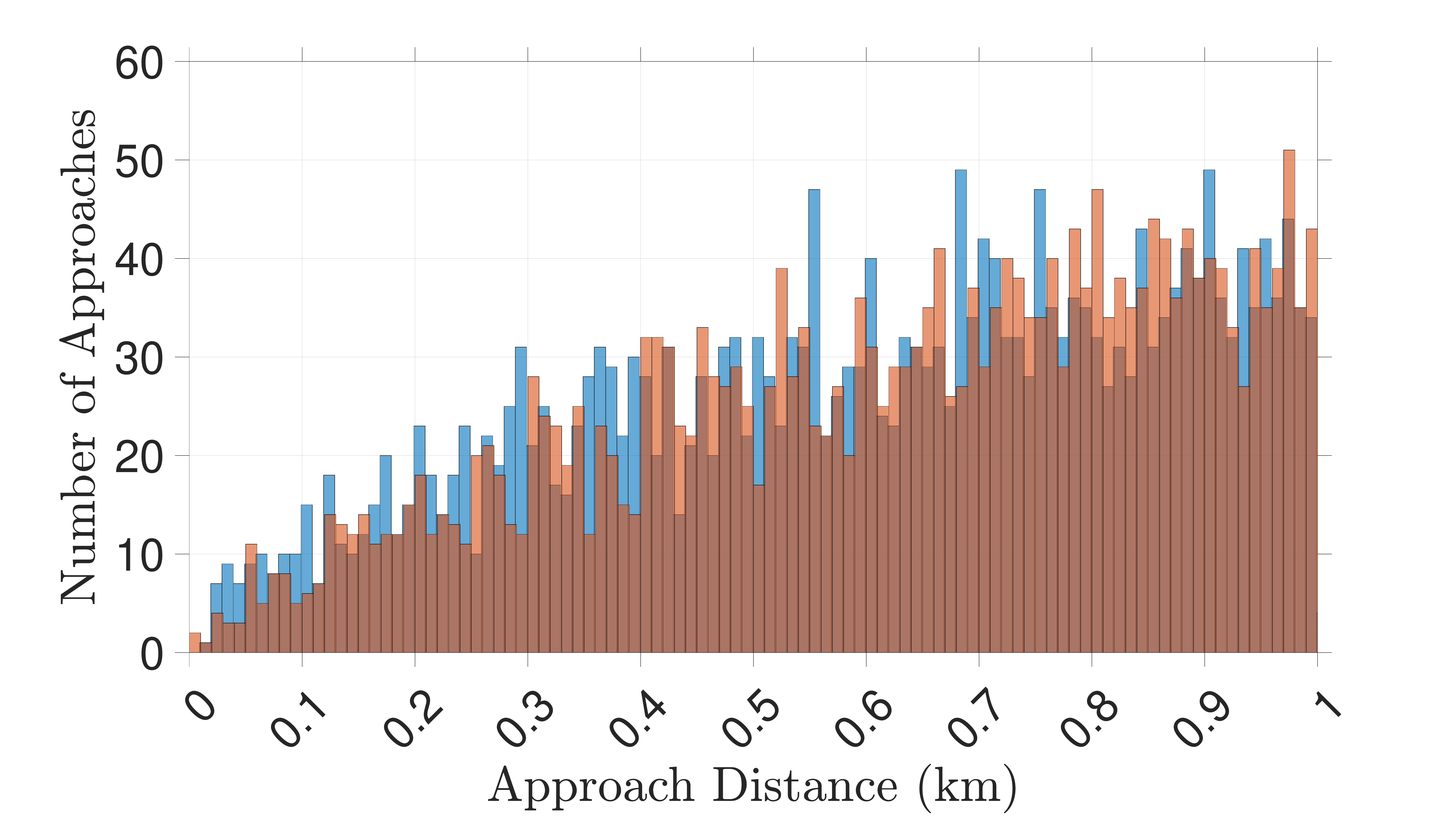}
	\includegraphics[trim = 0.7in 0.1in 2.2in 0.8in, clip,scale=0.5,width=0.4\textwidth]{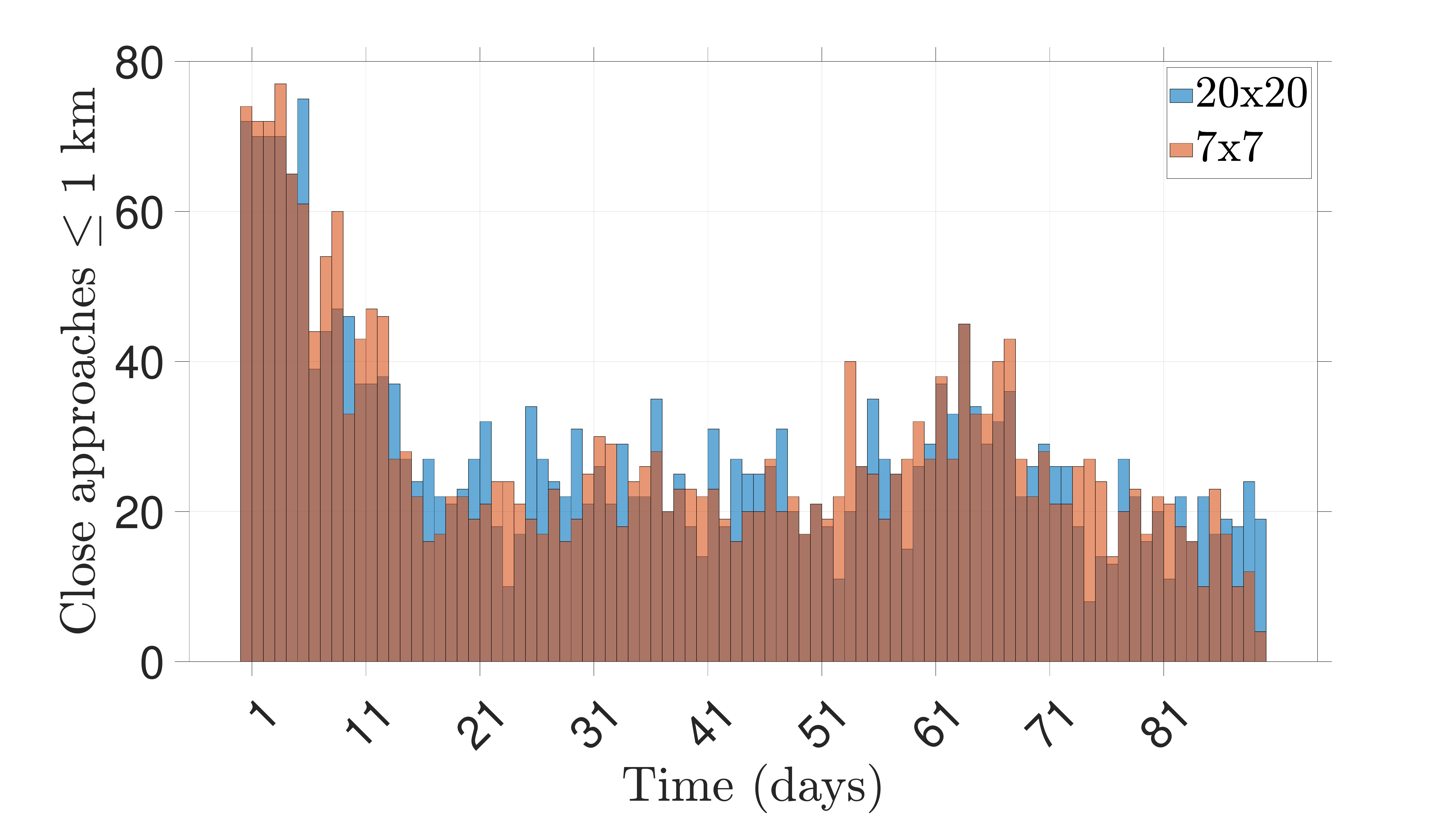}
	\includegraphics[trim = 1.1in 0.1in 2.2in 0.8in, clip,scale=0.5,width=0.4\textwidth]{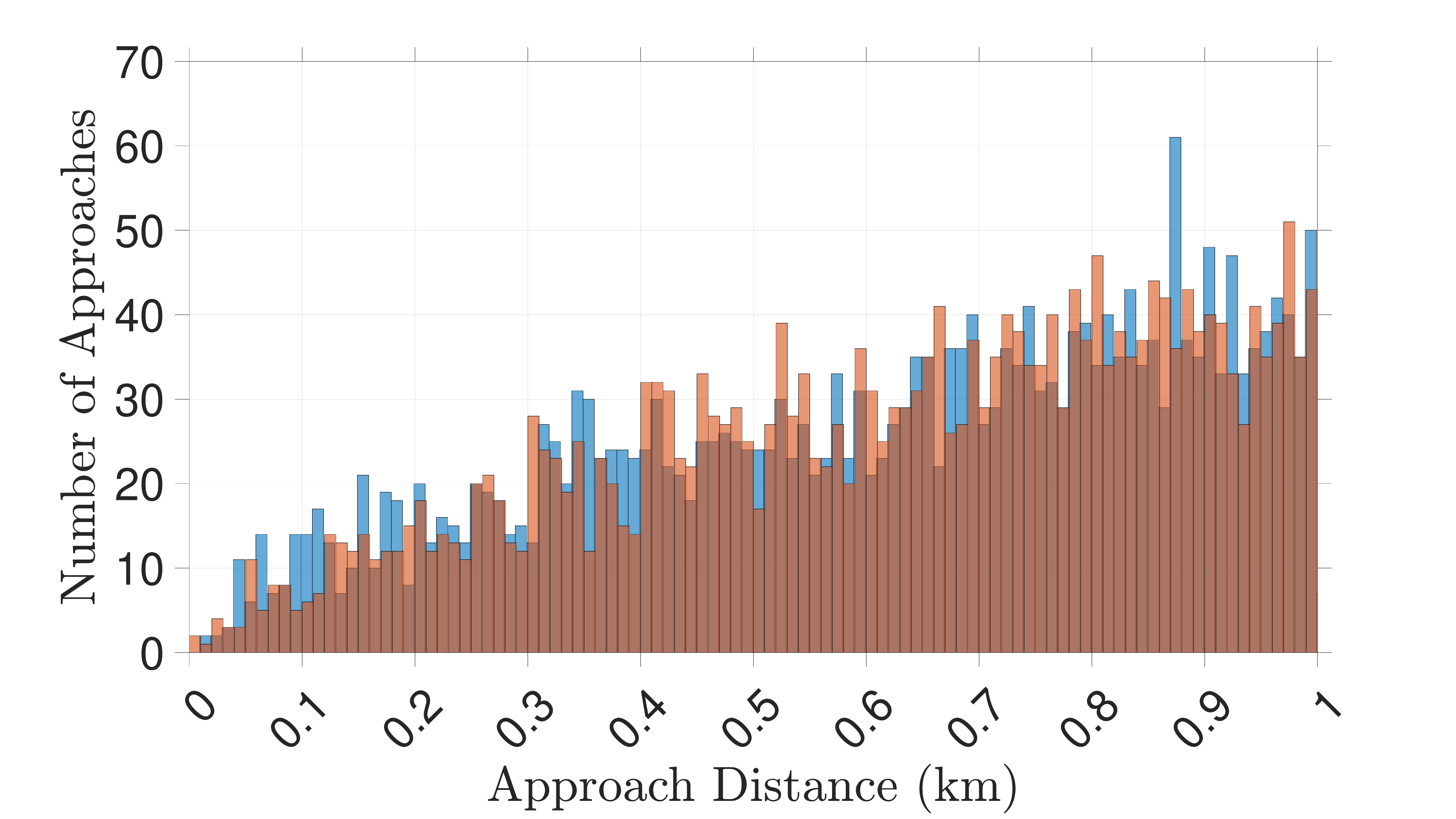}
	\includegraphics[trim = 0.7in 0.1in 2.2in 0.8in, clip,scale=0.5,width=0.4\textwidth]{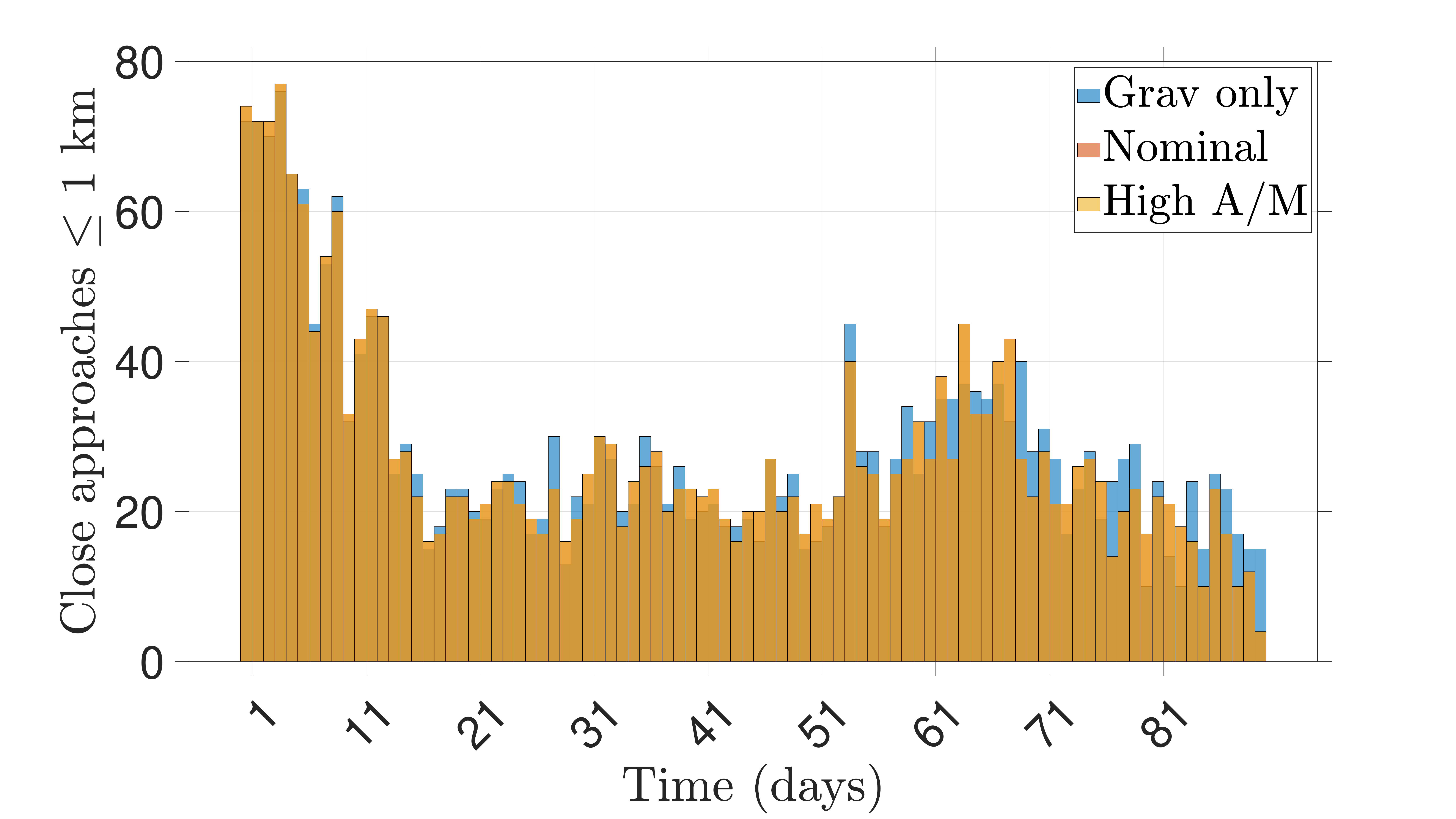}
	\includegraphics[trim = 1.1in 0.1in 2.2in 0.8in, clip,scale=0.5,width=0.4\textwidth]{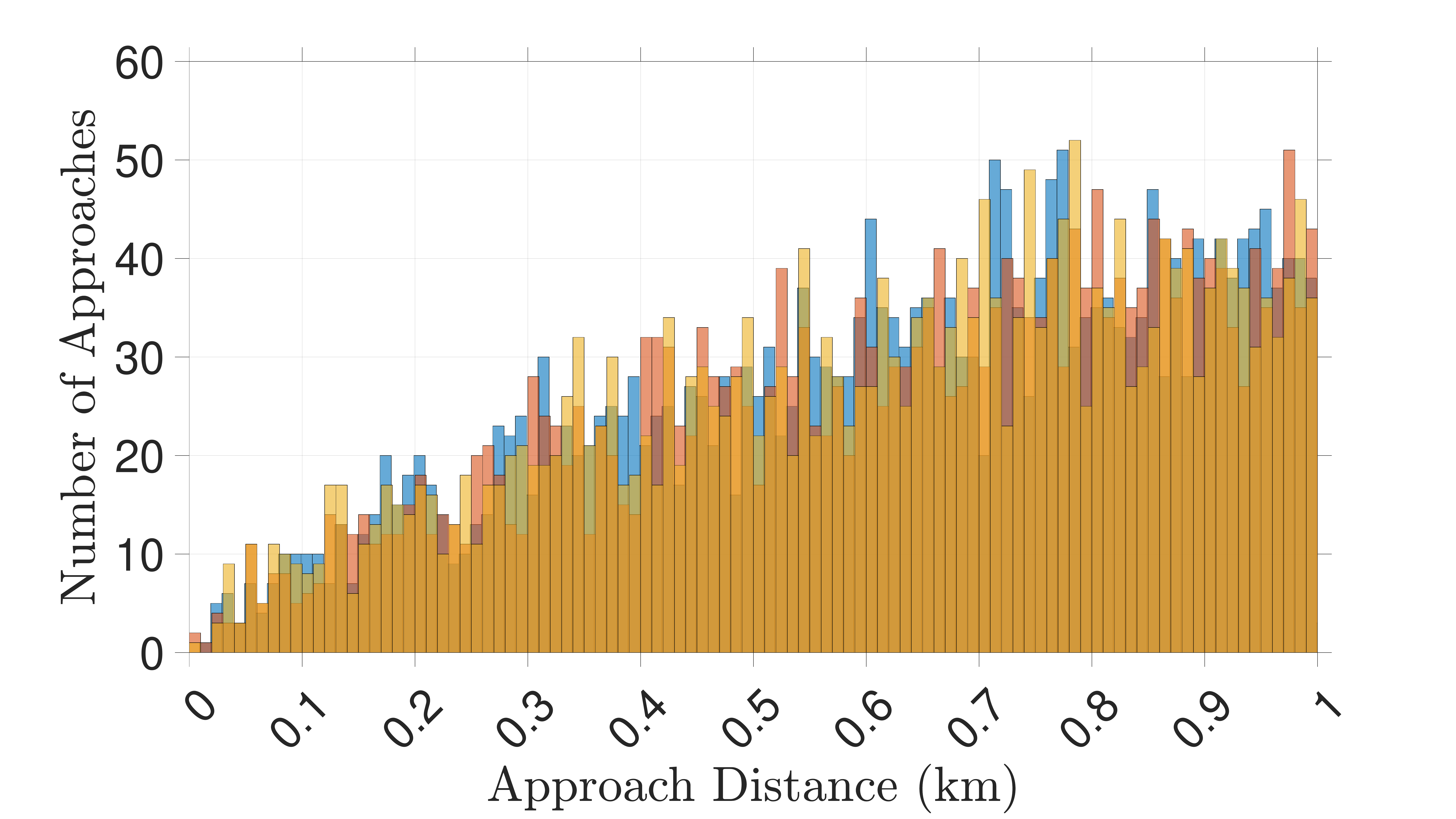}
	\caption{Frequency of close approaches within 1 km ({\it left}) and close approach distance ({\it right}) for the low accuracy ({\it top}), high accuracy ({\it middle}) models of the OneWeb LEO constellation compared against the nominal counterpart as well as a comparison of the gravitational-only, nominal, and high area to mass ratio models ({\it bottom}).}
	\label{fig:sensitivityResults}
\end{figure}

In order to determine to what effect variations in the force model would have on the close approaches predicted by \texttt{RICA}, a sensitivity analysis was conducted using five different force models in the \texttt{THALASSA} propagator. The first model was the nominal model used for all \texttt{RICA} and \texttt{JM} assessments previously mentioned in the work and consists of a 7x7 gravity field, Sun and Moon third-body gravitational perturbations, drag using the NRLMSISE-00 model, variable solar flux, and solar radiation pressure with a conical Earth shadow (Table~\ref{sec:physModel}). The ``low gravity'' and ``high gravity'' models are identical to the nominal model except that they use 3x3 and 20x20 gravity potentials, respectively. The ``gravity only'' model is the same as the nominal, but does not include any drag or solar radiation pressure forces. Finally a ``high area-to-mass ratio'' case was investigated where the the same force model as the nominal model was used, however the area-to-mass ratios ($A/m$) of all the target and field objects were inflated. In Fig.~\ref{fig:sensitivityResults} it can be seen that the difference in the frequency of approaches and in the experienced approach distances between the nominal (7x7), low gravity (3x3), and high gravity (20x20) are fairly small. Small differences are to be expected, but the the overall structure of the result is preserved. When comparing the gravity only, nominal, and high $A/M$ cases, we can see that the frequency of approaches on each day is extremely similar and the variation in close approach distances is comparable to what was observed in the comparison between different gravity potentials. Furthermore we notice that the variation between the nominal and High A/M ratio cases was slightly lower than the variation between the gravity only and nominal cases.

\newpage

\end{appendix}

\end{document}